\documentclass[prd, twocolumn, showpacs,aps,nofootinbib,floatfix,amsmath,amssymb]{revtex4}
\usepackage{graphicx}
\usepackage{soul}
\usepackage{color}
\usepackage[mathscr]{euscript}
\usepackage[usenames,dvipsnames]{xcolor}
\usepackage{amsmath}
\begin{document}

\makeatletter
\newbox\slashbox \setbox\slashbox=\hbox{$/$}
\newbox\Slashbox \setbox\Slashbox=\hbox{\large$/$}
\def\pFMslash#1{\setbox\@tempboxa=\hbox{$#1$}
  \@tempdima=0.5\wd\slashbox \advance\@tempdima 0.5\wd\@tempboxa
  \copy\slashbox \kern-\@tempdima \box\@tempboxa}
\def\pFMSlash#1{\setbox\@tempboxa=\hbox{$#1$}
  \@tempdima=0.5\wd\Slashbox \advance\@tempdima 0.5\wd\@tempboxa
  \copy\Slashbox \kern-\@tempdima \box\@tempboxa}
\def\FMslash{\protect\pFMslash}
\def\FMSlash{\protect\pFMSlash}
\def\miss#1{\ifmmode{/\mkern-11mu #1}\else{${/\mkern-11mu #1}$}\fi}

\newcommand{\psum}[1]{{\sum_{ #1}\!\!\!}'\,}
\makeatother

\title{Effects of universal extra dimensions on top-quark electromagnetic interactions}

\author{Javier Monta\~no$^{a}$, H\'ector Novales-S\'anchez$^b$, and J. Jes\'us Toscano$^b$}
\affiliation{$^{a}$CONACYT-Facultad de Ciencias F\'isico Matem\'aticas, Universidad Michoacana de San Nicol\'as de Hidalgo,
Av. Francisco J. M\'ugica s/n, C.~P. 58060, Morelia, Michoac\'an, M\'exico.\\$^b$Facultad de Ciencias F\'isico Matem\'aticas, Benem\'erita Universidad Aut\'onoma de Puebla, Apartado Postal 1152 Puebla, Puebla, M\'exico.}

\begin{abstract}
Universal extra dimensions, presumably observable at some high-energy scale, would modify low-energy observables, being particularly relevant for physical processes forbidden at tree level by the Standard Model. We address the Kaluza-Klein contributions from the 5-dimensional Standard Model to the anomalous magnetic moment and to branching ratios of electromagnetic decays of the top quark. In accordance with present bounds on the compactification scale, contributions to both quantities are found to be at least 3 orders of magnitude below Standard-Model predictions.
\end{abstract}

\pacs{11.10.Kk, 13.40.Em, 13.40.Hq, 14.65.Ha}

\maketitle


\section{Introduction}
\label{Int}
The use of extra dimensions in model building started with the works by G. Nordstr\"om and T. Kaluza, who attempted to unify electromagnetism and gravity by assuming the existence of a spatial extra dimension~\cite{Nordstrom,Kaluza}. Nevertheless, it was O. Klein who realized, for the first time, that compactification could be used to explain the lack of observations of extra dimensions~\cite{Klein}. The ulterior birth of {\it string theory}, as a theory of strong interactions~\cite{Veneziano,Nielsen,KoNi1,KoNi2,Nambu,Susskind1,Susskind2}, would eventually endow great relevance to formulations of extra dimensions. The original string-theory formulation already had this ingredient, as 26 spacetime dimensions were required to ensure unitarity~\cite{Lovelace}. The introduction of fermions in string theory~\cite{Ramond}, which came along with the discovery of {\it supersymmetry}~\cite{Ramond,WeZu}, and the presence of a massless particle of spin 2~\cite{SchSch}, to be identified as the {\it graviton}, were two main elements of {\it superstring theory} that motivated its use to achieve a quantum theory of gravity, always with the complicity of extra dimensions. Remarkably, the critical dimension of superstring theory turned out to be just 10, as it was shown by J. H. Schwarz~\cite{Schwarz}. The introduction of the {\it Green-Schwarz mechanism}~\cite{GS}, to eliminate quantum anomalies arising in string theory, then triggered the {\it first superstring revolution}, during which five consistent superstring formulations were given~\cite{GS1,GHMR1,GHMR2,GHMR3}. Furthermore, a connection, through compactification, between superstring theory, featuring a 6-dimensional {\it Calabi-Yau extra-dimensional manifold}~\cite{Yau}, and 4-dimensional supersymmetry was established~\cite{CHSW}. A {\it second superstring revolution} started with the emergence of the {\it M-theory}, by E. Witten~\cite{Witten}, who showed that the five superstring  formulations known at the time are limits of this single theory, which is a unifying fundamental theory set in 11 spacetime dimensions. The existence of {\it D-branes}, proposed by J. Polchinski~\cite{Polchinski} for the sake of string duality, was a major event. It was also shown that supergravity in 11 dimensions is a low-energy limit of the $M$-theory~\cite{HoWi1,HoWi2}. The {\it ADS/CFT correspondence}, which establishes a duality of 5-dimensional theories of gravity with gauge field theories set in 4 dimensions~\cite{Maldacena}, is a quite important result with remarkable practical advantages regarding nonperturbative physics. Among the events and advances experienced by string theory throughout the years, and a plethora of papers on the matter, we wish to emphasize that its development is the one that got modern physics used to extra dimensions. \\

Considerable interest in the phenomenology of extra dimensions arose because of the works by Antoniadis, Arkani-Hamed, Dimopoulos and Dvali~\cite{A,ADD,AADD}, who, motivated by the {\it hierarchy problem}, proposed the existence of {\it large extra dimensions}, responsible for the observed weakness of the gravitational interaction, at the stunning scale of a millimeter. 
Shortly after, L. Randall and R. Sundrum initiated an important branch of extra-dimensional models, the so-called {\it models of warped extra dimensions}, in which the hierarchy problem was tackled by introducing a spatial extra dimension and assuming that the associated 5-dimensional spacetime is characterized by an {\it Anti-de-Sitter} structure~\cite{RS1,RS2}. The present paper is developed within another well-known extra-dimensional framework, dubbed {\it universal extra dimensions}~\cite{ACD1}, proposed by Appelquist, Cheng and Dobrescu. A field theory with the structure of the 4-dimensional Standard Model (4DSM) is defined, rather, on a spacetime with compact spatial extra dimensions, where all the dynamic variables are assumed to propagate, thus leading to an infinite set of {\it Kaluza-Klein} (KK) {\it modes} per each extra-dimensional field\footnote{Models of universal extra dimensions have been reviewed in Refs.~\cite{HoPr,Servant}.}. In models of universal extra dimensions, conservation of extra-dimensional momentum yields, after integrating out the extra dimensions, 4-dimensional {\it KK effective field theories} in which {\it KK parity} is preserved, with the consequence that, from the perspective of the Feynman-diagrams approach, the very first effects from the KK modes on 4DSM Green's functions (and thus on 4DSM observables) occur at one loop~\cite{ACD1}. Such a feature is particularly relevant in the case of physical observables and processes that, within the context of the 4DSM, can take place exclusively at loop orders.
An appealing characteristic of these models is the small number of added parameters, which are a high-energy {\it compactification scale}, $R^{-1}$, and the number of extra dimensions, $n$. Moreover, universal-extra-dimensions models include {\it dark-matter candidates}~\cite{CMS,DDG,SeTa1,SeTa2}, which would be either the first {\it KK excited mode} of the photon or that corresponding to the neutrino. \\

Within the framework set by the Standard Model in 5 spacetime universal dimensions~\cite{CGNT}, we calculate new-physics effects, induced at one loop by the KK modes, on the anomalous magnetic moment (AMM) of a $u$-type quark. Of particular interest is the AMM of the top quark, whose 4DSM prediction is known to have a large value~\cite{BBGHLMR}. The paper also includes the calculation of KK contributions to the flavor-changing electromagnetic decay $u_\alpha^{(0)}\to A^{(0)}_\mu u_\beta^{(0)}$, with the greek indices $\alpha$ and $\beta$ labeling quark flavors and where all the initial- and final-state fields are assumed to be {\it KK zero modes}, that is, dynamic variables of the 4DSM. In the 4DSM, all these quantities receive contributions from loop Feynman diagrams, exclusively. Both calculations performed in the present paper comprehend the contributions from the whole set of {\it KK excited modes} of the KK effective Lagrangian emerged from the 5-dimensional Standard Model. Thus contributing one-loop diagrams with vector, pseudo-Goldstone and scalar KK excited modes circulating in loops are taken into account. Our analytic results are free of ultraviolet divergences
and they decouple as $R^{-1}\to\infty$. Moreover, by assuming that $R^{-1}$ is large, we find that dominant effects are small, being suppressed by a squared factor of the compactification scale. The analysis and estimation of the KK contributions to the top-quark AMM shows that the new-physics effects are smaller than those from the 4DSM by 3 to 4 orders of magnitude, as long as a compactification scale in the range $1.4\,{\rm TeV}<R^{-1}<5\,{\rm TeV}$ is considered. The impact of the extra-dimensional physics on the branching ratios ${\rm Br}(u_\alpha^{(0)}\to A^{(0)}_\mu u_\beta^{(0)})$ is also estimated. Contributions to such flavor-changing decays are further suppressed by the {\it Glashow-Iliopoulos-Maiani} (GIM) {\it mechanism}~\cite{GIMmech}. In particular, the new-physics contribution to ${\rm Br}(t^{(0)}\to A^{(0)}_\mu c^{(0)})$ turns out to be smaller than that from the 4DSM~\cite{EHS} by 3 to 4 orders of magnitude if $1.4\,{\rm TeV}<R^{-1}<5\,{\rm TeV}$. We compare our estimations for this decay with previous reports~\cite{KDandJ}. \\

Throughout Section~\ref{gth}, the KK effective theory that arises from the extra-dimensional Standard Model is discussed. The Yang-Mills and Higgs sectors are addressed in the general context of $n$ extra dimensions, as their particularization to $n=1$ is straightforward. The structure of the extra-dimensional fermion sector, dictated by Dirac spinors, depends on the number of extra dimensions, so the corresponding lagrangian terms are discussed for the specific case of 1 extra dimension. Analytical calculations of the $u$-type-quark electromagnetic vertex and the flavor-changing decay process we are interested in are presented in Section~\ref{anacalc}, where consistency of the results is emphasized and discussed. Furthermore, a large-compactification-scale scenario is considered, from which leading KK contributions are derived. The implementation of the expressions previously obtained is carried out in Section~\ref{nums}, where numerical analyses and discussions on the contributions from the whole KK theory are presented. We end the paper by presenting our conclusions in Section~\ref{concs}.\\


\section{Extra-dimensional Standard Model and its Kaluza-Klein theory}
\label{gth}
The definition of field theories, including those aimed at extending the Standard Model, rely on the choice of {\it symmetries} and {\it dynamic variables}~\cite{Wudka}. While a variety of symmetries relevant to this purpose is available, {\it spacetime} and {\it gauge symmetries}, in particular, turn out to be essential elements for the definition of field-theory descriptions. For instance, even though most models are set under the assumption that Lorentz symmetry holds, effective field theories with the ingredient of Lorentz-invariance nonconservation, inspired by the spontaneous breaking of Lorentz symmetry in string-theory formulations~\cite{KoSa,KoPo} and by the occurrence of Lorentz violation in {\it noncommutative field theory}~\cite{CHKLO}, have been propounded~\cite{CoKo1,CoKo2,Kostelecky}. On the other hand, the choice of the gauge-symmetry group has more often become the defining trademark of new-physics models. This is, for instance, the case of models that involve {\it left-right symmety}~\cite{PaSa,MoPa1,MoPa2,SenMo1,SenMo2,SenMo3}, which are based on the gauge group SU(2)$_L\times$SU(2)$_R\times$U(1)$_{B-L}$. Moreover, the main feature of the so-called {\it 331 models}~\cite{PlPi,Frampton} is a particular gauge group, in this case SU(3)$_{C}\times$SU(3)$_L\times$U(1)$_X$. Also, theories of {\it grand unification}, based on the symmetry group SU(5), were explored and thoroughly discussed~\cite{GeGl,GQW}. The possibility of defining field theories on spacetime manifolds with extra dimensions has opened alternative paths to go beyond the 4DSM\footnote{Field-theory models of extra dimensions extend the 4DSM in the direction of spacetime symmetry groups.}. The framework for the present investigation is the {\it Standard Model in 5 dimensions}, by which we mean some sort of replica of the 4DSM, but with all its field content and symmetries defined on the spacetime with the extra dimension. We assume the extra dimension to be spatial-like and universal. In this section, we briefly discuss this extra-dimensional Standard Model, with focus on those aspects that are relevant for the phenomenological calculation that we are about to tackle. 
Let us emphasize that the structure of tensors, in contrast with that of spinors, does not depend on the dimension on which they are defined~\cite{TdC1,TdC2}. We take advantage of this by developing our discussion on the gauge and scalar sectors in the general context of $n$ extra dimensions, which, of course, can be straightforwardly particularized to the case $n=1$. Fermion sectors, on the other hand, are developed for the case of only 1 extra dimension.
We spare the reader from the whole bunch of specific details characterizing this formulation, and suggest Refs.~\cite{CGNT,NT,LMNT1,LMNT2} for more detailed discussions on the matter.\\

\subsection{Gauge and scalar sectors}
In this subsection, we discuss the gauge and scalar sectors of the $(4+n)$-dimensional Standard Model. 
First consider, in general, a spacetime comprising 1 time-like dimension and $3+n$ spatial-like dimensions. Assume that, at some high-energy scale (short distances), this spacetime can be characterized by a $(4+n)$-dimensional manifold ${\cal M}^{4+n}$ with metric $g_{MN}={\rm diag}(1,-1,\ldots,-1)$. Here and in what follows, capital spacetime indices, like $M,N$, take the values\footnote{Note that a convention in which the first extra dimension is labeled by $M,N=5$ has been used.} $0,1,2,3,5,\ldots,4+n$. Think of a field-theory formulation defined on this manifold and governed by the extra-dimensional Poincar\'e group ${\rm ISO}(1,3+n)$. Imagine a process in which we study nature at increasing distances, starting from the aforementioned high-energy scale. While at certain range of high-energy scales the proper field-theory description is invariant with respect to ${\rm ISO}(1,3+n)$, we assume that the afore-described process leads us to a lower-energy scale at which $n$ out of the $3+n$ spatial-like dimensions display a compact nature. It is said that these $n$ dimensions are {\it compactified}. At this energy scale, also called {\it compactification scale}, an appropriate field-theory formulation is not governed by the $(4+n)$-dimensional Poincar\'e group anymore.
Besides being a theoretical possibility, the ingredient of compactification has the practical use of explaining the absence of measurements of extra dimensions~\cite{DFK,BDDM,BKP,BBBKP,HaWe,DGKN,CTW,APS,FKKMP}. A variety of geometries, suitable for compactified extra dimensions, are available~\cite{DoPo,LNielse,CDL,MNSY,CDD}. 
In general, all the symmetries and dynamic variables constituting a sensible physical description at the lower-energy scale are different. 
\\

From here on, $x$ and $\bar{x}$ denote, respectively, coordinates for the 4 standard spacetime dimensions and the $n$ extra dimensions. In this context, consider any dynamic variable, generically denoted by $\chi(x,\bar{x})$, which we assume to be a tensor field with respect to the $(4+n)$-dimensional Lorentz group. Now we break ${\rm ISO}(1,3+n)$ invariance by implementing compactification, for which we assume that each extra dimension is compactified on an orbifold $S^1/Z_2$ characterized by a radius $R_j$, with $j=1,2,\ldots,n$. This compactification scheme induces periodicity properties on $\chi$, with respect to the extra-dimensional coordinates $\bar{x}$. Moreover, it allows for the assignment to $\chi$ of definite-parity properties, with respect to reflections $\bar{x}\to-\bar{x}$. The field $\chi$ is then expanded in terms of a {\it complete set of orthogonal functions} $\{ f^{(\underline{k})}_{\rm E}(\bar{x}), f^{(\underline{k})}_{\rm O}(\bar{x}) \}$, which exclusively depend on extra-dimensional coordinates $\bar{x}$. Such an expansion runs over the multi-index $(\underline{k})=(k_1,k_2,\ldots,k_n)$, where any $k_j$ is an integer number. Furthermore, the  labels ``E'' and ``O'' mean that the corresponding function is even or odd under $\bar{x}\to-\bar{x}$. Once this expansion has been implemented on every field $\chi$, the extra-dimensional coordinates $\bar{x}$ no longer label degrees of freedom, which are now characterized by the {\it KK index} $(\underline{k})$. Each function $f^{(\underline{k})}_{\rm E}$ or $f^{(\underline{k})}_{\rm O}$, in the $\chi$ expansion, lies multiplied by a coefficient $\chi^{(\underline{k})}_{\rm E}(x)$ or $\chi^{(\underline{k})}_{\rm O}(x)$, respectively. These fields, which depend only on the 4-dimensional coordinates $x$ of the non-compact spacetime dimensions, are the new 4-dimensional dynamic variables, the KK modes, suitable for the physical description after compactification. Assume that a constant function, $f^{(\underline{0})}$, belongs to $\{ f^{(\underline{k})}_{\rm E}, f^{(\underline{k})}_{\rm O} \}$. This function, trivially even under $\bar{x}\to-\bar{x}$, comes with a 4-dimensional field $\chi^{(\underline{0})}(x)$.
Fields $\chi^{(\underline{0})}$, known as KK zero modes, are identified as the dynamic variables that constitute the low-energy description. The fields $\chi^{(\underline{k})}$, with $(\underline{k})\ne(\underline{0})$, are known as KK excited modes, and are interpreted as degrees of freedom that reflect the presence of extra dimensions from a 4-dimensional effective-theory viewpoint. Thus, only $\bar{x}$-even extra-dimensional fields $\chi$ yield low-energy dynamic variables. \\

The specific set $\{ f^{(\underline{k})}_{\rm E}, f^{(\underline{k})}_{\rm O} \}$ is determined, in part, by the geometry of the extra dimensions, but an extra-dimensional observable is also required to this end. This is the case of the {\it Casimir invariants} of ${\rm ISO}(n)$, among which we choose $\bar{P}^2$, with $\bar{P}$ the momentum operator along the extra dimensions. 
Being a hermitian operator, $\bar{P}^2$ has an associated set of eigenkets $\{ | \bar{p}^{(k)} \rangle \}$, with eigenvalues $(\bar{p}^{(\underline{k})})^2=\bar{p}^{(\underline{k})}\cdot\bar{p}^{(\underline{k})}$. The $\bar{P}^2$ eigenkets then define $\{ f^{(\underline{k})}_{\rm E}, f^{((\underline{k})}_{\rm O} \}$ from the wave-function relations $f^{(\underline{k})}_{\rm E,O}=\langle  \bar{x} | \bar{p}^{(\underline{k})} \rangle$.
Using such relations, together with appropriate boundary conditions, $f^{(\underline{k})}_{\rm E}$ and $f^{(\underline{k})}_{\rm O}$ are determined to be normalized trigonometric functions, so that the field $\chi$ is Fourier expanded,
with the following two disjoint cases:
\\ \\
{\it Even parity:}
\begin{eqnarray}
\chi(x,\bar{x})&=&\frac{1}{\sqrt{(2\pi)^n{\cal R}}}\,\,\chi^{(\underline{0})}(x)
\nonumber \\&& 
+\sum_{(\underline{k})}\sqrt{\frac{2}{(2\pi)^n{\cal R}}}\,\,\chi_{\rm E}^{(\underline{k})}(x)\cos\{\bar{p}^{(\underline{k})}\cdot\bar{x}\},
\label{efields}
\end{eqnarray}
\\
{\it Odd parity:}
\begin{equation}
\chi(x,\bar{x})=\sum_{(\underline{k})}\sqrt{\frac{2}{(2\pi)^n{\cal R}}}\,\,\chi_{\rm O}^{(\underline{k})}(x)\sin\{\bar{p}^{(\underline{k})}\cdot\bar{x}\}.
\label{ofields}
\end{equation}
In these equations, we denoted ${\cal R}=R_1\,R_2\cdots R_n$. 
We have defined discrete extra-dimensional momenta $\bar{p}^{(\underline{k})}=(k_1/R_1,k_2/R_2,\ldots,k_n/R_n)$ as well. The symbol $\sum_{(\underline{k})}=\sum_{k_1}\sum_{k_2}\cdots\sum_{k_n}$ represents a multiple sum that runs over every discrete vector $(\underline{k})$ labeling an independent field $\chi^{(\underline{k})}$, with the additional restriction that $(\underline{k})\ne(\underline{0})=(0,0,\ldots,0)$. 
Note that such an effective theory, often referred to as {\it KK theory}, is defined only in 4 dimensions of spacetime: once the extra dimensions have been compactified and the dynamic variables Fourier expanded, the whole dependence on extra dimensional coordinates in the action $S^{\rm SM}_{4+n}=\int d^{4+n}x\,{\cal L}^{\rm SM}_{4+n}(x,\bar{x})$ lies within trigonometric functions, which can be straightforwardly integrated out, leading to a Lagrangian ${\cal L}^{\rm SM}_{\rm KK}(x)=\int d^n\bar{x}\,{\cal L}^{\rm SM}_{4+n}(x,\bar{x})$ defined in 4 spacetime dimensions.\\

We start by considering a theory set on a $(4+n)$-dimensional spacetime with those features previously described. We also assume that such a formulation is invariant with respect to the extra-dimensional gauge group ${\rm SU}(3,{\cal M}^{4+n})_C\times{\rm SU}(2,{\cal M}^{4+n})_L\times{\rm U}(1,{\cal M}^{4+n})_Y$. The present discussion develops around gauge symmetry with respect to the subgroup ${\rm SU}(2,{\cal M}^{4+n})_L\times{\rm U}(1,{\cal M}^{4+n})_Y$, which introduces 4 gauge fields, denoted as ${\cal W}^j_M(x,\bar{x})$ and ${\cal B}_M(x,\bar{x})$, where $j=1,2,3$ is a gauge index. Besides the usual {\it Yang-Mills sector}, ${\cal L}^{\rm YM}_{4+n}$, given exclusively in terms of these gauge fields, we assume the presence in the theory of a {\it scalar sector}, ${\cal L}^{\rm S}_{4+n}$, defined in terms of an ${\rm SU}(2,{\cal M}^{4+n})_L$ doublet $\Phi(x,\bar{x})$, with hypercharge $Y_\Phi$. We also assume that this scalar sector includes a scalar potential, referred to as $V(\Phi,\Phi^\dag)$. Our theory contains the set of lagrangian terms
\begin{eqnarray}
{\cal L}^{\rm YM}_{4+n}+{\cal L}^{\rm S}_{4+n}&=&-\frac{1}{4}{\cal W}^j_{MN}{\cal W}^{jMN}-\frac{1}{4}{\cal B}_{MN}{\cal B}^{MN}
\nonumber \\&&
+(D_M\Phi)^\dag(D^M\Phi)-V(\Phi,\Phi^\dag).
\end{eqnarray}
This expression involves  the ${\rm SU}(2,{\cal M}^{4+n})$ Yang-Mills curvature components ${\cal W}^j_{MN}$ and the ${\rm U}(1,{\cal M}^{4+n})_Y$ tensor ${\cal B}_{MN}$, both defined as usual~\cite{PeSch}, and the ${\rm SU}(2,{\cal M}^{4+n})_L\times{\rm U}(1,{\cal M}^{4+n})_Y$ covariant derivative $D_M$, given in the representation of doublets. Coupling constants corresponding to the groups ${\rm SU}(2,{\cal M}^{4+n})_L$ and ${\rm U}(1,{\cal M}^{4+n})_Y$, which we respectively denote by $g_{4+n}$ and $g'_{4+n}$, are dimensionful, with units $({\rm mass})^{-n/2}$. The scalar potential is defined as 
\begin{equation}
V(\Phi,\Phi^\dag)=-\mu^2\Phi^\dag\Phi+\lambda_{4+n}(\Phi^\dag\Phi)^2,
\end{equation}
where $\mu^2$ is a positive quantity, with units $({\rm mass})^2$ whereas the units of the coupling constant $\lambda_{4+n}$ are $({\rm mass})^{-n}$.\\

Once defined the lagrangian terms ${\cal L}^{\rm YM}_{4+n}+{\cal L}^{\rm S}_{4+n}$, we implement compactification through a couple of canonical transformations to go from the $(4+n)$-dimensional perspective to the KK effective theory, set in 4 spacetime dimensions. Due to compactification, ${\cal W}^j_M$ and ${\cal B}_M$, which at first were $(4+n)$-vectors of ${\rm SO}(1,3+n)$, are split into the ${\rm SO}(1,3)$ 4-vectors ${\cal W}^j_\mu$ and ${\cal B}_\mu$, and the two sets of ${\rm SO}(1,3)$ scalar fields $\{{\cal W}^j_5,{\cal W}^j_6,\ldots,{\cal W}^j_{4+n}\}$ and $\{{\cal B}_5,{\cal B}_6,\ldots,{\cal B}_{4+n}\}$. From now on, we utilize greek indices like $\mu,\nu=0,1,2,3$ to denote 4-dimensional Lorentz indices and use indices $\bar{\mu},\bar{\nu}=5,6,\ldots,4+n$ to label extra-dimensions coordinates. The implementation of the afore-alluded splitting is a canonical transformation that maps covariant objects of ${\rm SO}(1,3+n)$ into covariant objects of ${\rm SO}(1,3)$~\cite{LMNT1,LMNT2}. In order to land on a KK effective Lagrangian consistently comprising the low-energy theory, namely the 4DSM, we assume that ${\cal W}^j_\mu$ and ${\cal B}_\mu$ are both even with respect to $\bar{x}\to-\bar{x}$ , but the definite parity of the scalar fields ${\cal W}^j_{\bar{\mu}}$ and ${\cal B}_{\bar{\mu}}$ under such a transformation is odd. Furthermore, we assume that the corresponding parity of the extra-dimensional scalar doublet $\Phi$ is even. Eqs.~(\ref{efields}) and (\ref{ofields}) embody a second canonical transformation~\cite{LMNT1,LMNT2} which, after implementation, yields sets of KK modes, recognized as dynamic variables of the KK Lagrangian. The whole set of KK modes from the gauge and scalar sectors, together with the two canonical maps generating them, is illustrated in Eq.~(\ref{canmaps}):
\begin{equation}
\begin{array}{rl}
{\cal W}^j_M(x,\bar{x})&\mapsto 
\left\{
\begin{array}{l}
{\cal W}^j_\mu(x,\bar{x})\mapsto
W^{(\underline{0})j}_\mu(x),\,
W^{(\underline{k})j}_\mu(x)
\vspace{0.3cm} \\ 
{\cal W}^j_{\bar{\mu}}(x,\bar{x})\mapsto W^{(\underline{k})j}_{\bar{\mu}}(x)
\end{array}
\right.
\\ \\
{\cal B}_M(x,\bar{x})&\mapsto
\left\{
\begin{array}{l}
{\cal B}_\mu(x,\bar{x})\mapsto 
B^{(\underline{0})}_\mu(x),\,
B^{(\underline{k})}_\mu(x)
\vspace{0.3cm} \\
{\cal B}_{\bar{\mu}}(x,\bar{x})\mapsto B^{(\underline{k})}_{\bar{\mu}}(x)
\end{array}
\right.
\\ \\
\Phi(x,\bar{x})&\mapsto\Phi(x,\bar{x})\mapsto 
\Phi^{(\underline{0})}(x),\,
\Phi^{(\underline{k})}(x)
\end{array}
\label{canmaps}
\end{equation}
\\

After usage of the canonical maps, and subsequent straightforward integration of the extra dimensions in the action, the 4-dimensional KK lagrangian terms ${\cal L}^{\rm YM}_{\rm KK}+{\cal L}^{\rm S}_{\rm KK}=\int d^n\bar{x} ({\cal L}^{\rm YM}_{4+n}+{\cal L}^{\rm S}_{4+n})$ arise. The effective-theory description provided by ${\cal L}^{\rm YM}_{\rm KK}+{\cal L}^{\rm S}_{\rm KK}$ is characterized by low-energy symmetries, among which 4-dimensional Poincar\'e symmetry is central. With respect to the Lorentz group ${\rm SO}(1,3)$, the KK fields $W^{(\underline{0})j}_\mu$, $W^{(\underline{k})j}_\mu$, $B^{(\underline{0})}_\mu$, and $B^{(\underline{k})}_\mu$ are 4-vectors, whereas $W^{(\underline{k})j}_{\bar{\mu}}$ and $B^{(\underline{k})}_{\bar{\mu}}$, as well as the components of $\Phi^{(\underline{0})}$ and $\Phi^{(\underline{k})}$, are scalars. About gauge symmetry, the effectuation of compactification entails the occurrence of {\it hidden symmetries}~\cite{LMNT1}. Originally characterized by the gauge group ${\rm SU}(2,{\cal M}^{4+n})_L\times{\rm U}(1,{\cal M}^{4+n})_Y$, set on $4+n$ spacetime dimensions, the $(4+n)$-dimensional Standard Model has been mapped into a KK theory that manifests gauge invariance corresponding to the low-energy group ${\rm SU}(2,{\cal M}^{4})_L\times{\rm U}(1,{\cal M}^{4})_Y$, defined on 4 spacetime dimensions. Collaterally, the gauge transformations of the $(4+n)$-dimensional connections ${\cal W}^j_M$ and ${\cal B}_M$ split into two disjoint sets of 4-dimensional gauge transformations~\cite{NT,CGNT,LMNT1,LMNT2}: {\it standard gauge transformations}, which constitute the gauge group ${\rm SU}(2,{\cal M}^{4})_L\times{\rm U}(1,{\cal M}^{4})_Y$ and with respect to which KK zero modes $W^{(\underline{0})j}_\mu$ and $B^{(\underline{0})}_\mu$ behave as gauge fields; {\it nonstandard gauge transformations}, under which KK excited modes $W^{(\underline{k})j}_\mu$ and $B^{(\underline{k})}_\mu$ are sort of like gauge fields, in the sense that they follow a transformation that is reminiscent of a gauge transformation, but which does not correspond to ${\rm SU}(2,{\cal M}^{4})_L\times{\rm U}(1,{\cal M}^{4})_Y$. 
Furthermore, let us remark that
KK excited modes $W^{(\underline{k})j}_{\mu}$ and $B^{(\underline{k})}_{\mu}$ are not connections of 
${\rm SU}(2,{\cal M}^{4})_L\times{\rm U}(1,{\cal M}^{4})_Y$, but, rather, they transform as matter fields, in the adjoint representation of this group~\cite{NT,CGNT,LMNT1,LMNT2}. Hence gauge symmetry governing the KK effective theory does not forbid the presence of mass terms for vector KK-excited-mode fields $W^{(\underline{k})j}_\mu$ and $B^{(\underline{k})}_\mu$. This is to be contrasted with the situation of KK zero modes $W^{(\underline{0})j}_\mu$ and $B^{(\underline{0})}_\mu$, which, being 4-dimensional gauge fields, are restricted to be massless. All scalar KK modes, on the other hand, transform as matter fields with respect to both sets of gauge transformations. In particular, the scalar fields $W^{(\underline{k})j}_{\bar{\mu}}$ and $B^{(\underline{k})}_{\bar{\mu}}$ are, in spite of their gauge origin, matter fields under ${\rm SU}(2,{\cal M}^{4})_L\times{\rm U}(1,{\cal M}^{4})_Y$, which opens the possibility for they to become massive. Moreover, the zero mode $\Phi^{(\underline{0})}$ and the excited modes $\Phi^{(\underline{k})}$ are ${\rm SU}(2,{\cal M}^4)_L$ doublets with hypercharge $Y_\Phi$. \\

A remarkable outcome of compactification is the occurrence of mass terms for the whole set of KK excited modes, which we refer to as the {\it KK mass-generating mechanism}, or {\it KK mechanism} for short. Any KK excited mode $\chi^{(\underline{k})}$, labeled by an specific multi-index $(\underline{k})$, acquires a {\it KK mass},
\begin{equation}
m_{(\underline{k})}=\sqrt{\left(\frac{k_1}{R_1}\right)^2+\left(\frac{k_2}{R_2}\right)^2+\cdots+\left(\frac{k_n}{R_n}\right)^2},
\end{equation}
no matter whether the original $(4+n)$-dimensional dynamic variable $\chi$ is a gauge or a scalar field. Mass terms for vector KK fields $W^{(\underline{k})j}_\mu$ and $B^{(\underline{k})}_\mu$ are found in a straightforward manner, which also happens with the components of the KK doublets $\Phi^{(\underline{k})}$. This contrasts with the case of scalar KK excited modes belonging to the set $\{W^{(\underline{k})j}_5,W^{(\underline{k})j}_6,\ldots,W^{(\underline{k})j}_{4+n}\}$, with fixed KK index $(\underline{k})$, since mixings among all the fields of such a set take place. And the same goes for the set of scalar fields $\{B^{(\underline{k})}_5,B^{(\underline{k})}_6,\ldots,B^{(\underline{k})}_{4+n}\}$, for any fixed $(\underline{k})$. Mixings for both sets of scalar KK modes are given by the same real and symmetric mixing matrix ${\cal M}^{(\underline{k})}$, with entries ${\cal M}^{(\underline{k})}_{\bar{\mu}\bar{\nu}}=m_{(\underline{k})}^2\,\delta_{\bar{\mu}\bar{\nu}}-\overline{p}^{(\underline{k})}_{\bar{\mu}}\overline{p}^{(\underline{k})}_{\bar{\nu}}$. Things can be conveniently arranged so that, denoting the orthogonal-diagonalization matrix of ${\cal M}^{(\underline{k})}$ by $R^{(\underline{k})}$, the diagonalization\footnote{In this equation, the repeated index $\bar{\mu}$ is not summed.} $(R^{(\underline{k}){\rm T}}{\cal M}^{(\underline{k})}R^{(\underline{k})})_{\bar{\mu}\bar{\nu}}=m_{(\underline{k})}^2\delta_{\bar{\mu}\bar{\nu}}(1-\delta_{\bar{\mu},4+n})$ can be executed. Note that the eigenvalues of $R^{(\underline{k}){\rm T}}{\cal M}^{(\underline{k})}R^{(\underline{k})}$ are $m_{(\underline{k})}^2$, except for that corresponding to $\bar{\mu}=\bar{\nu}=4+n$, which is 0. The null eigenvalue implies the presence of massless scalar KK excited modes, which we denote as $W^{(\underline{k})j}_{\rm G}$ and $B^{(\underline{k})}_{\rm G}$, and which turn out to be kind of pseudo-Goldstone bosons, in the sense that a nonstandard gauge transformation that eliminates them from the theory exists, indicating that such fields represent unphysical degrees of freedom. After the change of basis, induced by diagonalization, the resulting mass-eigenfields basis involves, for any fixed KK index $(\underline{k})$, the sets of scalar fields $\{ W'^{(\underline{k})j}_1,W'^{(\underline{k})j}_2,\ldots,W'^{(\underline{k})j}_{n-1} \}$ and $\{ B'^{(\underline{k})}_1,B'^{(\underline{k})}_2,\ldots,B'^{(\underline{k})}_{n-1} \}$, all of them with mass $m_{(\underline{k})}$, and the aforementioned pseudo-Goldstone bosons $W^{(\underline{k})j}_{\rm G}$ and $B^{(\underline{k})}_{\rm G}$. Such a diagonalization, with the associated set of resulting fields, is illustrated in Eq.~(\ref{diagYM}):
\begin{equation}
\begin{array}{rl}
\big\{W^{(\underline{k})j}_{\bar{\mu}}\big\}_{\bar{\mu}=5}^{4+n}\,
&\mapsto
W^{(\underline{k})j}_{\rm G},\,\big\{W'^{(\underline{k})j}_{\bar{n}}\big\}_{\bar{n}=1}^{n-1}
\vspace{0.3cm} \\
\big\{B^{(\underline{k})}_{\bar{\mu}}\big\}_{\bar{\mu}=5}^{4+n}&\mapsto
B^{(\underline{k})}_{\rm G},\,\big\{B'^{(\underline{k})}_{\bar{n}}\big\}_{\bar{n}=1}^{n-1}\,
\end{array}
\label{diagYM}
\end{equation}
\\

The KK-mechanism procedure bears features that evoke the {\it Englert-Higgs mechanism} (EHM)~\cite{EnBr,PWHiggs1,PWHiggs2}, responsible for mass generation in the 4DSM. A gauge-invariant scalar potential with degenerate minima, which can be characterized by the set of points constituting a hypersphere with radius determined by some {\it vacuum expectation value}, is the starting point of the EHM. The hypersphere points are connected to each other by gauge symmetry associated to some group $G$, of dimension $d_G$, so they represent physically equivalent vacuum states. To pick one of such minima, a specific constant vector, associated to a particular point on the hypersphere, is taken. Such a choice induces a map $G\mapsto H$ that breaks the gauge group $G$ down into one of its subgroups $H\subset G$, of dimension $d_H$. This procedure breaks $d_G-d_H$ generators of $G$, thus leaving $d_H$ unbroken generators. Any gauge field pointing towards the direction defined by a broken generator becomes massive, which yields the emergence of an associated pseudo-Goldstone boson. Hence the resulting set of fields involves $d_G-d_H$ massive gauge fields and the same number of pseudo-Goldstone bosons. On the other hand, the $d_H$ gauge fields pointing along directions corresponding to unbroken generators remain massless and are the connections of the gauge subgroup $H$, which governs the resultant theory. So, the remaining $d_H$ unbroken generators define the Lie algebra of $H$. Regarding the KK mechanism, note that the complete set of orthogonal functions $\{ f^{(\underline{k})}_{\rm E}, f^{((\underline{k})}_{\rm O} \}$ is not unique. In order to pick a particular set, an extra-dimensional observable, namely the ${\rm ISO}(n)$ Casimir invariant $\bar{P}^2$, was utilized, though other options, yielding different sets $\{ f^{(\underline{k})}_{\rm E}, f^{((\underline{k})}_{\rm O} \}$, are available. The definition of a such a set determines a canonical transformation that maps the extra-dimensional fields into the 4-dimensional KK modes, thus defining a theory governed by 4-dimensional Poincar\'e invariance. In other words, the map ${\rm ISO}(1,3+n)\to{\rm ISO}(1,3)$ takes place. Furthermore, while the extra-dimensional theory is invariant with respect to some gauge group defined on the spacetime with extra dimensions, after this map the resulting theory is manifestly governed a gauge group characterized by the same generators, though defined in 4 dimensions. Consider a connection of the gauge group in extra dimensions and assume that it has been mapped into its set of KK modes. The corresponding KK zero mode points along the direction of the constant function $f^{(\underline{0})}=\langle \bar{x} | \bar{p}^{(\underline{0})} \rangle$, determined by the $\bar{P}^2$ eigenket $| \bar{p}^{(\underline{0})} \rangle$. The zero mode remains massless and transforms as a gauge field with respect to the 4-dimensional gauge group, which resembles what happens with the gauge fields pointing towards the directions associated to unbroken generators in the EHM. Moreover, the remaining $2^n-1$ eigenkets $| \bar{p}^{(\underline{k})} \rangle$, with $(\underline{k})\ne(\underline{0})$, are analogues of the broken gauge-group generators from the EHM, in the sense that they define independent directions $f^{(\underline{k})}_{\rm O}$ and $f^{(\underline{k})}_{\rm E}$ along which vector fields with masses acquired by the KK mechanism are directed, with the presence of the same number of associated pseudo-Goldstone bosons. It is worth emphasizing that, in contrast with the case of the EHM, the KK mechanism does not involve broken gauge generators, since the the extra-dimensional and the 4-dimensional gauge groups share the same generators.
\\

Implementation of the EHM takes place in the next step: the electroweak group ${\rm SU}(2,{\cal M}^4)_L\times{\rm U}(1,{\cal M}^4)_Y$ is {\it spontaneously broken} down into the electromagnetic group ${\rm U}(1,{\cal M}^4)_e$, for which the potential-minimizing critical point $\Phi^{(\underline{0}){\rm T}}_0=(0,v/ \sqrt{2})$, for the zero-mode doublet $\Phi^{(\underline{0})}$, is chosen, with $v$ the vacuum expectation value of the Higgs field. The 4DSM charged bosons $W^{(\underline{0})\pm}_\mu=(W^{(\underline{0})1}_\mu\mp i\,W^{(\underline{0})2}_\mu)/\sqrt{2}$ and the neutral vector boson $Z^{(\underline{0})}=c_W W^{(\underline{0})3}_\mu-s_W B^{(\underline{0})}_\mu$ respectively get masses $m_{W^{(\underline{0})}}=gv/2$ and $m_{Z^{(\underline{0})}}=m_{W^{(\underline{0})}}/c_W$ by this mean. Here, $g=g_{4+n}/\sqrt{(2\pi)^n{\cal R}}$ is the ${\rm SU}(2,{\cal M}^4)_L$ coupling constant, and the notation $s_W=\sin\theta_W$, $c_W=\cos\theta_W$ has been utilized, with $\theta_W$ the {\it weak mixing angle}. On the other hand, the zero mode $A^{(\underline{0})}_\mu=s_W W^{(\underline{0})3}_\mu+c_W B^{(\underline{0})}_\mu$, to be interpreted as the electromagnetic field, remains massless. Within this context, the Higgs scalar field, $h^{(\underline{0})}$, gets a mass $m_{h^{(\underline{0})}}=\sqrt{2\mu^2}$.\\

The EHM also induces mass-term contributions for KK excited modes, so that, at the end of the day, KK masses are the result of two contributions originated in two mass-generating mechanisms. KK vector fields $W^{(\underline{k})\pm}_\mu=(W^{(\underline{k})1}_\mu\mp iW^{(\underline{k})2}_\mu)/\sqrt{2}$, $Z^{(\underline{k})}_\mu=c_WW^{(\underline{k})3}_\mu-s_WB^{(\underline{k})}_\mu$, and $A^{(\underline{k})}_\mu=s_WW^{(\underline{k})3}_\mu+c_WB^{(\underline{k})}_\mu$ get mass-term contributions adding to those mass terms previously generated by the KK mechanism, thus resulting in masses given by $m_{W^{(\underline{k})}}^2=m_{W^{(\underline{0})}}^2+m_{(\underline{k})}^2$, $m_{Z^{(\underline{k})}}^2=m_{Z^{(\underline{0})}}^2+m_{(\underline{k})}^2$, and $m_{A^{(\underline{k})}}^2=m_{(\underline{k})}^2$, respectively. A mass contribution for the KK scalar field $h^{(\underline{k})}$ is also generated, which turns out to be given by $m^2_{h^{(\underline{k})}}=m^2_{h^{(\underline{0})}}+m^2_{(\underline{k})}$. Moreover, the sets of scalar KK fields $\{ W'^{(\underline{k})\pm}_{1},W'^{(\underline{k})\pm}_{2},\ldots,W'^{(\underline{k})\pm}_{n-1} \}$, $\{ Z'^{(\underline{k})}_{1},Z'^{(\underline{k})}_{2},\ldots,Z'^{(\underline{k})}_{n-1} \}$ and $\{ A'^{(\underline{k})}_{1},A'^{(\underline{k})}_{2},\ldots,A'^{(\underline{k})}_{n-1} \}$ are defined by $W'^{(\underline{k})\pm}_{\bar{n}}=(W'^{(\underline{k})1}_{\bar{n}}\mp iW'^{(\underline{k})2}_{\bar{n}})/\sqrt{2}$, $Z'^{(\underline{k})}_{\bar{n}}=c_WW'^{(\underline{k})3}_{\bar{n}}-s_WB'^{(\underline{k})}_{\bar{n}}$, and $A'^{(\underline{k})}_{\bar{n}}=s_WW'^{(\underline{k})3}_{\bar{n}}+c_WB'^{(\underline{k})}_{\bar{n}}$. These fields respectively acquire masses $m_{W^{(\underline{k})}}$, $m_{Z^{(\underline{k})}}$ and $m_{A^{(\underline{k})}}$. Mass contributions for the remaining scalar KK-excitation fields are generated as well, but bilinear mixings arise, so that a mass-eigenfields basis is to be defined. Excited-mode doublets $\Phi^{(\underline{k}){\rm T}}=(S_W^{(\underline{k})+},(h^{(\underline{k})}+iS_Z^{(\underline{k})})/\sqrt{2})$, involve neutral fields $h^{(\underline{k})}$ and $S_Z^{(\underline{k})}$, and charged scalar fields $S_W^{(\underline{k})\pm}$ as well. The pseudo-Goldstone bosons $W^{(\underline{k})1}_{\rm G}$ and $W^{(\underline{k})2}_{\rm G}$, originated in the $(4+n)$-dimensional Yang-Mills sector, define the charged pseudo-Goldstone bosons $W^{(\underline{k})\pm}_{\rm G}=(W^{(\underline{k})1}_{\rm G}\mp iW^{(\underline{k})2}_{\rm G})/\sqrt{2}$. For any fixed KK index $(\underline{k})$, mixings among pseudo-Goldstone bosons $W^{(\underline{k})\pm}_{\rm G}$ and scalar fields $S_W^{(\underline{k})\pm}$ take place, which is characterized by a hermitian mixing matrix. Then a unitary diagonalization, parametrized by the mixing angle $\xi^{(\underline{k})}=\tan^{-1}(m_{W^{(\underline{0})}}/m_{(\underline{k})})$, yields massless pseudo-Goldstone bosons $G_W^{(\underline{k})\pm}=\cos\xi^{(\underline{k})}W^{(\underline{k})\pm}_{\rm G}\pm i\sin\xi^{(\underline{k})}S_W ^{(\underline{k})\pm}$ and physical scalars $W^{(\underline{k})\pm}=\sin\xi^{(\underline{k})}W^{(\underline{k})\pm}_{\rm G}\mp i\cos\xi^{(\underline{k})}S_W^{(\underline{k})\pm}$ with mass $m_{W^{(\underline{k})}}$. On the other hand, neutral pseudo-Goldstone bosons are defined as $Z^{(\underline{k})}_{\rm G}=c_WW^{(\underline{k})3}_{\rm G}-s_WB^{(\underline{k})}_{\rm G}$ and $A^{(\underline{k})}_{\rm G}=s_WW^{(\underline{k})3}_{\rm G}+c_WB^{(\underline{k})}_{\rm G}$. A mixing involving $Z^{(\underline{k})}_{\rm G}$ and $S_Z^{(\underline{k})}$ emerges, while $A^{(\underline{k})}_{\rm G}$, connected with the electromagnetic field $A^{(\underline{0})}_\mu$ and, so, completely unrelated to the EHM, does not mix. An orthogonal diagonalization, with mixing angle $\eta^{(\underline{k})}=\tan^{-1}(m_{Z^{(\underline{0})}}/m_{(\underline{k})})$, defines the pseudo-Goldstone boson $G^{(\underline{k})}_Z=\sin\eta^{(\underline{k})}S_Z^{(\underline{k})}+\cos\eta^{(\underline{k})}Z^{(\underline{k})}_{\rm G}$ and the massive scalar $Z^{(\underline{k})}=\cos\eta^{(\underline{k})}S_Z^{(\underline{k})}-\sin\eta^{(\underline{k})}Z^{(\underline{k})}_{\rm G}$, with mass $m_{Z^{(\underline{k})}}$. Eqs.~(\ref{kkvec}), (\ref{kksc}), and (\ref{kkscypgb}) have been used to illustrate the resulting KK-excitation field content, in the mass-eigenfields basis.
\\ \\
{\it KK vectors:}
\begin{equation}
\begin{array}{rl}
\left.
\begin{array}{r}
W^{(\underline{k})j}_\mu
\vspace{0.2cm}\\
B^{(\underline{k})}_\mu
\end{array}
\right\}&\mapsto
\left\{
\begin{array}{l}
W^{(\underline{k})\pm}_\mu
\vspace{0.2cm} \\
Z^{(\underline{k})}_\mu
\vspace{0.2cm} \\
A^{(\underline{k})}_\mu
\end{array}
\right.
\end{array}
\label{kkvec}
\end{equation}

\noindent
{\it KK scalars and pseudo-Goldstone bosons:}
\begin{equation}
\begin{array}{rl}
\left.
\begin{array}{r}
W'^{(\underline{k})j}_{\bar{n}}
\vspace{0.3cm}\\
B'^{(\underline{k})}_{\bar{n}}
\end{array}
\right\}&\mapsto
\left\{
\begin{array}{l}
W'^{(\underline{k})\pm}_{\bar{n}}
\vspace{0.2cm} \\
Z'^{(\underline{k})}_{\bar{n}}
\vspace{0.2cm}\\
A'^{(\underline{k})}_{\bar{n}}
\end{array}
\right.
\end{array}
\label{kksc}
\end{equation}
\\

\begin{equation}
\left.
\begin{array}{rl}
\left.
\begin{array}{l}
W^{(\underline{k})j}_{\rm G}
\vspace{0.2cm}
\\
B^{(\underline{k})}_{\rm G}
\end{array}
\right\}
\mapsto&
\left\{
\begin{array}{l}
W^{(\underline{k})\pm}_{\rm G}
\vspace{0.2cm}
\\
Z^{(\underline{k})}_{\rm G}
\vspace{0.2cm}
\\
A^{(\underline{k})}_{\rm G}
\end{array}
\right.
\vspace{0.3cm}
\\
\Phi^{(\underline{k})}
&\left\{
\begin{array}{l}
S^{(\underline{k})\pm}_W
\vspace{0.2cm}
\\
S^{(\underline{k})}_Z
\vspace{0.2cm}
\\
h^{(\underline{k})}
\end{array}
\right.
\end{array}
\right\}
\left\{
\begin{array}{l}
\hspace{0.17cm}h^{(\underline{k})}
\\
\left.
\begin{array}{c}
W^{(\underline{k})\pm}_{\rm G}
\vspace{0.2cm}
\\
S^{(\underline{k})\pm}_W
\end{array}
\right\}
\mapsto
\left\{
\begin{array}{l}
G_W^{(\underline{k})\pm}
\vspace{0.2cm}\\
W^{(\underline{k})\pm}
\end{array}
\right.
\vspace{0.2cm} \\
\left.
\begin{array}{c}
Z^{(\underline{k})}_{\rm G}
\vspace{0.2cm} \\
S^{(\underline{k})}_Z
\end{array}
\right\}
\mapsto
\left\{
\begin{array}{l}
G^{(\underline{k})}_Z
\vspace{0.2cm}\\
Z^{(\underline{k})}
\end{array}
\right.
\vspace{0.2cm} \\
\hspace{0.15cm}A^{(\underline{k})}_{\rm G}
\end{array}
\right.
\label{kkscypgb}
\end{equation}
\\

\subsection{Gauge fixing} 
Field formulations aimed at furnishing sensible quantum descriptions of nature are usually built on the grounds of gauge symmetry. The essence of gauge symmetry resides in the presence of more degrees of freedom than those strictly required by some given system for its description~\cite{HenTe}. Gauge transformations link a whole family of different mathematical configurations which, in order for gauge symmetry to make physical sense, must lead to the exact same physical results. In other words, any observable intended to be genuinely physical must be {\it gauge independent}. Even though gauge symmetry is a main element for the definition of field theories, it turns out that quantization requires {\it gauge fixing} to be carried out, which means to choose a specific gauge, thus resulting in a formulation that is not gauge invariant anymore. \\

Being associated to local symmetry groups, gauge transformations are defined by functions, known as {\it gauge parameters}, which depend on spacetime coordinates. The selection of a set of specific spacetime-dependent functions to play the role of gauge parameters fixes the gauge, establishing a particular gauge configuration. A systematic path to pick a gauge, among the so-called {\it linear gauges}, was developed long ago by the authors of Ref~\cite{FLS}. In their approach, gauge fixing is parametrized by a {\it gauge-fixing parameter}, usually denoted as $\xi$, whose different values correspond to different gauges. In such an approach, the {\it Landau gauge}, $\xi=0$, and the {\it Feynman-'t Hooft gauge}, $\xi=1$, are commonly utilized. Another customary choice is the {\it unitary gauge}, which, in this scheme, is obtained by taking the limit as $\xi\to\infty$. \\

The {\it field-antifield formalism} and the {\it Becchi-Rouet-Stora-Tyutin} (BRST) {\it symmetry} constitute an efficacious mean through which the quantization of gauge systems can be achieved~\cite{GPS,BaVi1,BaVi2,BaVi3,BaVi4,BaVi5,BRS1,BRS2,Tyutin}. In this framework, the field content defining some gauge theory gets systematically extended. First, a set of {\it ghost} and {\it antighost fields} is added to the theory; more precisely, per each gauge parameter participating in the theory, a ghost-antighost pair is introduced. Also, a set of {\it auxiliary fields} is included. Then, a further enlargement of the field content takes place by the incorporation of antifields, one per each field already defined. Moreover, a {\it symplectic structure}, known as {\it the antibracket} is defined, with each field-antifield pair being {\it canonical conjugate variables}. The resultant increased set of fields is then understood to define an {\it extended action}, which is assumed to satisfy the {\it Batalin-Vilkovisky master equation}. BRST transformations, which include gauge transformations, are generated by the extended action, governed by BRST symmetry. Once established the master equation, the main objective is the determination of a {\it proper solution}, which is distinguished from other extended actions by suitable boundary conditions connecting it with the original action, previous to incrementation of the field content. The next goal is gauge fixing, which is nontrivially performed through the definition of a fermionic functional aimed at the elimination of the whole set of antifields. The idea is to kill two birds with one stone by getting rid of antiflields and, collaterally, fix the gauge. This process ends with the emergence of a {\it quantum action}, which depends on general gauge-fixing functions. At this point, gauge invariance has been completely removed in a general framework in which sets of {\it ad hoc} gauge-fixing functions, with minimal restrictions, can be defined to establish a particular gauge configuration.\\

With the above discussion in mind, our next objective is gauge fixing in the KK theory, which we address within the framework of the BRST formalism. The extended-action proper solution for the 4-dimensional gauge group ${\rm SU}(N)$ has been discussed in detail in Ref.~\cite{GPS}, while a generalization to 5 spacetime dimensions and the corresponding KK theory are found in Ref.~\cite{NT}. In this approach, gauge fixing in the Standard Model in 5 dimensions has been discussed in Ref.~\cite{CGNT}. The implementation of these techniques to the $(4+n)$-dimensional Standard Model and its KK effective description yields the quantum Lagrangian, ${\cal L}^{\rm SM}_{\rm QKK}={\cal L}^{\rm SM}_{\rm KK}+{\cal L}_{\rm KK}^{\rm G}+{\cal L}_{\rm KK}^{\rm GF}$. Here ${\cal L}^{\rm SM}_{\rm KK}$ is the KK Lagrangian produced by the whole $(4+n)$-dimensional Standard Model. The lagrangian term ${\cal L}^{\rm G}_{\rm KK}$ is the KK {\it ghost-antighost sector}, given in terms of KK modes of ghost and antighost fields. The last term, ${\cal L}^{\rm GF}_{\rm KK}$, is the {\it gauge-fixing sector}, which we write as ${\cal L}^{\rm GF}_{\rm SM}={\cal L}^{{\rm GF}(\underline{0})}_{\rm SM}+{\cal L}^{{\rm GF}(\underline{k})}_{\rm SM}$. In this expression, ${\cal L}_{\rm SM}^{{\rm GF}(\underline{0})}$ is a gauge-fixing lagrangian term defined, exclusively, by KK zero modes, thus being meant for the specification of a gauge configuration among those defined by the symmetry group ${\rm SU}(2,{\cal M}^4)_L\times{\rm U}(1,{\cal M}^4)_Y$. On the other hand,
\begin{equation}
{\cal L}^{{\rm GF}(\underline{k})}_{\rm SM}=
-\frac{1}{2\xi}\sum_{(\underline{k})}\Big(f^{(\underline{k})j}f^{(\underline{k})j}+f^{(\underline{k})}f^{(\underline{k})}\Big)
\end{equation}
is a gauge-fixing Lagrangian made of both zero- and excited-mode KK fields, and which is defined by gauge-fixing functions $f^{(\underline{k})}$ and $f^{(\underline{k})j}$, with $j=1,2,3$ an ${\rm SU}(2,{\cal M}^4)_L$ gauge index. The purpose of ${\cal L}^{{\rm GF}(\underline{k})}_{\rm SM}$ is to pick a gauge configuration allowed by invariance associated to nonstandard gauge transformations.\\

Symmetry with respect to nonstandard gauge transformations can be removed from the KK effective Lagrangian ${\cal L}^{\rm SM}_{\rm QKK}$ without touching the gauge group ${\rm SU}(2,{\cal M}^4)_L\times{\rm U}(1,{\cal M}^4)_Y$. The trick lies in noticing that only standard gauge transformations are associated to this 4-dimensional gauge group. In this context, a set of gauge-fixing functions $f^{(\underline{k})j}$, $f^{(\underline{k})}$, suitably defined to transform covariantly under ${\rm SU}(2,{\cal M}^4)_L\times{\rm U}(1,{\cal M}^4)_Y$, shall get the job done. So we use the following gauge-fixing functions:
\begin{eqnarray}
f^{(\underline{k})j}&=&{\cal D}^{(\underline{0})jm}_\mu W^{(\underline{k})m\mu}-\xi m_{(\underline{k})}W^{(\underline{k})j}_{\rm G}
\nonumber \\ &&
+ig\xi\Big( \Phi^{(\underline{k})\dag}\frac{\sigma^j}{2}\Phi^{(\underline{0})}-\Phi^{(\underline{0})\dag}\frac{\sigma^j}{2}\Phi^{(\underline{k})} \Big),
\label{gfc1}
\\ \nonumber \\ %
f^{(\underline{k})}&=&\partial_\mu B^{(\underline{k})\mu}-\xi m_{(\underline{k})}B^{(\underline{k})}_{\rm G}
\nonumber \\ &&
+\frac{ig'Y_\phi}{2}\xi\Big( \Phi^{(\underline{k})\dag}\Phi^{(\underline{0})}-\Phi^{(\underline{0})\dag}\Phi^{(\underline{k})} \Big),
\label{gfc2}
\end{eqnarray}
where ${\cal D}^{(\underline{0})jm}_\mu$ is the covariant derivative of ${\rm SU}(2,{\cal M}^{4})_L$, in the adjoint representation. Our choice of functions $f^{(\underline{k})j}$, $f^{(\underline{k})}$, given in Eqs.~(\ref{gfc1}) and (\ref{gfc2}), thus leaves the issue of zero-mode gauge-fixing to the lagrangian term ${\cal L}_{\rm SM}^{{\rm GF}(\underline{0})}$, which is to be used to establish a gauge configuration among those connected by ${\rm SU}(2,{\cal M}^4)_L\times{\rm U}(1,{\cal M}^4)_Y$.\\


\subsection{Yukawa and currents sectors}
\label{YandCsects}
The present subsection is devoted to lagrangian terms in which fermions are involved. Differently from our general treatment of the gauge and scalar sectors, in the sense of the number of extra dimensions, this discussion on fermion sectors is developed in the context of the Standard Model defined on a spacetime with only 1 spatial extra dimension. Under such circumstances, the geometry of the compact extra dimension is assumed to be that of an orbifold $S^1/Z_2$, characterized by a radius $R$.\\

Chirality is not defined for odd-dimensional spacetimes, which includes the case of 1 extra dimension: given the set of 5 gamma matrices $\{\Gamma^\mu=\gamma^\mu,\Gamma^5=i\gamma^5\}$, which satisfy the Dirac algebra $\{ \Gamma^M,\Gamma^N \}=2g^{MN}$ and which we use from here on, a proper chiral matrix, say $\Gamma_6$, does not exist. After compactification, any 5-dimensional spinor $\Psi(x,\bar{x})$ can be Fourier expanded, yielding three types of KK Dirac spinors: a zero mode $\hat{\psi}^{(0)}(x)$; excited modes $\hat{\psi}^{(k)}(x)$, multiplied by cosines; and excited modes $\tilde{\psi}^{(k)}(x)$, which multiply sines. These KK spinors, defined on 4 spacetime dimensions, can be decomposed into chiral spinors as usual. From the transformation law of 5-dimensional spinors under space reflection $\bar{x}\to-\bar{x}$, and because of orbifold compactification, parity-even and parity-odd 5-dimensional spinors are respectively expanded as
\\ \\
{\it Odd parity:}
\begin{eqnarray}
\Psi(x,\bar{x})&=&\frac{1}{\sqrt{2\pi R}}\hat{\psi}^{(0)}_L(x)
+\sum_{k=1}^\infty\frac{1}{\sqrt{\pi R}}\Big[ \hat{\psi}_L^{(k)}(x)\cos\{ \bar{p}^{(k)}\bar{x} \}
\nonumber \\ \nonumber \\ &&
+\tilde{\psi}_R^{(k)}(x)\sin\{ \bar{p}^{(k)}\bar{x} \} \Big],
\label{oddfer}
\end{eqnarray}
\\
{\it Even parity:}
\begin{eqnarray}
\Psi(x,\bar{x})&=&\frac{1}{\sqrt{2\pi R}}\hat{\psi}^{(0)}_R(x)
+\sum_{k=1}^\infty\frac{1}{\sqrt{\pi R}}\Big[ \hat{\psi}_R^{(k)}(x)\cos\{ \bar{p}^{(k)}\bar{x} \}
\nonumber \\ \nonumber \\ &&
+\tilde{\psi}_L^{(k)}(x)\sin\{ \bar{p}^{(k)}\bar{x} \} \Big],
\label{evenfer}
\end{eqnarray}
with extra-dimensional momentum given by $\bar{p}^{(k)}=k/R$.
\\

To establish the lagrangian terms constituting the fermion sector of the Standard Model in 5 dimensions, we first define the spinor-field content. We introduce six ${\rm SU}(2,{\cal M}^5)_L$ doublets
\begin{equation}
L_\alpha=
\left(
\begin{array}{c}
\tilde{\nu}_\alpha(x,\bar{x})
\vspace{0.13cm}\\
\tilde{l}_\alpha(x,\bar{x})
\end{array}
\right),
\hspace{0.2cm}
Q_\beta=
\left(
\begin{array}{c}
\tilde{u}_\beta(x,\bar{x})
\vspace{0.13cm}\\
\tilde{d}_\beta(x,\bar{x})
\end{array}
\right),
\label{ferdoublets}
\end{equation}
with $\alpha=e,\mu,\tau$ and $\beta=u,c,t$. All lepton doublets $L_\alpha$ are assumed to share the same hypercharge $Y^l_L$, with respect to ${\rm U}(1,{\cal M}^5)_Y$. Similarly, an ${\rm U}(1,{\cal M}^5)_Y$ hypercharge $Y^q_L$ is assumed to characterize the three quark doublets $Q_\beta$. We also assume the presence of twelve ${\rm SU}(2,{\cal M}^5)_L$ singlets 
\begin{equation}
\nu_\alpha(x,\bar{x}),\hspace{0.2cm}l_\alpha(x,\bar{x}),\hspace{0.2cm}u_\beta(x,\bar{x}),\hspace{0.2cm}d_\beta(x,\bar{x}),
\label{singlets}
\end{equation}
where, again, $\alpha=e,\mu,\tau$ and $\beta=u,c,t$. Neutrino fields $\nu_\alpha$ are also assumed to be singlets with respect to ${\rm U}(1,{\cal M}^5)_Y$, so these fields are singlets of the whole gauge group ${\rm SU}(2,{\cal M}^5)_L\times{\rm U}(1,{\cal M}^5)_Y$, with the consequence that zero-mode-neutrino masses arise from the EHM~\cite{GiKi}. On the other hand, ${\rm U}(1,{\cal M}^5)_Y$ hypercharge assignments $Y^l_R$, $Y^u_R$, $Y^d_R$ for fields $l_\alpha$, $u_\beta$, $d_\beta$ are respectively assumed.\\

The 5-dimensional {\it currents} and {\it Yukawa sectors} are defined in Eqs.~(\ref{lepsec}) and (\ref{quasec}).
\\ \\
{\it Lepton sector:}
\begin{eqnarray}
{\cal L}^l_{\rm Y}+{\cal L}^l_{\rm C}=
\sum_{\alpha,\beta}\Big[
-\mathscr{Y}^l_{5,\alpha\beta}\bar{L}_\alpha\Phi l_\beta-\mathscr{Y}^\nu_{5,\alpha\beta}\bar{L}_\alpha\tilde{\Phi} \nu_\beta+{\rm H.c.}
\nonumber \\
+\bar{L}_\alpha i\Gamma^MD_ML_\alpha+\bar{l}_\alpha i\Gamma^MD_Ml_\alpha+\bar{\nu}_\alpha i\Gamma^M\partial_M\nu_\alpha
\Big],\nonumber\\
\label{lepsec}
\end{eqnarray}
{\it Quark sector:}
\begin{eqnarray}
{\cal L}^q_{\rm Y}+{\cal L}^q_{\rm C}=
\sum_{\alpha,\beta}\Big[
-\mathscr{Y}^d_{5,\alpha\beta}\bar{Q}_\alpha\Phi\,d_\beta-\mathscr{Y}^u_{5,\alpha\beta}\bar{Q}_\alpha\tilde{\Phi}\,u_\beta+{\rm H.c.}
\nonumber \\
+\bar{Q}_\alpha i\Gamma^MD_MQ_\alpha+\bar{d}_\alpha i\Gamma^MD_Md_\alpha+\bar{u}_\alpha i\Gamma^MD_Mu_\alpha
\Big],\nonumber\\
\label{quasec}
\end{eqnarray}
where $\tilde{\Phi}=i\sigma^2\Phi^*$. The 5-dimensional Yukawa constants $\mathscr{Y}^\nu_{5,\alpha\beta}$, $\mathscr{Y}^l_{5,\alpha\beta}$, $\mathscr{Y}^u_{5,\alpha\beta}$, $\mathscr{Y}^d_{5,\alpha\beta}$, characterizing the Yukawa terms in Eqs.~(\ref{lepsec}) and (\ref{quasec}), are dimensionful, with units $({\rm mass})^{-1/2}$. In what follows, we shorten our notation by utilizing $f=\nu,l,u,d$; so, for instance, the aforementioned Yukawa constants are generically denoted as $\mathscr{Y}^f_{5,\alpha\beta}$. Aiming at a sensible 4-dimensional effective description, a consistent connection with low-energy physics is established through the assumption that ${\rm SU}(2,{\cal M}^4)_Y$ doublets $L_\alpha$ and $Q_\alpha$, defined in Eq.~(\ref{ferdoublets}), have odd parity with respect to $\bar{x}\to-\bar{x}$, which means that their Fourier expansions are provided by Eq.~(\ref{oddfer}). With the same goal in mind, ${\rm SU}(2,{\cal M}^5)_L$ singlets $\nu_\alpha$, $l_\alpha$, $u_\alpha$, and $d_\alpha$, given in Eq.~(\ref{singlets}), are assumed to be parity even, with their Fourier expansions thus determined by Eq.~(\ref{evenfer}). The use of such KK expansions, and the ulterior integration of the extra dimension in the action, yields KK lagrangian terms, where, of course, the dynamic variables are 4-dimensional KK fields. \\

In accordance with Eq.~(\ref{oddfer}), doublets $L_\alpha(x,\bar{x})$ unfold into three kinds of KK ${\rm SU}(2,{\cal M}^4)_L$ doublets, denoted by $L^{(0)}_{\alpha,L}(x)$, $L^{(k)}_{\alpha,L}(x)$, and $L^{(k)}_{\alpha,R}(x)$. The zero-mode doublet $L^{(0)}_{\alpha,L}$, made of 4-dimensional chiral spinors $\nu^{(0)}_{\alpha,L}$ and $l^{(0)}_{\alpha,L}$, is identified as the standard lepton ${\rm SU}(2,{\cal M}^4)_L$ doublet of the 4DSM. KK-excitation lepton doublets $L^{(k)}_{\alpha,L}$ and $L^{(k)}_{\alpha,R}$ combine to define the ${\rm SU}(2,{\cal M}^4)_L$ doublet $L^{(k)}_\alpha=L^{(k)}_{\alpha,L}+L^{(k)}_{\alpha,R}$, with 4-dimensional non-chiral spinor components $\tilde{\nu}^{(k)}_\alpha$ and $\tilde{l}^{(k)}_\alpha$. By inspection of Eq.~(\ref{evenfer}), note that 5-dimensional singlets $\nu_\alpha(x,\bar{x})$ and $l_{\alpha}(x,\bar{x})$ respectively yield KK ${\rm SU}(2,{\cal M}^4)_L$ singlets $\nu^{(0)}_{\alpha,R}(x)$, $\nu^{(k)}_{\alpha,L}(x)$, $\nu^{(k)}_{\alpha,R}(x)$ and $l^{(0)}_{\alpha,R}(x)$, $l^{(k)}_{\alpha,L}(x)$, $l^{(k)}_{\alpha,R}(x)$. Zero modes $l^{(0)}_{\alpha,R}$ play the role of standard ${\rm SU}(2,{\cal M}^4)_L$ singlets, introduced for the 4DSM. Right-handed zero-mode neutrino fields $\nu^{(0)}_{\alpha,R}$, which are sterile with respect to the whole 4-dimensional gauge group ${\rm SU}(2,{\cal M}^4)\times{\rm U}(1,{\cal M}^4)_Y$, allow for neutrino masses in the so-called {\it minimally extended Standard Model}~\cite{GiKi}, in 4 dimensions. Moreover, from the resulting sets of KK-excitation fields, non-chiral spinors $\nu^{(k)}_\alpha=\nu^{(k)}_{\alpha,L}+\nu^{(k)}_{\alpha,R}$ and $l^{(k)}_\alpha=l^{(k)}_{\alpha,L}+l^{(k)}_{\alpha,R}$ are defined. The field content of KK leptons is illustrated in Eqs.~(\ref{ldfc}) and (\ref{lsfc}):
\begin{equation}
L_\alpha(x,\bar{x})
\mapsto
\left\{
\begin{array}{l}
\,\,\,L^{(0)}_{\alpha,L}(x)
\left\{
\begin{array}{c}
\nu^{(0)}_{\alpha,L}(x)
\vspace{0.15cm} \\
l^{(0)}_{\alpha,L}(x)
\end{array}
\right.
\vspace{0.2cm} \\
\left.
\begin{array}{c}
L^{(k)}_{\alpha,L}(x)
\vspace{0.15cm} \\
L^{(k)}_{\alpha,R}(x)
\end{array}
\right\}
\mapsto
L^{(k)}_\alpha(x)
\left\{
\begin{array}{c}
\tilde{\nu}^{(k)}_\alpha(x)
\vspace{0.15cm} \\
\tilde{l}^{(k)}_\alpha(x)
\end{array}
\right.
\end{array}
\right.
\label{ldfc}
\end{equation}

\begin{equation}
\begin{array}{c}
\nu_\alpha(x,\bar{x})
\mapsto
\left\{
\begin{array}{l}
\,\,\,\nu^{(0)}_{\alpha,R}(x)
\vspace{0.2cm} \\
\left.
\begin{array}{c}
\nu^{(k)}_{\alpha,L}(x)
\vspace{0.1cm} \\
\nu^{(k)}_{\alpha,R}(x)
\end{array}
\right\}
\mapsto
\nu^{(k)}_\alpha
\end{array}
\right.
\vspace{0.3cm} \\ 
l_\alpha(x,\bar{x})
\mapsto
\left\{
\begin{array}{l}
\,\,\,l^{(0)}_{\alpha,R}(x)
\vspace{0.2cm} \\
\left.
\begin{array}{c}
l^{(k)}_{\alpha,L}(x)
\vspace{0.1cm} \\
l^{(k)}_{\alpha,R}(x)
\end{array}
\right\}
\mapsto
l^{(k)}_\alpha
\end{array}
\right.
\end{array}
\label{lsfc}
\end{equation}
The discussion for the quark sector goes exactly the same, so we just illustrate the process in Eqs.~(\ref{qdfc}) and (\ref{qsfc}):
\begin{equation}
Q_\alpha(x,\bar{x})
\mapsto
\left\{
\begin{array}{l}
\,\,\,Q^{(0)}_{\alpha,L}(x)
\left\{
\begin{array}{c}
u^{(0)}_{\alpha,L}(x)
\vspace{0.15cm} \\
d^{(0)}_{\alpha,L}(x)
\end{array}
\right.
\vspace{0.2cm} \\
\left.
\begin{array}{c}
Q^{(k)}_{\alpha,L}(x)
\vspace{0.15cm} \\
Q^{(k)}_{\alpha,R}(x)
\end{array}
\right\}
\mapsto
Q^{(k)}_\alpha(x)
\left\{
\begin{array}{c}
\tilde{u}^{(k)}_\alpha(x)
\vspace{0.15cm} \\
\tilde{d}^{(k)}_\alpha(x)
\end{array}
\right.
\end{array}
\right.
\label{qdfc}
\end{equation}

\begin{equation}
\begin{array}{c}
u_\alpha(x,\bar{x})
\mapsto
\left\{
\begin{array}{l}
\,\,\,u^{(0)}_{\alpha,R}(x)
\vspace{0.2cm} \\
\left.
\begin{array}{c}
u^{(k)}_{\alpha,L}(x)
\vspace{0.1cm} \\
u^{(k)}_{\alpha,R}(x)
\end{array}
\right\}
\mapsto
u^{(k)}_\alpha
\end{array}
\right.
\vspace{0.3cm} \\ 
d_\alpha(x,\bar{x})
\mapsto
\left\{
\begin{array}{l}
\,\,\,d^{(0)}_{\alpha,R}(x)
\vspace{0.2cm} \\
\left.
\begin{array}{c}
d^{(k)}_{\alpha,L}(x)
\vspace{0.1cm} \\
d^{(k)}_{\alpha,R}(x)
\end{array}
\right\}
\mapsto
d^{(k)}_\alpha
\end{array}
\right.
\end{array}
\label{qsfc}
\end{equation}
As it happened with the gauge and scalar KK fields, every KK excited fermion mode gets the same mass contribution $m_{(k)}=k/R$ as a consequence of the KK mechanism. It is worth commenting that the corresponding mass-term contributions are not Yukawa like, but they come from the currents sector instead. Also worth of attention is the presence of a wrong sign in KK mass terms of neutrino and up-quark fields; this issue is solved later by following Ref.~\cite{PapaSan}.\\

After compactification, the resultant KK Yukawa sector depends on the zero-mode scalar ${\rm SU}(2,{\cal M}^4)_L$ doublet $\Phi^{(0)}$, thus being directly affected by spontaneous symmetry breaking occurring at the energy scale $v$. The EHM produces, at a first stage, bilinear mixings among fermion zero modes, driven by the dimensionless Yukawa matrices $\mathscr{Y}^f$, with entries $\mathscr{Y}^f_{\alpha\beta}=\mathscr{Y}^f_{5,\alpha\beta}/\sqrt{2\pi R}$. Biunitary diagonalizations $V_L^{f\dag}\mathscr{Y}^fV_R^{f}=\mathscr{Y}'^f$, where $\mathscr{Y}'^f$ is a real diagonal matrix with positive diagonal entries,
take place, thus inducing the changes of bases $f'^{(0)}_{\alpha,L}=(V^{f\dag}_L)_{\alpha\beta}\,f^{(0)}_{\beta,L}$ and $f'^{(0)}_{\alpha,R}=(V^{f\dag}_R)_{\alpha\beta}\,f^{(0)}_{\beta,R}$. While these expressions apply for $f=\nu,l,u,d$, note that each case determines a set of labels upon which the greek index $\alpha$ runs: if $f=\nu$, then $\alpha\equiv j=1,2,3$; $f=l$ corresponds to $\alpha=e,\mu,\tau$; for $f=u$ we have $\alpha=u,c,t$; and for the case $f=d$, the corresponding labels are $\alpha=d,s,b$. This notation is also used for our discussion about KK excited modes, later in this paper. After diagonalization of the Yukawa matrices, non-chiral zero-mode mass eigenfields $f^{(0)}_\alpha=f'^{(0)}_{\alpha,L}+f'^{(0)}_{\alpha,R}$ are defined\footnote{Notice that the prime symbol, present in the mass-eigenfields chiral spinors, has been suppressed in the definition of the non-chiral spinor $f^{(0)}_\alpha$.}, and Dirac mass terms carrying masses $m_{f^{(0)}_\alpha}=\mathscr{Y}'^f_{\alpha} v/\sqrt{2}$ are identified, where $\mathscr{Y}'^f_\alpha\equiv\mathscr{Y}'_{\alpha\alpha}$ are the diagonal entries of the diagonalized Yukawa matrices $\mathscr{Y}'$. \\

The implementation of spontaneous symmetry breaking also affects KK excited-mode fermion fields lying in the Yukawa sector. After compactification, Yukawa terms involving Yukawa constants $\mathscr{Y}^f_{\alpha\beta}$ and fermion KK excited modes $\tilde{f}^{(k)}_\alpha$ and $f^{(k)}_\alpha$ naturally arise.
Among other things, such terms produce, through the EHM, terms that are quadratic in these two types of KK fields, which include mixings.
On the other hand, the biunitary diagonalizations of the Yukawa matrices $\mathscr{Y}^f$, previously defined, trigger changes of bases on such KK fields. The corresponding transformations, given by the same unitary matrices that yield the zero-mode fermion eigenfields, are $\tilde{f}'^{(k)}=V^{f\dag}_L\tilde{f}^{(k)}$ and $f'^{(k)}=V^{f\dag}_Rf^{(k)}$. The effectuation of these changes of bases does not eliminate mixings among the different types of KK excited-mode spinors, which are now $\tilde{f}'^{(k)}_\alpha$ and $f'^{(k)}_\alpha$, so a further diagonalization is to be carried out. For any fixed $(k)$, a fermion mixing is given by a real and symmetric $2\times2$ matrix, ${\cal M}_{f^{(k)}_\alpha}$, which is diagonalized by an orthogonal matrix, $P_{f^{(k)}_\alpha}$, characterized by the mixing angle $\theta_{f_\alpha^{(k)}}=\tan^{-1}\big[(m_{f_\alpha^{(k)}}+m_{(k)})/(m_{f_\alpha^{(k)}}-m_{(k)})\big]^{1/2}$. Here, KK masses given by $m^2_{f^{(k)}_\alpha}=m^2_{f^{(0)}_\alpha}+m^2_{(k)}$ have been defined. This diagonalization yields the field definitions $f^{(k)}_{1,\alpha}=\cos\theta_{f^{(k)}_\alpha} f'^{(k)}_\alpha+\sin\theta_{f^{(k)}_\alpha}\tilde{f}'^{(k)}_\alpha$ and $f'^{(k)}_{2,\alpha}=-\sin\theta_{f^{(k)}_\alpha} f'^{(k)}_\alpha+\cos\theta_{f^{(k)}_\alpha}\tilde{f}'^{(k)}_\alpha$. The eigenvalues of the mixing matrix ${\cal M}_{f^{(k)}_\alpha}$ are $\pm m_{f^{(k)}_\alpha}$, meaning that the wrong mass-term sign, pointed out before, still remains. Specifically, it affects mass terms for the primed KK fermion fields $f'^{(k)}_{2,\alpha}$. As the authors of Ref.~\cite{PapaSan} showed, the redefinition $f'^{(k)}_{2,\alpha}=\gamma_5f^{(k)}_{2,\alpha}$ suffices to put things right. At the end of the day, KK excited-mode fermion fields $f^{(k)}_{1,\alpha}$ and $f^{(k)}_{2,\alpha}$ turn out to have the same mass $m_{f^{(k)}_\alpha}$. The final set of KK excited-mode fermion dynamic variables is illustrated in Eq.~(\ref{kkendfermions}):
\begin{equation}
\left.
\begin{array}{l}
\tilde{f}^{(k)}_\alpha\mapsto \tilde{f}'^{(k)}_\alpha
\vspace{0.2cm} \\
f^{(k)}_\alpha\mapsto f^{(k)}_\alpha
\end{array}
\right\}
\mapsto
\left\{
\begin{array}{l}
f^{(k)}_{1,\alpha}
\left\{
\begin{array}{l}
\nu^{(k)}_{1,j}
\vspace{0.13cm} \\
l^{(k)}_{1,\beta}
\vspace{0.13cm} \\
u^{(k)}_{1,\gamma}
\vspace{0.13cm} \\
d^{(k)}_{1,\delta}
\end{array}
\right.
\vspace{0.25cm}\\
f'^{(k)}_{2,\alpha}\mapsto f^{(k)}_{2,\alpha}
\left\{
\begin{array}{c}
\nu^{(k)}_{2,j}
\vspace{0.13cm} \\
l^{(k)}_{2,\beta}
\vspace{0.13cm} \\
u^{(k)}_{2,\gamma}
\vspace{0.13cm} \\
d^{(k)}_{2,\delta}
\end{array}
\right.
\end{array}
\right.
\label{kkendfermions}
\end{equation}
where $j=1,2,3$, $\beta=e,\mu,\tau$, $\gamma=u,c,t$, and $\delta=d,s,b$.
\\

The 4DSM, defined exclusively by KK zero-mode fields, includes charged currents in which quark flavor is not preserved, a feature that emerges after biunitary diagonalizations of Yukawa matrices and which is characterized by the {\it Cabibbo-Kobayashi-Maskawa matrix}~\cite{NCabibbo,KoMa,Wolfenstein,ChKe,PDG}, $\kappa=V^{u\dag}_LV^d_L$. After compactification, but previous to spontaneous symmetry breaking, our KK Lagrangian comprises, among its dynamic variables, sterile right-handed zero-mode neutrino fields $\nu^{(0)}_{\alpha,R}$. A  consequence of this is the appearance, after implementation of spontaneous symmetry breaking, of lepton charged currents in which lepton flavor changes, with such an effect described by the {\it Pontecorvo-Maki-Nakagawa-Sakata matrix}~\cite{Pontecorvo,Kamiokande,SNO,DayaBay,RENO}, given by the matrix product $V^{l\dag}_LV^\nu_L$. An important characteristic of both favor-changing matrices is the incorporation of $CP$-violating effects, carried by complex phases. The phenomenon of $CP$ non-conservation has great relevance due to its role in {\it baryon asymmetry}, according to the {\it Sakharov conditions}~\cite{Sakharov}. Flavor-changing charged currents in which KK excited modes participate occur as well in both the lepton and the quark sectors, and, as it is the case of the 4DSM, the characterization of such effects are also given by the Cabibbo-Kobayashi-Maskawa and the Pontecorvo-Maki-Nakagawa-Sakata matrices.

\subsection{Selected lagrangian terms}
The full KK effective lagrangian, found after implementation of compactification and the EHM, includes the whole 4DSM, but also contains a plethora of couplings in which KK excited modes take part. In this subsection, we provide explicit expressions of those couplings that are required for the main calculation to be executed. Of course, there are also 4DSM couplings generating low-energy effects, but the corresponding lagrangian terms and/or Feynman rules are available in the literature~\cite{CheLi,Langacker}, so we rather focus on lagrangian terms in which KK excited-mode fields participate. \\

From the sum ${\cal L}^{\rm YM}_{\rm KK}+{\cal L}_{\rm KK}^{\rm S}+{\cal L}_{\rm KK}^{{\rm GF}(k)}$, which combines the KK gauge, scalar, and excited-mode gauge-fixing sectors, the lagrangian terms
\begin{eqnarray}
{\cal L}_{A^{(0)}W^{(k)}W^{(k)}}=ie\,F^{(0)}_{\mu\nu}W^{(k)+\mu}W^{(k)-\nu}
\nonumber  \\ 
+\Big[ie\,A^{(0)}_\mu\big( W^{(k)-\mu\nu}W^{(k)+}_\nu
\nonumber \\
-\frac{1}{\xi}W^{(k)-\mu}\partial_\nu W^{(k)+\nu}\big)+{\rm H.c.}\Big],
\label{AWnWn}
\end{eqnarray}
\begin{equation}
{\cal L}_{A^{(0)}G_W^{(k)}G_W^{(k)}}=ie\,A^{(0)}_\mu G_W^{(k)-}\partial^\mu G_W^{(k)+}+{\rm H.c.},
\label{AGwnGwn}
\end{equation}
\begin{eqnarray}
{\cal L}_{A^{(0)}W^{(k)}W^{(k)}}=ie\,A^{(0)}_\mu W^{(k)-}\partial^\mu W^{(k)+}+{\rm H.c.}
\label{AWsnWsn}
\end{eqnarray}
emerge, with the definitions $W^{(k)\pm}_{\mu\nu}\equiv\partial_\mu W^{(k)\pm}_\nu-\partial_\nu W^{(k)\pm}_\mu$ and where $F^{(0)}_{\mu\nu}$ is the 4-dimensional electromagnetic tensor. By inspection of Eq.~(\ref{AWnWn}), terms proportional to the inverse gauge-fixing parameter $\xi^{-1}$ can be noticed. They proceed from the gauge-fixing lagrangian ${\cal L}^{{\rm GF}(k)}_{\rm KK}$, with the choice of gauge-fixing functions displayed in Eqs.~(\ref{gfc1}) and (\ref{gfc2}). Eqs.~(\ref{AGwnGwn}) and (\ref{AWsnWsn}) show that KK couplings $A_\mu^{(0)}G_W^{(k)}G_W^{(k)}$ and $A^{(0)}_\mu W_{\rm s}^{(k)}W_{\rm s}^{(k)}$, with $W^{(k)}_{\rm s}$ denoting physical charged KK scalars $W^{(k)\pm}$, have been generated. Worth of mention are fine cancellations whose occurrence eliminates the couplings $A_\mu^{(0)}W_\nu^{(k)}G_W^{(k)}$, $A_\mu^{(0)}W^{(k)}_{\rm s}G_W^{(k)}$, and $A_\mu^{(0)}W_\nu^{(k)}W^{(k)}_{\rm s}$ from the theory. In particular, the cancellation of contributions to $A_\mu^{(0)}W_\nu^{(k)}G_W^{(k)}$ is a consequence of gauge fixing, defined by Eqs.~(\ref{gfc1}) and (\ref{gfc2}).\\

The sum ${\cal L}_{\rm KK}^{\rm Y}+{\cal L}_{\rm KK}^{\rm C}$, of the KK Yukawa and currents sectors, defines the lagrangian terms
\begin{equation}
{\cal L}_{A^{(0)}d^{(k)}_\alpha d^{(k)}_\alpha}=eN_d\,A^{(0)}_\mu\Big[ \bar{d}^{(k)}_{1,\alpha}\gamma^\mu d^{(k)}_{1,\alpha}+\bar{d}^{(k)}_{2,\alpha}\gamma^\mu d^{(k)}_{2,\alpha} \Big],
\label{Adkdk}
\end{equation}
\begin{eqnarray}
{\cal L}_{A^{(0)}u^{(k)}_\alpha u^{(k)}_\alpha}=eN_u\,A^{(0)}_\mu\Big[ \bar{u}^{(k)}_{1,\alpha}\gamma^\mu u^{(k)}_{1,\alpha}+\bar{u}^{(k)}_{2,\alpha}\gamma^\mu u^{(k)}_{2,\alpha} \Big],
\label{Aukuk}
\end{eqnarray}
\begin{eqnarray}
{\cal L}_{u^{(0)}_\beta d^{(k)}_\alpha W^{(k)}}=\frac{g\,\kappa_{\beta\alpha}}{\sqrt{2}}W^{(k)+}_\mu\bar{u}^{(0)}_\beta\gamma^\mu P_L
\Big[\sin\theta_{d^{(k)}_\alpha}d^{(k)}_{1,\alpha}
\nonumber \\
-\cos\theta_{d^{(k)}_\alpha}d^{(k)}_{2,\alpha} 
\Big]
+{\rm H.c.},
\label{udkWk}
\end{eqnarray}
\begin{eqnarray}
{\cal L}_{u^{(0)}_\beta d^{(k)}_\alpha G_W^{(k)}}=\frac{ig\,m_{(k)}\kappa_{\beta\alpha}}{\sqrt{2}\,m_{W^{(k)}}}G^{(k)+}_W\bar{u}^{(0)}_\beta P_R
\Big[\sin\theta_{d^{(k)}_\alpha}d^{(k)}_{1,\alpha}
\nonumber \\
+\cos\theta_{d^{(k)}_\alpha}d^{(k)}_{2,\alpha}
\Big]
+{\rm H.c.},
\label{udkGwk}
\end{eqnarray}
\begin{eqnarray}
{\cal L}_{u^{(0)}_\beta d^{(k)}_\alpha W^{(k)}_{\rm s}}=\frac{ig\,m_{W^{(0)}}\kappa_{\beta\alpha}}{\sqrt{2}\,m_{W^{(k)}}}W^{(k)+}\bar{u}^{(0)}_\beta P_R
\nonumber \\
\times\Big[
\sin\theta_{d^{(k)}_\alpha}d^{(k)}_{1,\alpha}
+\cos\theta_{d^{(k)}_\alpha}d^{(k)}_{2,\alpha}
\Big]
+{\rm H.c.},
\label{udkWsk}
\end{eqnarray}
\begin{eqnarray}
&&
{\cal L}_{u^{(0)}_\alpha u^{(k)}_\alpha A^{(k)}}=eN_uA_\mu^{(k)}\bar{u}^{(0)}_\alpha\gamma^\mu
\nonumber \\ &&
\times\Big[ \big(P_L\sin\theta_{u^{(k)}_\alpha}
+P_R\cos\theta_{u^{(k)}_\alpha}\big)u^{(k)}_{1,\alpha}
\nonumber \\ &&
-\big(P_L\cos\theta_{u^{(k)}_\alpha}
+P_R\sin\theta_{u^{(k)}_\alpha}\big)u^{(k)}_{2,\alpha} \Big]+{\rm H.c.},
\label{uukAk}
\end{eqnarray}
\begin{eqnarray}
&&
{\cal L}_{u^{(0)}_\alpha u^{(k)}_\alpha A^{(k)}_{\rm G}}=ieN_uA^{(k)}_{\rm G}\bar{u}^{(0)}_\alpha
\nonumber \\ &&
\times\Big[ \big(P_R\sin\theta_{u^{(k)}_\alpha}-P_L\cos\theta_{u^{(k)}_\alpha}\big)u^{(k)}_{1,\alpha}
\nonumber \\ &&
+\big( P_R\cos\theta_{u^{(k)}_\alpha}-P_L\sin\theta_{u^{(k)}_\alpha} \big)u^{(k)}_{2,\alpha} \Big]+{\rm H.c.},
\label{uukAGk}
\end{eqnarray}
\begin{eqnarray}
&&
{\cal L}_{u^{(0)}_\alpha u^{(k)}_\alpha Z^{(k)}}=\frac{g}{2c_W}Z^{(k)}_\mu\bar{u}^{(0)}_\alpha\gamma^\mu
\nonumber \\ &&
\times\Big[ \big( P_Rh^u_0\cos\theta_{u^{(k)}_\alpha}
+P_Lh^u_1\sin\theta_{u^{(k)}_\alpha}
 \big)u^{(k)}_{1,\alpha}
\nonumber \\ &&
-\big(P_Rh^u_0\sin\theta_{u^{(k)}_\alpha}
+P_Lh^u_1\cos\theta_{u^{(k)}_\alpha} \big)u^{(k)}_{2,\alpha} \Big]+{\rm H.c.},
\label{uukZk}
\end{eqnarray}
\begin{eqnarray}
&&
{\cal L}_{u^{(0)}_\alpha u^{(k)}_\alpha G^{(k)}_Z}=\frac{-ig}{2c_Wm_{Z^{(k)}}}G_Z^{(k)}\bar{u}^{(0)}_\alpha
\nonumber \\ &&
\times\Big[ \big( P_L\big( m_{u^{(k)}_\alpha}+h^u_1m_{(k)} \big)\cos\theta_{u^{(k)}_\alpha}
\nonumber \\&&
-P_R( m_{u^{(k)}_\alpha}+h^u_0m_{(k)} )\sin\theta_{u^{(k)}_\alpha}
\big)u^{(k)}_{1,\alpha} 
\nonumber \\ &&
-\big( P_L\big( m_{u^{(k)}_\alpha}-h^u_1m_{(k)} \big)\sin\theta_{u^{(k)}_\alpha}
\nonumber \\ &&
-P_R\big( m_{u^{(k)}_\alpha}-h^u_0m_{(k)} \big)\cos\theta_{u^{(k)}_\alpha} \big)u^{(k)}_{2,\alpha}
\Big]+{\rm H.c.},
\label{uukGzk}
\end{eqnarray}
\begin{eqnarray}
&&
{\cal L}_{u^{(0)}_\alpha u^{(k)}_\alpha Z^{(k)}_{\rm s}}=\frac{-ig}{2c_W\,m_{Z^{(0)}}m_{Z^{(k)}}}Z^{(k)}\bar{u}^{(0)}_\alpha
\nonumber \\ &&
\times\Big[
\Big( P_L\big( m_{(k)}(m_{(k)}+m_{u^{(k)}_\alpha})-h^u_0m^2_{Z^{(0)}} \big)\cos\theta_{u^{(k)}_\alpha}
\nonumber \\ &&
+P_R\big( m_{(k)}(m_{(k)}-m_{u^{(k)}_\alpha})+h^u_1m^2_{Z^{(0)}} \big)\sin\theta_{u^{(k)}_\alpha}\Big)u^{(k)}_{1,\alpha}
\nonumber \\ &&
+\Big( P_L\big( m_{(k)}(m_{(k)}-m_{u^{(k)}_\alpha})-h^u_0m^2_{Z^{(0)}} \big)\sin\theta_{u^{(k)}_\alpha}
\nonumber \\ &&
+P_R\big( m_{(k)}(m_{(k)}+m_{u^{(k)}_\alpha})+h^u_1m^2_{Z^{(0)}} \big)\cos\theta_{u^{(k)}_\alpha} \Big)u^{(k)}_{2,\alpha}
\Big]
\nonumber \\&&
+{\rm H.c.}
\label{uukZsk}
\end{eqnarray}
Here, $eN_u=+2e/3$ and $eN_d=-e/3$ are, respectively, the electric charges of any $u$-type quark and any $d$-type quark. Moreover, $P_L=({\bf 1}-\gamma_5)/2$ and $P_R=({\bf 1}+\gamma_5)/2$ are the {\it chiral projection operators}. The factors $h^u_j=j-2N_us_W^2$ have been also defined. The lagrangian terms given in Eqs.~(\ref{Adkdk})-(\ref{udkWk}), (\ref{uukAGk}), and (\ref{uukZk}) are solely generated by the currents sector, differently from what happened in the cases of the couplings given by Eqs.~(\ref{udkGwk}), (\ref{udkWsk}), (\ref{uukGzk}), and (\ref{uukZsk}) which are the result of combining contributions from both fermion sectors. \\


\section{Analytic calculation of anomalous magnetic moments and flavor-changing decays}
\label{anacalc}
A feature of phenomenological significance characterizing models with universal extra dimensions is that their very first contributions to low-energy Green's functions, and thus to low-energy observables, are produced by loop Feynman diagrams. For that reason, physical processes forbidden by the 4DSM at tree level are important to this kind of beyond-the-Standard-Model physics. For example, this is the case of the {\it oblique parameters}~\cite{ACD1} and the  {\it muon anomalous magnetic moment}~\cite{AppDo}, it concerns {\it flavor-changing processes} from the fermion sector as well~\cite{ADW}, and it is interesting, in this sense, for the gluon-fusion Higgs-boson production mechanism $h\to gg$ and the Higgs decays $h\to\gamma\gamma$ and $h\to\gamma Z$~\cite{Petriello,NoTo1}.\\

In this section, we use the KK theory previously discussed to calculate one-loop contributions from KK modes to the AMMs of $u$-type quarks and to branching ratios ${\rm Br}\big(u^{(0)}_\alpha\to A^{(0)}_\mu u^{(0)}_\beta \big)$. We calculate such contributions analytically, in an exact manner, by means of the {\it Passarino-Veltman tensor reduction method}~\cite{PassVe}, for which the software {\it Mathematica}, by Wolfram, and the package {\it Feyncalc}~\cite{MBD} are utilized. Then, we consider a scenario characterized by a very small extra dimension and obtain analytic expressions, in terms of elementary mass-dependent functions. Consistency of results with respect to 
renormalization and decoupling is discussed in this section as well.\\

At one loop, AMMs of $u$-type quarks receive contributions from four sorts of Feynman diagrams, distinguished of each other by which virtual KK tensor modes, among those of the photon, the $Z$ boson, the Higgs boson or the $W$ boson, circulate in the loop. In the case of decays $u^{(0)}_\alpha\to A^{(0)}_\mu u^{(0)}_\beta$, contributions exclusively arise from diagrams with loop KK modes of the $W$ boson, as the required changes of quark flavor are only allowed by charged currents. The present discussion does not comprehend the calculation of contributions from the 4DSM to AMMs, since the specific value of this quantity, at least for the case of the top quark, is available in the literature~\cite{BBGHLMR}. Nevertheless, specific expressions of 4DSM amplitudes are required to analyze the decay $u^{(0)}_\alpha\to A^{(0)}_\mu u^{(0)}_\beta$, so we do calculate such contributions. Then bear in mind that any reference to virtual KK zero modes, that is $(k)=(0)$, appertains only to contributions to this decay process.\\

\begin{figure}[!ht]
\center
\includegraphics[width=4cm]{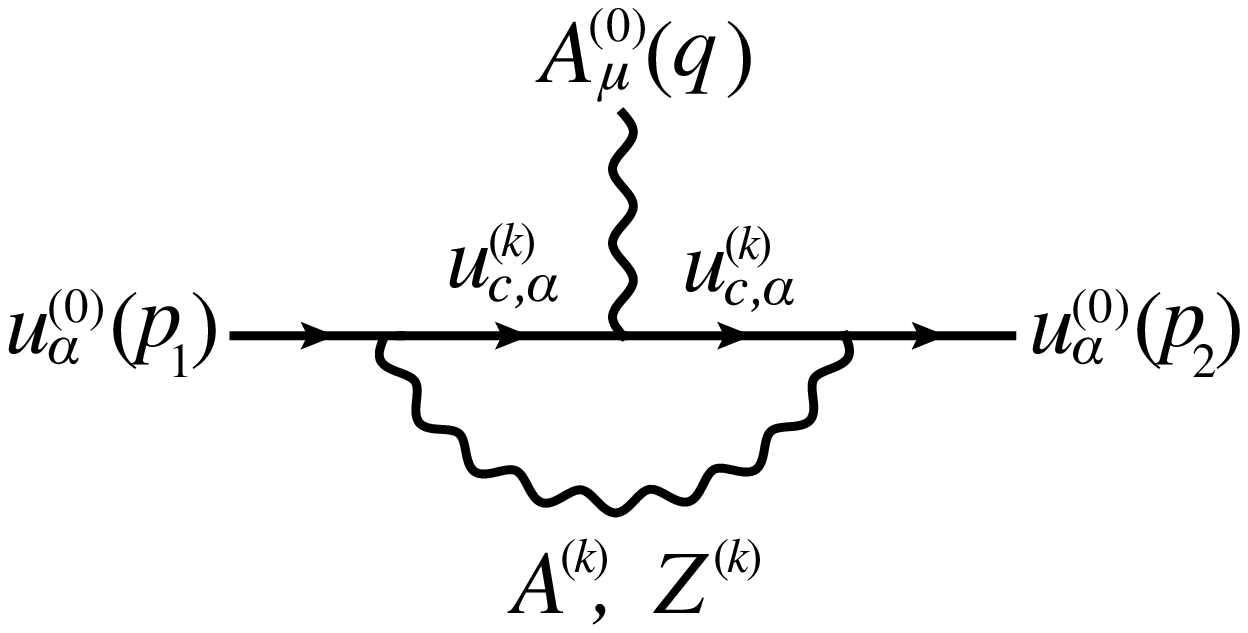}
\hspace{0.4cm}
\includegraphics[width=4cm]{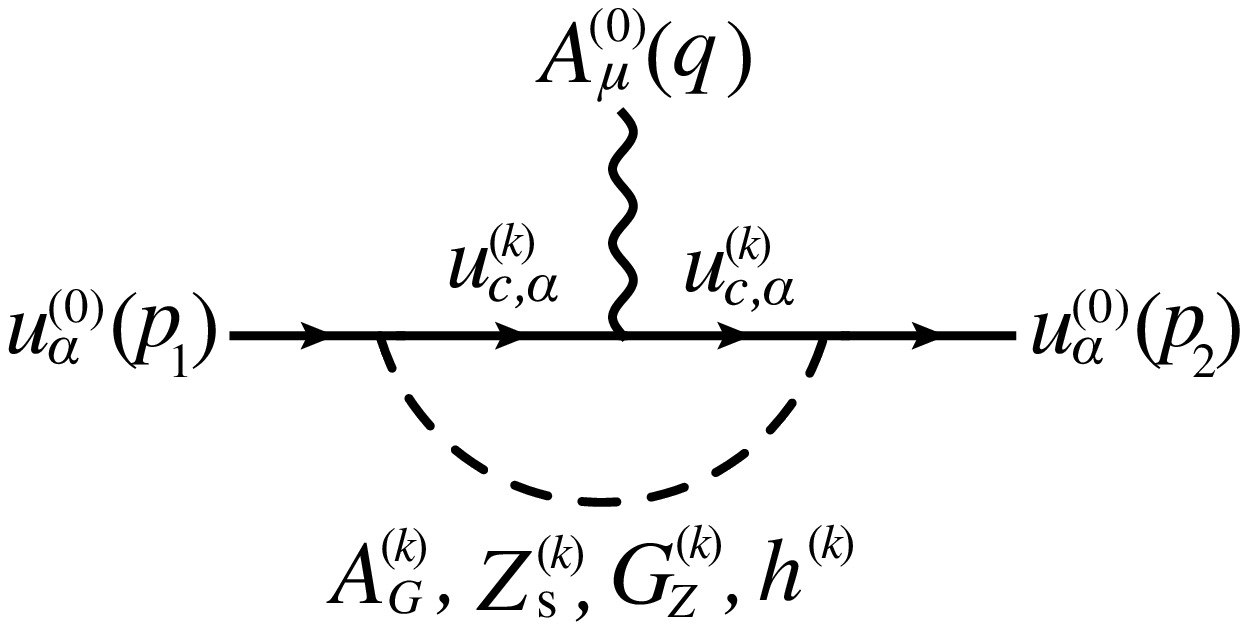}
\caption{\label{NeutContMMAs} \footnotesize{KK neutral-field diagrams contributing to $A^{(0)}_\mu u^{(0)}_\alpha u^{(0)}_\alpha$ at one loop. No zero-mode contributions are considered, so $(k)\ne(0)$. Internal KK quarks carry an index $c=1,2$, which denotes two kinds of KK spinors (see Subsection~\ref{YandCsects}).}}
\end{figure}

\begin{figure}[!ht]
\centering
\includegraphics[width=4cm]{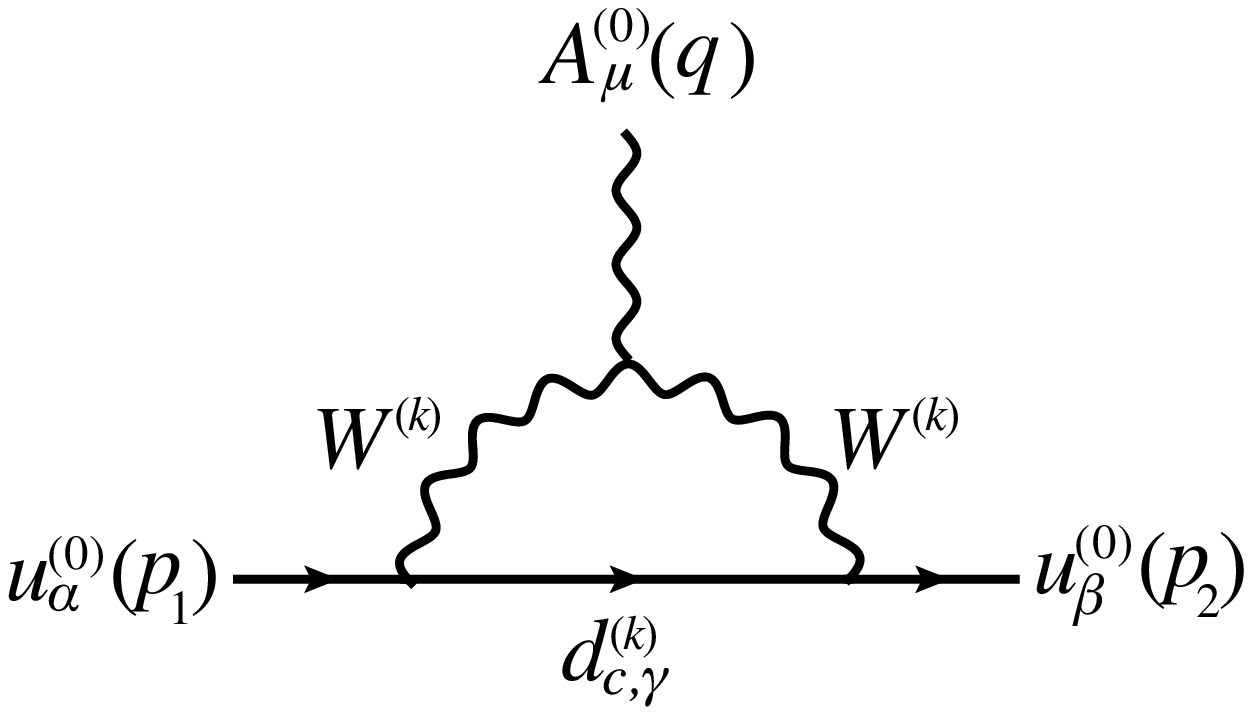}
\hspace{0.4cm}
\includegraphics[width=4cm]{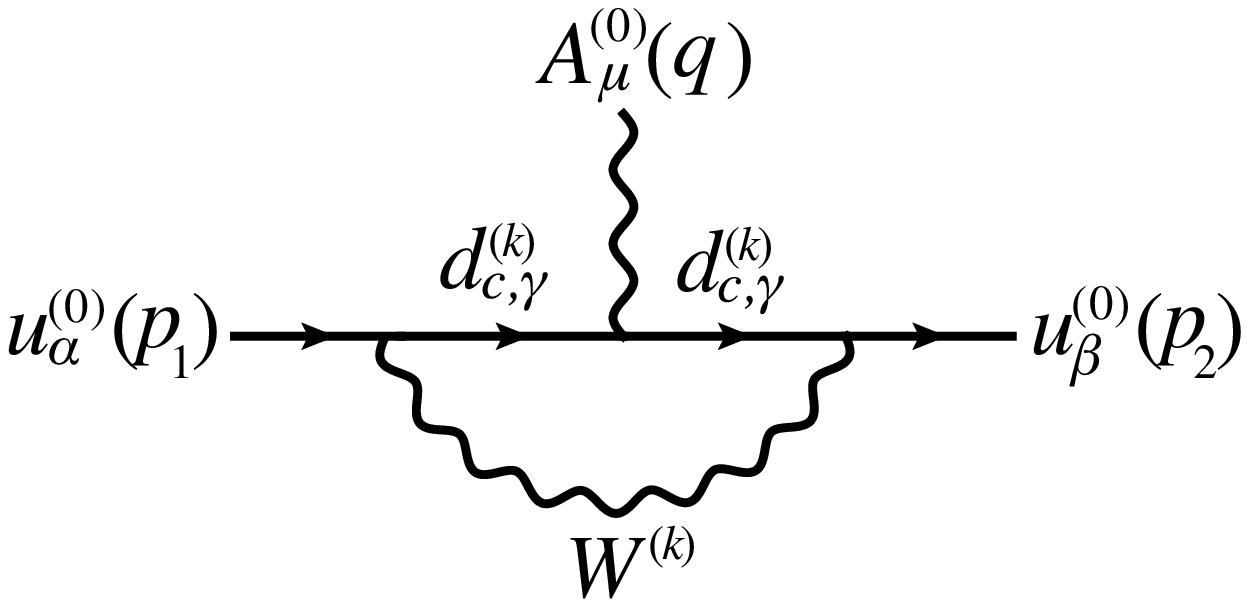}
\vspace{0.3cm} \\
\includegraphics[width=4cm]{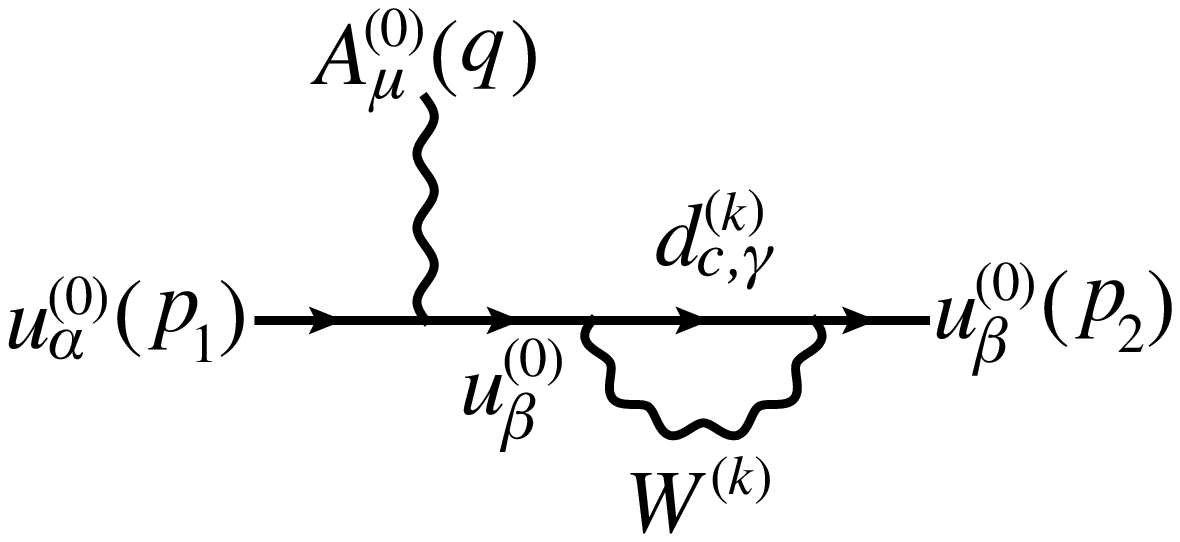}
\hspace{0.4cm}
\includegraphics[width=4cm]{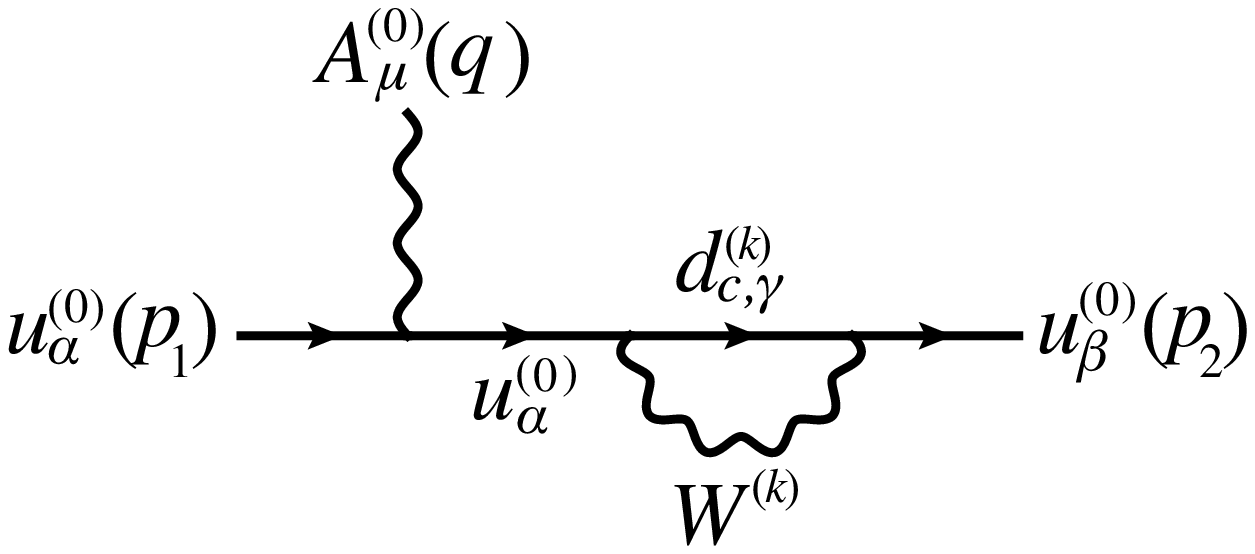}
\caption{\label{vdiags} {\footnotesize KK charged-vector-field diagrams contributing to $A^{(0)}_\mu u^{(0)}_\alpha u^{(0)}_\alpha$ and $u^{(0)}_\alpha\to u^{(0)}_\beta A^{(0)}_\mu$ at one loop. 4DSM contributing diagrams correspond to $(k)=(0)$. Internal KK quarks carry an index $c=1,2$, which denotes two kinds of KK spinors (see Subsection~\ref{YandCsects}).}}
\end{figure}

\begin{figure}
\centering
\includegraphics[width=4cm]{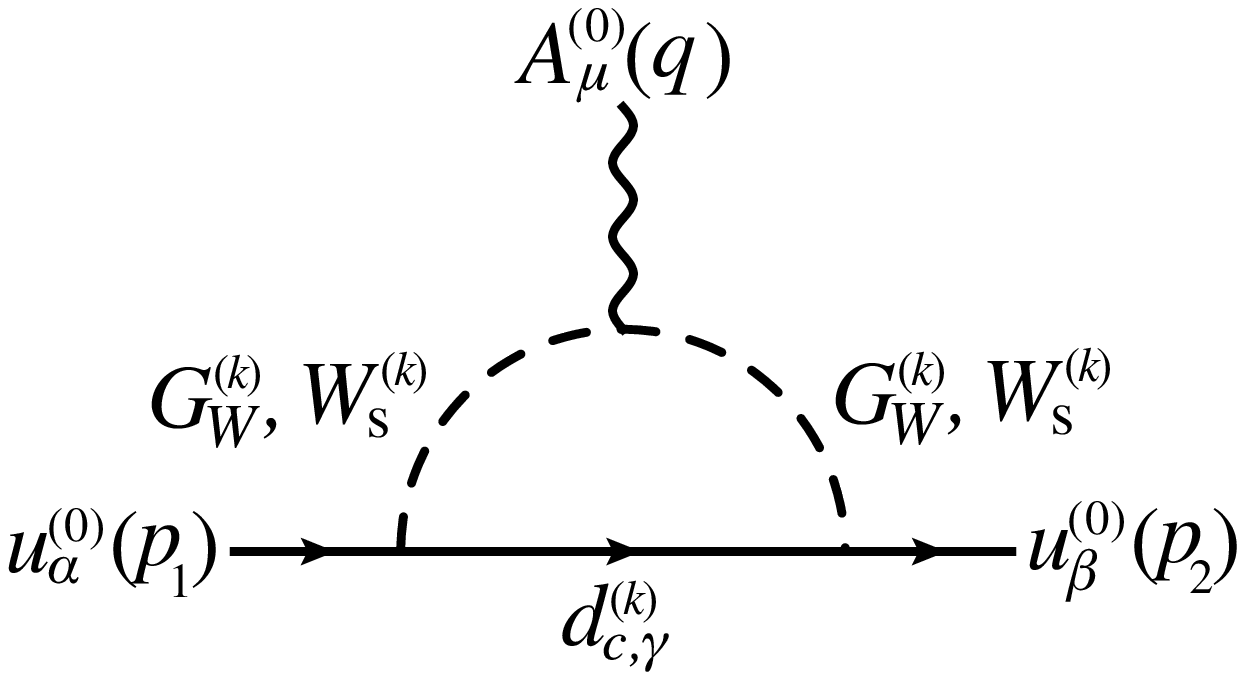}
\hspace{0.4cm}
\includegraphics[width=4cm]{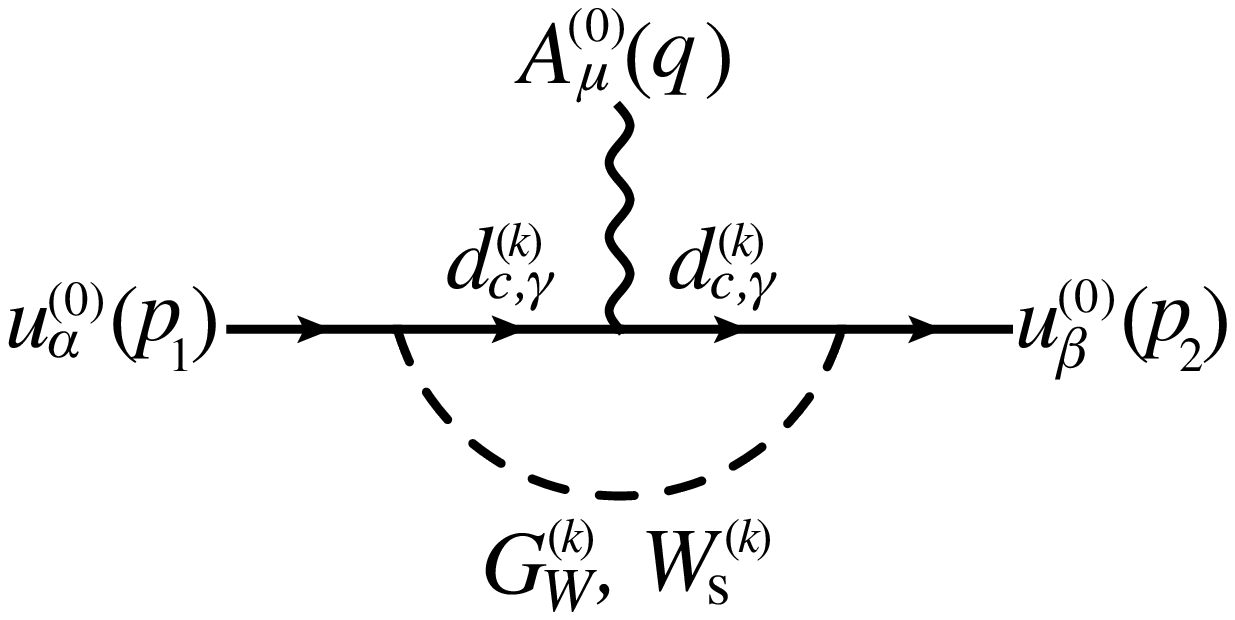}
\vspace{0.3cm} \\
\includegraphics[width=4cm]{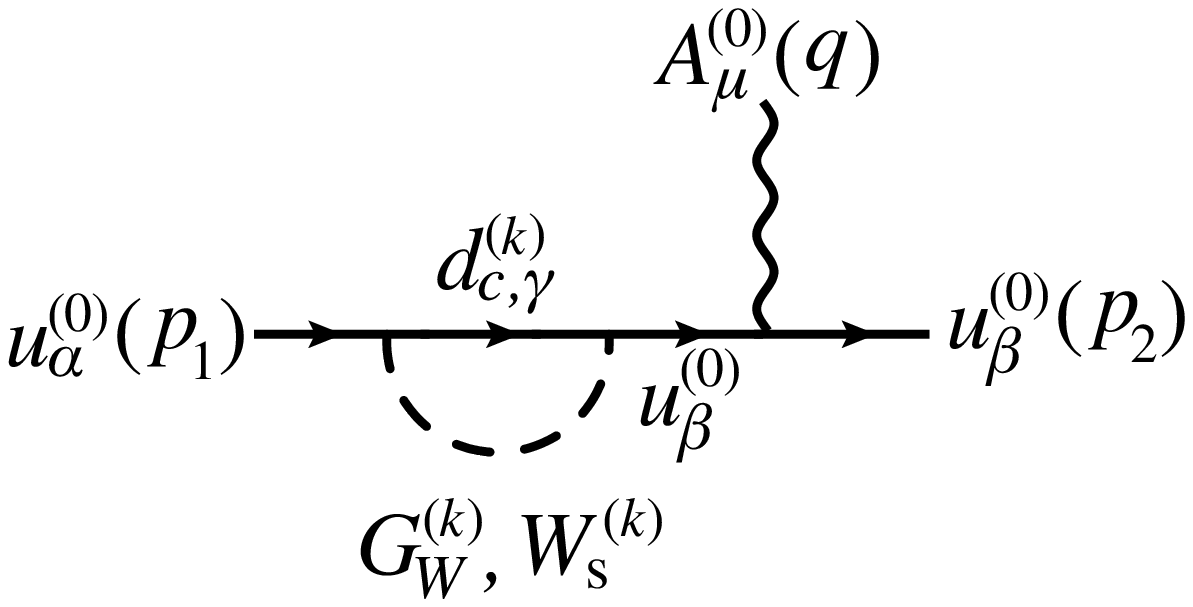}
\hspace{0.4cm}
\includegraphics[width=4cm]{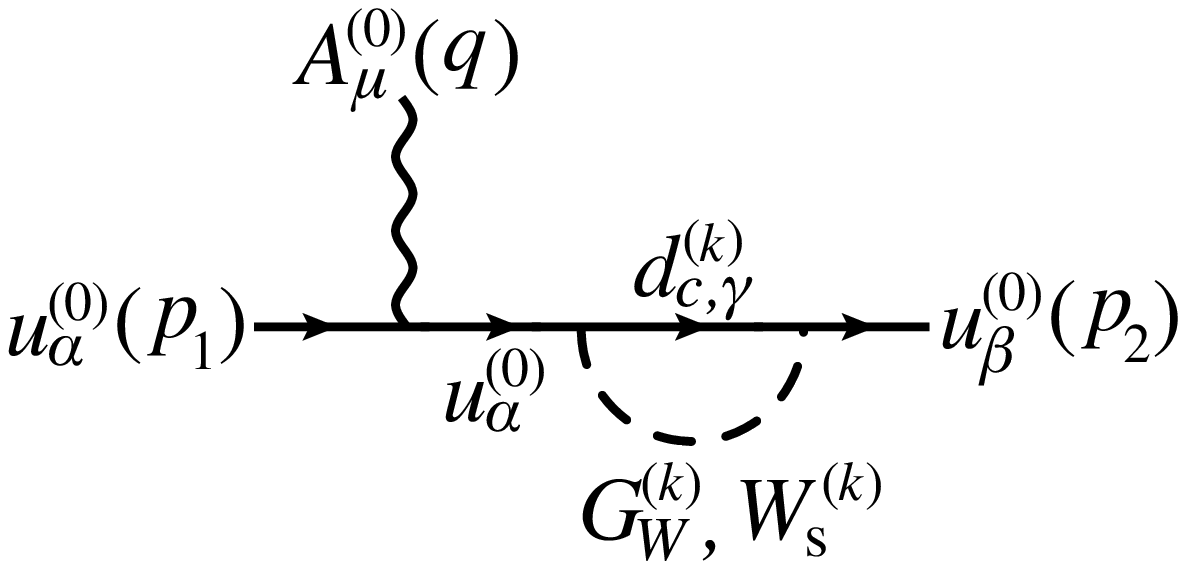}
\caption{\label{sdiags} {\footnotesize KK charged-scalar-field diagrams contributing to $A^{(0)}_\mu u^{(0)}_\alpha u^{(0)}_\alpha$ and $u^{(0)}_\alpha\to A^{(0)}_\mu u^{(0)}_\beta$ at one loop. Scalar contributions are associated to either pseudo-Goldstone bosons $G_W^{(k)\pm}$ or physical scalars $W^{(k)\pm}_{\rm s}$. Internal KK quarks carry an index $c=1,2$, which denotes two kinds of KK spinors (see Subsection~\ref{YandCsects}).}}
\end{figure}

With the whole spectrum of KK fields already defined, Figs.~\ref{NeutContMMAs}-\ref{sdiags} display the full set of one-loop Feynman diagrams that generate KK contributions to both the electromagnetic vertex $A^{(0)}_\mu u^{(0)}_\alpha u^{(0)}_\alpha$ and the decay $u^{(0)}_\alpha\to A^{(0)}_\mu u^{(0)}_\beta$. 
Diagrams in which neutral KK tensor fields take part are shown in Fig.~\ref{NeutContMMAs}.
The first Feynman diagram of Fig.~\ref{NeutContMMAs} stands for any contributing diagram in which vector-boson virtual lines correspond to either a KK excited-mode field  $A^{(k)}_\mu$ or $Z^{(k)}_\mu$. The second diagram in this figure involves a scalar loop line instead, which generically represents pseudo-Goldstone bosons $A^{(k)}_G$ or $G^{(k)}_Z$, or physical KK scalars $Z^{(k)}_{\rm s}$ or $h^{(k)}$. In these diagrams, loop KK modes of $u$-type quarks include the index $c=1,2$, which labels the two sorts of KK excited-mode spinor fields that eventuate from compactification, as discussed in Section~\ref{YandCsects}. Contributing Feynman diagrams with loop $W$-boson KK modes are exhibited in Figs.~\ref{vdiags} and \ref{sdiags}. Note that contributions to the flavor-changing decay process $u^{(0)}_\alpha\to A^{(0)}_\mu u^{(0)}_\beta$ are exclusively generated by this set of Feynman diagrams. Three types of contributing diagrams are comprised by Figs.~\ref{vdiags} and \ref{sdiags}: diagrams that involve virtual vector modes $W^{(0)\pm}_\mu$ or $W^{(k)\pm}_\mu$, with both cases taken into account in Fig.~\ref{vdiags}; contributions from diagrams in which charged pseudo-Goldstone bosons $G^{(0)\pm}_W$ or $G^{(k)\pm}_W$ participate, which are displayed in Fig.~\ref{sdiags}; and diagrams with KK physical scalars $W^{(k)\pm}$ circulating in loops, which are exhibited in Fig.~\ref{sdiags}. Note that, for fixed $(k)$ and $\gamma=d,s,b$, the presence of the index $c=1,2$, in the KK loop spinors $d^{(k)}_{c,\gamma}$, doubles the number of diagrams in Figs.~\ref{vdiags} and \ref{sdiags}.\\

As discussed before, in the gauge-fixing approach developed in the present paper the removal of symmetries with respect to ${\rm SU}(2,{\cal M}^4)_L\times{\rm U}(1,{\cal M}^4)_Y$ and to nonstandard gauge transformations is achieved by gauge choices that are independent of each other. In short, gauge fixing for KK zero modes is unattached to gauge fixing for excited modes. Therefore, even though the gauge-fixing functions given in Eqs.~(\ref{gfc1}) and (\ref{gfc2}) have been already used to eliminate invariance under nonstandard gauge transformations, we calculate the contributions from the 4DSM in the unitary gauge instead. Under such circumstances, no diagrams from Fig.~\ref{sdiags} with loop zero modes exist; the only contributions from the 4DSM come from diagrams of Fig.~\ref{vdiags}, with $(k)=(0)$.\\

Let us concentrate, for a moment, on those diagrams of Fig.~\ref{vdiags} with $(k)\ne(0)$, in which KK excited-mode vector fields $W^{(k)\pm}_\mu$ participate. The total contribution produced by the eight diagrams included by this figure, with either virtual KK quarks $d^{(k)}_{1,\gamma}$ or $d^{(k)}_{2,\gamma}$, behaves as the contribution from just four diagrams with one sole KK quark $d^{(k)}_{\gamma}$, not associated to any $\theta_{d^{(k)}_\gamma}$ mixing. Aiming at grasping such an assertion, we first point out that, according to the lagrangian term ${\cal L}_{u^{(0)}_\beta d^{(k)}_\alpha W^{(k)}_\mu}$, Eq.~(\ref{udkWk}), KK couplings $u^{(0)}_\beta d^{(k)}_{1,\alpha}W^{(k)}_\mu$ and $u^{(0)}_\beta d^{(k)}_{2,\alpha}W^{(k)}_\mu$ differ from their 4DSM counterparts $u^{(0)}_\beta d^{(0)}_\alpha W^{(0)}_\mu$ only by global factors~\cite{CheLi,Langacker}, not present in zero-mode couplings and which are $\sin\theta_{d^{(k)}_\alpha}$ for $u^{(0)}_\beta d^{(k)}_{1,\alpha}W^{(k)}_\mu$ and $-\cos\theta_{d^{(k)}_\alpha}$ for $u^{(0)}_\beta d^{(k)}_{2,\alpha}W^{(k)}_\mu$. As a result, the sum of diagrams with virtual KK quark fields $d^{(k)}_{1,\gamma}$ involves the global factor $\sin^2\theta_{d^{(k)}_{\gamma}}$. Similarly, the sum of diagrams with loop KK quarks $d^{(k)}_{2,\gamma}$ has the global factor $\cos^2\theta_{d^{(k)}_{\gamma}}$. Such trigonometric factors incarnate the only $\theta_{d^{(k)}_\gamma}$ dependence of these contributions. The sum of all these diagrams can be schematically expressed as
\begin{equation}
\begin{gathered}
\vspace{0.15cm}
\includegraphics[width=1.16cm]{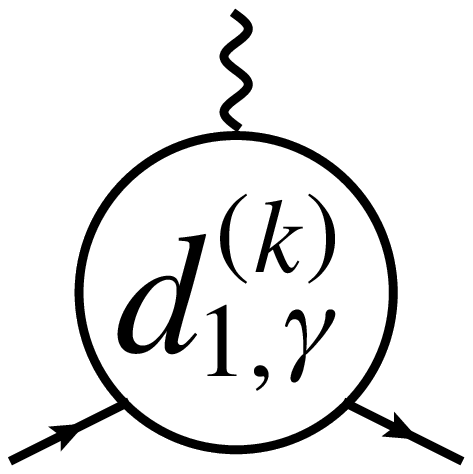}
\end{gathered}
+
\begin{gathered}
\vspace{0.15cm}
\includegraphics[width=1.16cm]{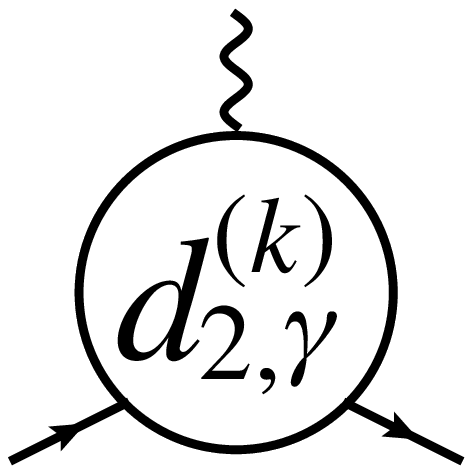}
\end{gathered}
=\big(\sin^2\theta_{d^{(k)}_\gamma}+\cos^2\theta_{d^{(k)}_\gamma}\big)
\begin{gathered}
\vspace{0.15cm}
\includegraphics[width=1.16cm]{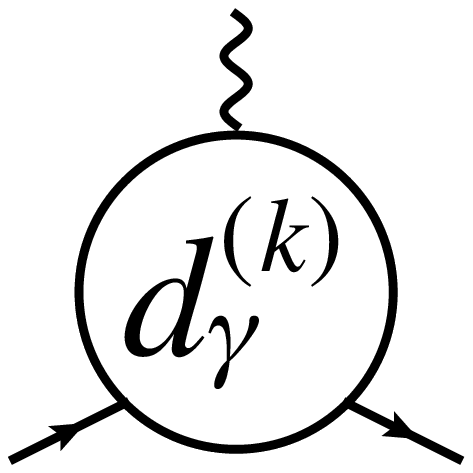}
\end{gathered}.
\label{sumdiags}
\end{equation}
The first term of the left-hand side of Eq.~(\ref{sumdiags}), labeled by $d^{(k)}_{1,\gamma}$, represents the sum of all those diagrams of Fig.~\ref{vdiags} in which the KK quarks circulating in loops are $d^{(k)}_{1,\gamma}$. Likewise, the second term, with label $d^{(k)}_{2,\gamma}$, stands for the sum of all the diagrams of Fig.~\ref{vdiags} with virtual KK quark fields $d^{(k)}_{2,\gamma}$. Regarding the right-hand side of Eq.~(\ref{sumdiags}), the factor labeled by $d^{(k)}_\gamma$ symbolizes the sum of the four diagrams shown in Fig.~\ref{vdiags}, but with the Feynman rules for KK couplings $u^{(0)}_\beta d^{(k)}_{\gamma}W^{(k)}_\mu$ taken without $\theta_{d^{(k)}_\gamma}$ dependence, thus having the same structure as the analogue 4DSM couplings. Moreover, the trigonometric factor in the right-hand side of Eq.~(\ref{sumdiags}) shows that any $\theta_{d^{(k)}_\gamma}$ dependence emerged from diagrams of Fig.~\ref{vdiags} vanishes when considering the total contribution. By the same token, the total contribution from diagrams of Fig.~\ref{sdiags} involving KK pseudo-Goldstone bosons $G_W^{(k)\pm}$ is $\theta_{d_\gamma^{(k)}}$ independent. By contrast, contributions from diagrams of Fig.~\ref{sdiags} with virtual KK physical scalars $W^{(k)\pm}$ do not combine in this manner, so they do depend on the mixing angle $\theta_{d^{(k)}_\gamma}$.\\

By considering all the Feynman diagrams of Figs.~\ref{NeutContMMAs}-\ref{sdiags}, with all the external particles {\it on shell}, and adding them together, the total one-loop contribution $i{\cal M}^{\beta\alpha}=i\,\bar{u}(p_2,m_{u^{(0)}_\beta})\,\Gamma^{\beta\alpha}_\mu (q)\,u(p_1,m_{u^{(0)}_\alpha})\,\epsilon^\mu(q,\lambda)$ is found, where
\begin{equation}
\Gamma^{\beta\alpha}_\mu=F_{\rm V}^{\beta\alpha}\gamma_\mu+F_{\rm A}^{\beta\alpha}\gamma_\mu\gamma_5+M^{\beta\alpha}\sigma_{\mu\nu}q^\nu+E^{\beta\alpha}\sigma_{\mu\nu}q^\nu\gamma_5
\label{totamp}
\end{equation}
involves the {\it magnetic form factor} $M^{\beta\alpha}(q^2=0)$ and the {\it electric form factor} $E^{\beta\alpha}(q^2=0)$~\cite{NPR,BGS}. Note that a contributing Feynman diagram exists for every KK index $(k)$, and all such diagrams must be summed together:
\begin{equation}
i{\cal M}^{\alpha\beta}=
\begin{gathered}
\vspace{0.1cm}
\includegraphics[width=1.16cm]{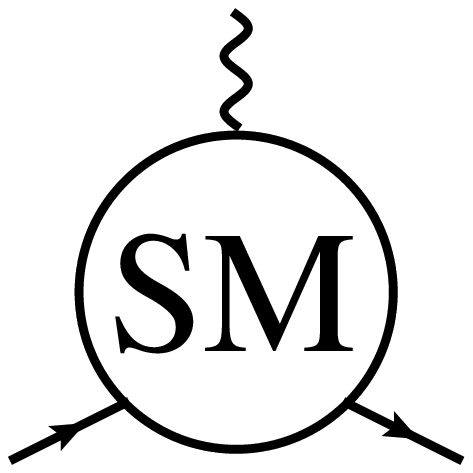}
\end{gathered}
+\sum_{k=1}^\infty\,\,
\begin{gathered}
\vspace{0.1cm}
\includegraphics[width=1.16cm]{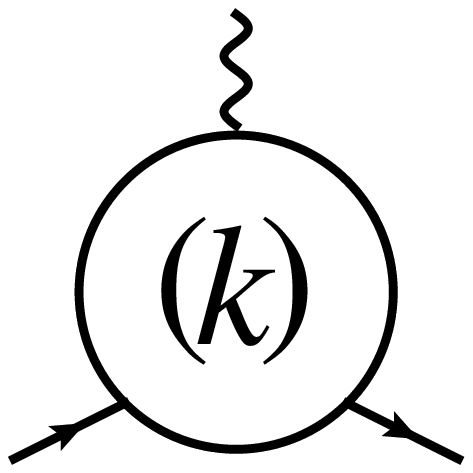}\,\,\, .
\end{gathered}
\label{ampKKsum}
\end{equation}
Thus the form factors defining Eq.~(\ref{totamp}) are given as sums over the complete set of KK contributions. Keep in mind that the cases $\alpha=\beta$ and $\alpha\ne\beta$ yield results which are qualitatively different of each other, so in practice they are treated separately. For $\alpha=\beta$, we have explicitly verified that the electric form factor $E^{\alpha\alpha}$ vanishes, which indicates that the KK contributions preserve $CP$ symmetry. Nevertheless, the form factors $F^{\alpha\alpha}_{\rm V}$ and $F^{\alpha\alpha}_{\rm A}$ are nonzero. If, on the other hand, $\beta\ne\alpha$, we find that $F^{\beta\alpha}_{\rm V}=0$ and $F^{\beta\alpha}_{\rm A}=0$, whereas a nonzero contribution $E^{\beta\alpha}$ arises. 
In the presence of Lorentz invariance\footnote{As shown in Refs.~\cite{MNTT1,MNTT2}, nonconservation of Lorentz symmetry allows for a richer structure of this parametrization.}, the standard parametrization of the $Vff$ vertex, with $f$ a fermion and $V$ either the photon or the $Z$ boson, reads~\cite{HIRSS,Schwartzbuch}
\begin{equation}
\Gamma^{Vff}_\mu=-ie\Big\{ \gamma_\mu\big[ V^V_f-A^V_f\gamma_5 \big]-\sigma_{\mu\nu}q^\nu\Big[ i\frac{a^V_f}{2m_f}-\frac{d^V_f}{e}\gamma_5 \Big] \Big\},
\label{genAffpar}
\end{equation}
with $q$ the outgoing momentum of the $V$-boson external line. Here, $V^V_f(q^2)$ and $A^V_f(q^2)$ respectively parametrize the vector and axial-vector currents. If the vector boson $V$ is assumed to be an on-shell photon, the corresponding factors $a^\gamma_f(q^2=0)$ and $d^\gamma_f(q^2=0)$ are the {\it anomalous magnetic moment} and the {\it electric dipole moment} of $f$, respectively. In the case $\alpha=\beta$, Eq.~(\ref{totamp}) is straightforwardly written as Eq.~(\ref{genAffpar}), from which contributions to AMMs are identified. Recall that, in our case, the electric dipole moment, known to be connected to $CP$ violation, vanishes exactly. \\

In general, the implementation of the Passarino-Veltman method in a given calculation reduces {\it tensor loop integrals} into expressions given exclusively in terms of {\it scalar loop integrals}~\cite{PassVe}, also referred to as {\it Passarino-Veltman scalar functions}, or {\it scalar functions} for short. Scalar functions are determined by quadratic masses of dynamic variables involved in the calculation. Using this method, we have written the KK form factors $M^{\beta\alpha}$ and $E^{\beta\alpha}$ in terms of 2-point scalar functions, $B^j_0$, and 3-point scalar functions, $C^j_0$. 
The explicit expressions of such KK contributions are provided in Appendixes~\ref{AppffAMM} and \ref{AppffFCD}. \\

The magnetic and electric form factors are free of ultraviolet divergences. To understand how these divergences are eradicated, first let us generically denote any form-factor contribution $M^{\beta\alpha}$ or $E^{\beta\alpha}$ by $F^{\beta\alpha}$. Any such form factor is found to have the general structure $F^{\beta\alpha}=\sum_{j}\chi_2^jB^j_0+\sum_k\chi_3^kC^k_0+\chi_0$, where $\chi_2^j$, $\chi_3^k$, $\chi_0$ are functions of masses, while $B^j_0$ are 2-point scalar functions and $C^k_0$ are 3-point scalar functions. We remark that $B^j_0$ functions are ultraviolet divergent, whereas $C^k_0$ functions are finite~\cite{HooVe}. Using {\it dimensional regularization}~\cite{PeSch,BoGi}, any 2-point function $B^j_0$ is split into a sum of a divergent term $\Delta_{\rm div}$ and a finite term $f^j_{\rm fin}$, that is, $B^j_0=\Delta_{\rm div}+f^j_{\rm fin}$. The essential observation is that all the scalar functions $B^j_0$ share the exact same divergent term $\Delta_{\rm div}$. So, if we consider any two 2-point functions, say $B^j_0$ and $B^i_0$, the difference $B^j_0-B^i_0$ cancels the divergent terms, thus yielding an ultraviolet-finite expression. We have verified that the whole dependence of $F^{\beta\alpha}$ on 2-point functions $B^j_0$ can be written as a sum of terms, all of them proportional to a difference like $B^j_0-B^i_0$. The form-factor contributions given in Appendixes~\ref{AppffAMM} and \ref{AppffFCD} show this explicitly. Thus we conclude that the form factors $M^{\beta\alpha}$ and $E^{\beta\alpha}$ are finite.\\

Next we split the electromagnetic form factors as $F^{\beta\alpha}=F^{\beta\alpha}_W+F^{\beta\alpha}_{Z,A,H}$, where the term $F^{\beta\alpha}_W$ denotes the contributions from diagrams with virtual KK excited modes of the $W$ boson, comprised by Figs.~\ref{vdiags} and \ref{sdiags}, whereas the total contribution from the whole set of diagrams with virtual KK excited modes of the $Z$ boson, the photon, and the Higgs boson, all of them displayed in Fig.~\ref{NeutContMMAs}, has been represented by $F^{\beta\alpha}_{Z,A,H}$. The form-factor contributions from $W$-boson KK excited modes are then expressed as
\begin{eqnarray}
M^{\beta\alpha}_W&=&\sum_{\gamma=d,s,b}\sum_{k=0}^\infty\kappa_{\beta\gamma}\,\kappa^*_{\alpha\gamma}\,M^{(k)\beta\alpha}_{W,\gamma},
\label{fMff}
\\
E^{\beta\alpha}_{W}&=&\sum_{\gamma=d,s,b}\sum_{k=0}^\infty\kappa_{\beta\gamma}\,\kappa^*_{\alpha\gamma}\,E^{(k)\beta\alpha}_{W,\gamma}.
\label{fEff}
\end{eqnarray}
In these equations, any form factor $M^{(k)\beta\alpha}_{W,\gamma}$ or $E^{(k)\beta\alpha}_{W,\gamma}$, with $\gamma$ and $(k)$ fixed, represents a contribution from a specific $d$-type quark flavor with a particular KK index $(k)$, with the case $(k)=(0)$ included for $\alpha\ne\beta$. Let us write the KK contributions $M^{(k)\beta\alpha}_{W,\gamma}$ and $E^{(k)\beta\alpha}_{W,\gamma}$ as
\begin{eqnarray}
M^{(k)\beta\alpha}_{W,\gamma}&=&G^{(k)\beta\alpha}_{M,\gamma}+P^{(k)\beta\alpha}_{M,\gamma}+\hat{S}^{(k)\beta\alpha}_{M,\gamma}+\tilde{S}^{(k)\beta\alpha}_{M,\gamma},
\label{contsfsM}
\\ \nonumber \\
E^{(k)\beta\alpha}_{W,\gamma}&=&G^{(k)\beta\alpha}_{E,\gamma}+P^{(k)\beta\alpha}_{E,\gamma}+\hat{S}^{(k)\beta\alpha}_{E,\gamma}+\tilde{S}^{(k)\beta\alpha}_{E,\gamma}.
\label{contsfsE}
\end{eqnarray}
In these equations, the terms $G^{(k)\beta\alpha}_{M,\gamma}$ and $G^{(k)\beta\alpha}_{E,\gamma}$ stand for the form-factor contributions generated by the diagrams with virtual vector-fields $W^{(k)\pm}_\mu$, shown in Fig.~\ref{vdiags}. Contributions produced by Feynman diagrams with loop KK pseudo-Goldstone bosons $G_W^{(k)\pm}$, included in Fig.~\ref{sdiags}, are represented by the the terms $P^{(k)\beta\alpha}_{M,\gamma}$ and $P^{(k)\beta\alpha}_{E,\gamma}$. The contributions from diagrams with KK quarks $d^{(k)}_{1,\gamma}$ and physical scalars $W^{(k)\pm}$ circulating in loops, also shown in Fig.~\ref{sdiags}, correspond to the terms $\hat{S}^{(k)\beta\alpha}_{M,\gamma}$ and $\hat{S}^{(k)\beta\alpha}_{E,\gamma}$. Finally, the terms $\tilde{S}^{(k)\beta\alpha}_{M,\gamma}$ and $\tilde{S}^{(k)\beta\alpha}_{E,\gamma}$ are the contributions from diagrams with virtual KK quarks $d^{(k)}_{2,\gamma}$ and physical scalars $W^{(k)\pm}$, which are included in Fig.~\ref{sdiags} as well. No terms $P^{(0)\beta\alpha}_{M,\gamma}$ and $P^{(0)\beta\alpha}_{E,\gamma}$ exist, since the 4DSM contributions have been calculated in the unitary gauge. Moreover, the 4DSM has no analogues for the scalars $W^{(k)\pm}$, so neither terms $\hat{S}^{(0)\beta\alpha}_{M,\gamma}$, $\hat{S}^{(0)\beta\alpha}_{E,\gamma}$, $\tilde{S}^{(0)\beta\alpha}_{M,\gamma}$, or $\tilde{S}^{(0)\beta\alpha}_{E,\gamma}$ arise. \\

As it can be appreciated from Eqs.~(\ref{fMff}) and (\ref{fEff}), the total contribution to any form factor $M^{\beta\alpha}_W$ or $E^{\beta\alpha}_W$ includes a sum, $\sum_\gamma$, over quark flavors. Each term of this sum incorporates a product $\kappa_{\beta\gamma}\,\kappa^*_{\alpha\gamma}$, of entries of the Cabibbo-Kobayashi-Maskawa mixing matrix $\kappa$. This matrix is unitary, so the GIM mechanism~\cite{GIMmech} operates. In this context, 
the quark-flavor sum must be consistently implemented, since individual contributions, corresponding to the different flavors $\gamma$, may display a nondecoupling behavior in the limit as $R^{-1}\to\infty$, in which case the GIM mechanism would render the total contribution decoupling. Such an implementation is carried out by using unitarity of $\kappa$, for which the cases $\alpha=\beta$ and $\alpha\ne\beta$ are treated separately, thus yielding the following expressions:
\begin{enumerate}
\item Contributions to {\it diagonal form factors} ($\alpha=\beta$), with
\begin{equation}
M^{\alpha\alpha}_W=\sum_{k=0}^\infty
\Big[
M^{(k)\alpha\alpha}_{W,d}
+\sum_{\gamma=s,b}|\kappa_{\alpha \gamma}|^2
\big(M^{(k)\alpha\alpha}_{W,\gamma}-M^{(k)\alpha\alpha}_{W,d}\big)
\Big].
\label{GIMinfMamm}
\end{equation}
\item Contributions to {\it transition form factors} ($\alpha\ne\beta$), with
\begin{eqnarray}
M^{\beta\alpha}_W&=&\sum_{k=0}^\infty\sum_{\gamma=s,b}\kappa_{\beta \gamma}\kappa^*_{\alpha \gamma}
\big(M^{(k)\beta\alpha}_{W,\gamma}-M^{(k)\beta\alpha}_{W,d}\big),
\label{GIMinfM}
\\
E^{\beta\alpha}_W&=&\sum_{k=0}^\infty\sum_{\gamma=s,b}\kappa_{\beta \gamma}\kappa^*_{\alpha \gamma}
\big(E^{(k)\beta\alpha}_{W,\gamma}-E^{(k)\beta\alpha}_{W,d}\big).
\label{GIMinfE}
\end{eqnarray}
\end{enumerate}
We emphasize that, in these equations, quark-flavor sums only run over two values: $\gamma=s,b$.\\

With the exact expressions of the form-factor contributions at hand, we continue our discussion in the context of a scenario characterized by a large compactification scale $R^{-1}$. Regarding the limits on the compactification scale, most results have been reported for the case of only one extra dimension. In the minimal version of these models, supersymmetry-searches data from the Large Hadron Collider were taken advantage of to derive the lower bound $1.4\,{\rm TeV}\lesssim R^{-1}$~\cite{DFK}. A bound $1\,{\rm TeV}\lesssim R^{-1}$, also obtained from Large-Hadron-Collider data, was recently estimated by the authors of Ref.~\cite{BDDM}. The lower limit $1.3\,{\rm TeV}\lesssim R^{-1}$ was established in Ref.~\cite{BKP} through the investigation of the contributions from KK {\it dark matter} to {\it relic density}. Large-Hadron-Collider data from searches of the 4DSM Higgs boson, analyzed in Ref.~\cite{BBBKP}, provided the less-stringent bound $0.5\,{\rm TeV}\lesssim R^{-1}$. The decay process $\bar{B}\to X_s\gamma$ has also been considered in order to bound the compactification scale, resulting in the limit $0.6\,{\rm TeV}\lesssim R^{-1}$~\cite{HaWe}. In the context of a non-minimal model of universal extra dimensions, enriched by the presence {\it boundary localized kinetic terms}~\cite{DGKN,CTW,APS}, the authors of Ref.~\cite{FKKMP} were able to give a more stringent bound on the compactification scale: $2.4\,{\rm TeV}\lesssim R^{-1}$.\\

As illustrated by Eq.~(\ref{ampKKsum}), a sum over the whole set of KK indices is required in order to achieve the total new-physics contribution, so form factors $F^{\beta\alpha}$ can be expressed as $F^{\beta\alpha}=\sum_{k=0}^\infty F^{(k)\beta\alpha}$, where $F^{(k)\beta\alpha}$ represents the contribution from all the KK-mode fields with KK index $(k)$. Under the assumption of a large compactification scale $R^{-1}$, we express the contributions from KK excited modes, that is with $(k)\ne(0)$, to form factors $F^{\beta\alpha}$ as series with respect to the compactification radius $R$. For fixed KK index $(k)$, we find contributions with the general structure 
\begin{equation}
F^{(k)\beta\alpha}=\sum_{j=1}\eta^{\beta\alpha}_{2j}\,(R/k)^{2j}.
\label{decouplingffs}
\end{equation}
In this equation, the whole dependence on the KK index $(k)$ has been factorized, together with the compactification radius, so factors $\eta^{\beta\alpha}_{2j}$ depend only on zero-mode masses. Therefore, KK sums $\sum_{k=1}^\infty$ are straightforwardly turned into {\it Riemann zeta functions} $\zeta(2j)$. Since the sum starts at $j=1$, notice that $R^2$ is the smallest power of the compactification radius, which yields the conclusion that, for fixed KK index $(k)$, contributions $F^{(k)\beta\alpha}$ vanish as the compactification scale $R^{-1}$ becomes larger, decoupling in the limit as $R^{-1}\to\infty$.\\

\subsection{Anomalous magnetic moments}
The full set of Feynman diagrams shown in Figs.~\ref{NeutContMMAs}-\ref{sdiags} produces contributions to AMMs of KK zero modes of $u$-type quarks. The contributions from diagrams with virtual KK modes of the $W$ boson, the $Z$ boson, the photon, and the Higgs boson are respectively denoted by $a^{\rm KK}_{W,\alpha}$, $a^{\rm KK}_{Z,\alpha}$, $a^{\rm KK}_{A,\alpha}$, and $a^{\rm KK}_{h,\alpha}$. Within the context of a large compactification scale $R^{-1}$, the following expressions for leading KK excited-mode contributions are determined:
\begin{eqnarray}
a^{\rm KK}_{W,\alpha}
=
-R^2\frac{\pi\alpha}{432s_W^2}\frac{m_{u^{(0)}_\alpha}^2}{m_{W^{(0)}}^2}\Big(11m_{W^{(0)}}^2-m_{u^{(0)}_\alpha}^2
\nonumber \\ 
+4m_{d^{(0)}}^2-4 \eta_{\alpha\alpha}\Big)+{\cal O}(R^4),
\label{ammWcont}
\end{eqnarray}
\begin{eqnarray}
a^{\rm KK}_{Z,\alpha}=R^2\frac{\pi\alpha}{15552 s_W^2}\frac{m_{u^{(0)}_\alpha}^2}{m^2_{W^{(0)}}}\bigg( 
m^2_{W^{(0)}}\bigg(\frac{53}{c_W^{2}}-8
\nonumber \\ 
-32s_W^2\bigg)+54m_{u^{(0)}_\alpha}^2
\bigg)+{\cal O}(R^4),
\label{ammZcont}
\end{eqnarray}
\begin{eqnarray}
a^{\rm KK}_{A,\alpha}=R^2\frac{\pi\alpha}{486}m_{u^{(0)}_\alpha}^2+{\cal O}(R^4),
\label{ammAcont}
\end{eqnarray}
\begin{eqnarray}
a^{\rm KK}_{h,\alpha}=-R^2\frac{5\pi  \alpha}{864s_W^2}\frac{m_{u^{(0)}_\alpha}^4}{m_{W^{(0)}}^2}+{\cal O}(R^4),
\label{ammHcont}
\end{eqnarray}
for which the factor
\begin{equation}
\eta_{\beta\alpha}=\sum_{\gamma=s,b}\kappa_{\beta \gamma}\kappa^*_{\alpha \gamma}\big(m_{d_\gamma^{(0)}}^2-m_{d^{(0)}}^2\big)
\label{etaab}
\end{equation}
has been defined. The presence of the global factor $m^2_{u^{(0)}_\alpha}$, in Eqs.~(\ref{ammWcont})-(\ref{ammHcont}), implies that, by far, the largest $u$-type-quark AMM generated by the KK modes is the one corresponding to the top quark. Moreover, Eqs.~(\ref{ammWcont})-(\ref{ammHcont}) show that negative contributions to the top-quark AMM arise from $a^{\rm KK}_{W,\alpha}$ and $a^{\rm KK}_{h,\alpha}$, whereas $a^{\rm KK}_{Z,\alpha}$, and $a^{\rm KK}_{A,\alpha}$ turn out to be positive.

\subsection{The decay process $u^{(0)}_\alpha\to A^{(0)}_\mu u^{(0)}_\beta$}
Recall Eqs.~(\ref{contsfsM}) and (\ref{contsfsE}), which define the form-factor contributions $F^{(k)\beta\alpha}_{W,\gamma}$ as sums of terms $G^{(k)\beta\alpha}_{M,\gamma}$, $P^{(k)\beta\alpha}_{M,\gamma}$, $\hat{S}^{(k)\beta\alpha}_{M,\gamma}$, $\tilde{S}^{(k)\beta\alpha}_{M,\gamma}$, $G^{(k)\beta\alpha}_{E,\gamma}$, $P^{(k)\beta\alpha}_{E,\gamma}$, $\hat{S}^{(k)\beta\alpha}_{E,\gamma}$, and $\tilde{S}^{(k)\beta\alpha}_{E,\gamma}$. Each of such individual contributions has the large-$R^{-1}$ decoupling structure already pointed out for $F^{(k)\beta\alpha}$, in Eq.~(\ref{decouplingffs}), so their smallest compactification-scale suppression is of order $R^2$. A further suppression is introduced by the GIM mechanism, which we implement to individual contributions in the same manner as that shown in Eqs.~(\ref{GIMinfM}) and (\ref{GIMinfE}). Regarding vector-field contributions $G^{(k)\beta\alpha}_{M,\gamma}$ and $G^{(k)\beta\alpha}_{E,\gamma}$, this mechanism eliminates all $R^2$-order terms exactly, thus leaving dominant contributions of order $R^4$. Furthermore, the same GIM suppression takes place in the case of pseudo-Goldstone boson contributions $P^{(k)\beta\alpha}_{M,\gamma}$ and $P^{(k)\beta\alpha}_{E,\gamma}$.
Explicitly, the corresponding contributions read
\begin{eqnarray}
&&
\sum_{\gamma=d,s,b}\sum_{k=1}^\infty\kappa_{\beta\gamma}\kappa^*_{\alpha\gamma}\Big( G^{(k)\beta\alpha}_{M,\gamma}+P^{(k)\beta\alpha}_{M,\gamma} \Big)
\nonumber \\ &&
=R^4\frac{i \pi ^{7/2} \alpha ^{3/2}}{s_W^2}\frac{26m_{u^{(0)}_\alpha}-7m_{u^{(0)}_\beta}}{129600}\eta_{\beta\alpha}+{\cal O}(R^6),
\label{GPMR4}
\end{eqnarray}
\begin{eqnarray}
&&
\sum_{\gamma=d,s,b}\sum_{k=1}^\infty\kappa_{\beta\gamma}\kappa^*_{\alpha\gamma}\Big( G^{(k)\beta\alpha}_{E,\gamma}+P^{(k)\beta\alpha}_{E,\gamma} \Big)
\nonumber \\ &&
=R^4\frac{i \pi ^{7/2} \alpha ^{3/2}}{s_W^2}\frac{26m_{u^{(0)}_\alpha}-7m_{u^{(0)}_\beta}}{129600}\eta_{\beta\alpha}+{\cal O}(R^6),
\label{GPER4}
\end{eqnarray}
where the definition of $\eta_{\beta\alpha}$, given Eq.~(\ref{etaab}), has been utilized. About physical-scalar contributions $\hat{S}^{(k)\beta\alpha}_{M,\gamma}$, $\tilde{S}^{(k)\beta\alpha}_{M,\gamma}$, $\hat{S}^{(k)\beta\alpha}_{E,\gamma}$, and $\tilde{S}^{(k)\beta\alpha}_{E,\gamma}$, they get suppressed by the GIM mechanism as well, but $R^2$-order contributions remain, with the consequence that these are the most important KK excited-mode contributions. The following expressions are found:
\begin{eqnarray}
&&
\sum_{\gamma=d,s,b}\sum_{k=1}^\infty\kappa_{\beta\gamma}\kappa^*_{\alpha\gamma}\Big( \hat{S}^{(k)\beta\alpha}_{M,\gamma}+\tilde{S}^{(k)\beta\alpha}_{M,\gamma} \Big)
\nonumber \\ &&
=\frac{-R^2}{m_{W^{(0)}}^2}\frac{i\pi^{3/2}\alpha ^{3/2}}{s_W^2}\frac{5m_{u^{(0)}_\alpha}+3m_{u^{(0)}_\beta}}{864}\eta_{\beta\alpha}+{\cal O}(R^4),
\nonumber \\
\label{S12MR2}
\end{eqnarray}
\begin{eqnarray}
&&
\sum_{\gamma=d,s,b}\sum_{k=1}^\infty\kappa_{\beta\gamma}\kappa^*_{\alpha\gamma}\Big( \hat{S}^{(k)\beta\alpha}_{E,\gamma}+\tilde{S}^{(k)\beta\alpha}_{E,\gamma} \Big)
\nonumber \\ &&
=\frac{-R^2}{m_{W^{(0)}}^2}\frac{i\pi^{3/2}\alpha^{3/2}}{s_W^2}\frac{5m_{u^{(0)}_\alpha}-3m_{u^{(0)}_\beta}}{864}\eta_{\beta\alpha}+{\cal O}(R^4).
\nonumber \\ 
\label{S12ER2}
\end{eqnarray}
\\

Now we separate the form-factor contributions generated by the 4DSM from those produced by the whole set of KK excited modes. We write the magnetic and electric form-factor contributions as $M^{\beta\alpha}_W=M^{(0)\beta\alpha}_W+M^{{\rm KK},\beta\alpha}_W$ and $E^{\beta\alpha}_W=E^{(0)\beta\alpha}_W+E^{{\rm KK},\beta\alpha}_W$. In accordance with Eqs.~(\ref{GIMinfM}) and (\ref{GIMinfE}), contributions from the 4DSM have been denoted as $M^{(0)\beta\alpha}_W=\sum_\gamma\kappa_{\beta\gamma}\kappa^*_{\alpha\gamma}(M^{(0)\beta\alpha}_{W,\gamma}-M^{(0)\beta\alpha}_{W,d})$ and $E^{(0)\beta\alpha}_W=\sum_\gamma\kappa_{\beta\gamma}\kappa^*_{\alpha\gamma}(M^{(0)\beta\alpha}_{W,\gamma}-E^{(0)\beta\alpha}_{W,d})$, with $\gamma=s,b$. Moreover, KK excited-mode contributions are given by $M^{{\rm KK},\beta\alpha}_W=\sum_\gamma\sum_{k=1}^\infty\kappa_{\beta\gamma}\kappa^*_{\alpha\gamma}(M^{(k)\beta\alpha}_{W,\gamma}-M^{(k)\beta\alpha}_{W,d})$ and $E^{{\rm KK},\beta\alpha}_W=\sum_\gamma\sum_{k=1}^\infty\kappa_{\beta\gamma}\kappa^*_{\alpha\gamma}(E^{(k)\beta\alpha}_{W,\gamma}-E^{(k)\beta\alpha}_{W,d})$, where again $\gamma=s,b$. Thus, the decay rate for $u^{(0)}_\alpha\to A_\mu^{(0)}u^{(0)}_\beta$ is expressed as
\begin{eqnarray}
&&
\Gamma_{u^{(0)}_\alpha\to A_\mu^{(0)}u^{(0)}_\beta}
=\frac{\big(\Delta m^2_{u^{(0)}_{\alpha\beta}}\,\big)^3}{8\pi\,m_{u^{(0)}_\alpha}^6}\Big(|M^{(0)\beta\alpha}_W|^2+|E^{(0)\beta\alpha}_W|^2
\nonumber \\ &&
\hspace{1cm}+2{\rm Re}\Big\{ M^{{\rm KK},\beta\alpha}_WM^{(0)\beta\alpha*}_W+E^{{\rm KK},\beta\alpha}_WE^{(0)\beta\alpha*}_W \Big\}
\nonumber \\ &&
\hspace{1cm}+|M^{{\rm KK},\beta\alpha}_W|^2+|E^{{\rm KK},\beta\alpha}_W|^2\Big),
\label{DR1to2}
\end{eqnarray}
where the difference $\Delta m^2_{u^{(0)}_{\alpha\beta}}=m^2_{u^{(0)}_\alpha}-m^2_{u^{(0)}_\beta}$, of squared masses, has been defined. The first line of the right-hand side of Eq.~(\ref{DR1to2}) determines the total contribution produced by zero modes, that is, by fields from the 4DSM. The second line of this equation represents the interference between the low-energy theory and the KK excited-mode fields. Finally, the third line corresponds to effects exclusively associated to KK excited modes.

\section{Numerical estimations and discussion of results}
\label{nums}
In this section, the analytical expressions previously derived in the paper are implemented to estimate the extra-dimensional contributions to the AMM of the top quark and to the zero-mode decay processes $u^{(0)}_\alpha\to A_\mu^{(0)}u^{(0)}_\beta$. Following the lower bound reported in Ref.~\cite{DFK}, we consider, for our discussion, values of the compactification scale $R^{-1}>1.4\,{\rm TeV}$.

\subsection{Anomalous magnetic moment of the top quark}
First, we estimate the total contribution from the KK theory to the AMM of the top quark. The physics of the top quark is usually stressed; its strong connection to electroweak symmetry breaking, evidenced by its large mass, has fed the belief that TeV-scale new physics is likely to manifest through the physics of this particle. The 4DSM contribution, at the two-loop order, to the AMM of the top quark was calculated and estimated in Ref.~\cite{BBGHLMR}, with the predicted value $a_t=2\times10^{-2}$. In the same context, that paper also provided estimations of order $10^{-2}$, at two different renormalization scales, for the bottom-quark AMM. The AMMs of the electron and the muon have been thoroughly investigated, from both the experimental and the theoretical sides, reaching results with a remarkable level of precision~\cite{MuonColl,HFG,HHG,AHKNe,AHKNm,Kinoshita}. The top-quark AMM, on the other hand, lies beyond current experimental sensitivity, but measurements of such a quantity are getting closer, specially with the advent of new-generation colliders and increasing data from the Large Hadron Collider, which points towards an upcoming era of precision measurements. Using measurements of the branching ratio and $CP$ asymmetry of $B\to X_s\gamma$, as well as data on $t\bar{t}\gamma$ production from the CDF Collaboration~\cite{CDFColl}, the authors of Ref.~\cite{BoLa1} established the constraint $-3<a_t<0.45$ on the top-quark AMM, but pointed out and illustrated that $t\bar{t}\gamma$ production at the Large Hadron Collider would improve constraints on such a quantity. The importance of $t\bar{t}\gamma$ production at hadron colliders to bound the top-quark AMM was first stated in Ref.~\cite{BJOR}. The same authors of Ref.~\cite{BoLa1} determined, in Ref.~\cite{BoLa2}, that the measurement of $\sigma(\gamma e\to t\bar{t})$ at the Large Hadron Electron Collider may further improve sensitivity as $|a_t|<7.5\times10^{-2}$. As discussed in Refs.~\cite{FaGe,FTM,KBG}, more sensitivity improvements in searches of the top-quark AMM are expected from future linear colliders and from physical processes occurring at the Large Hadron Collider as well.\\

\begin{figure}[!ht]
\center
\includegraphics[width=8.5cm]{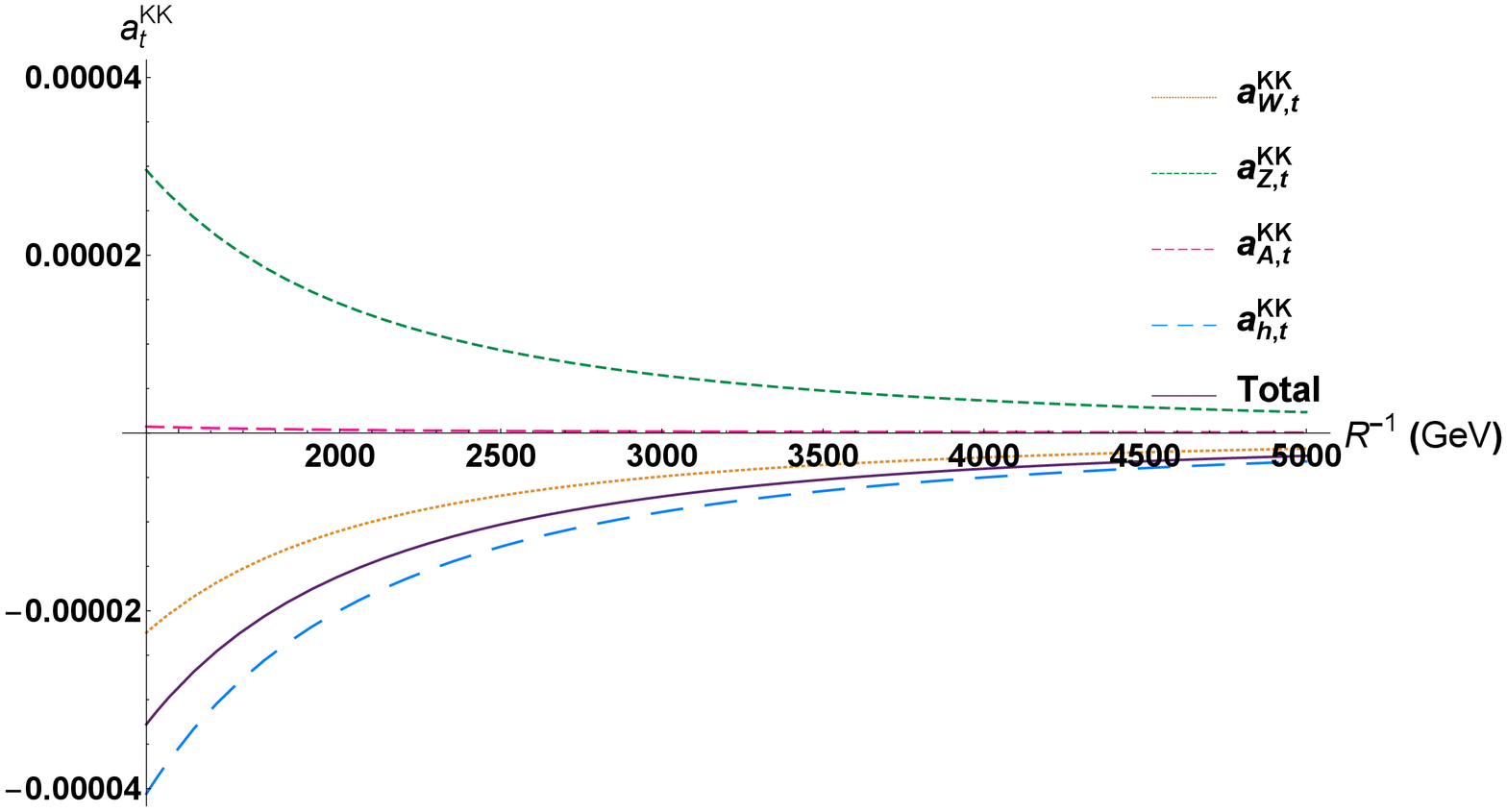}
\\ \vspace{0.7cm}
\includegraphics[width=8.5cm]{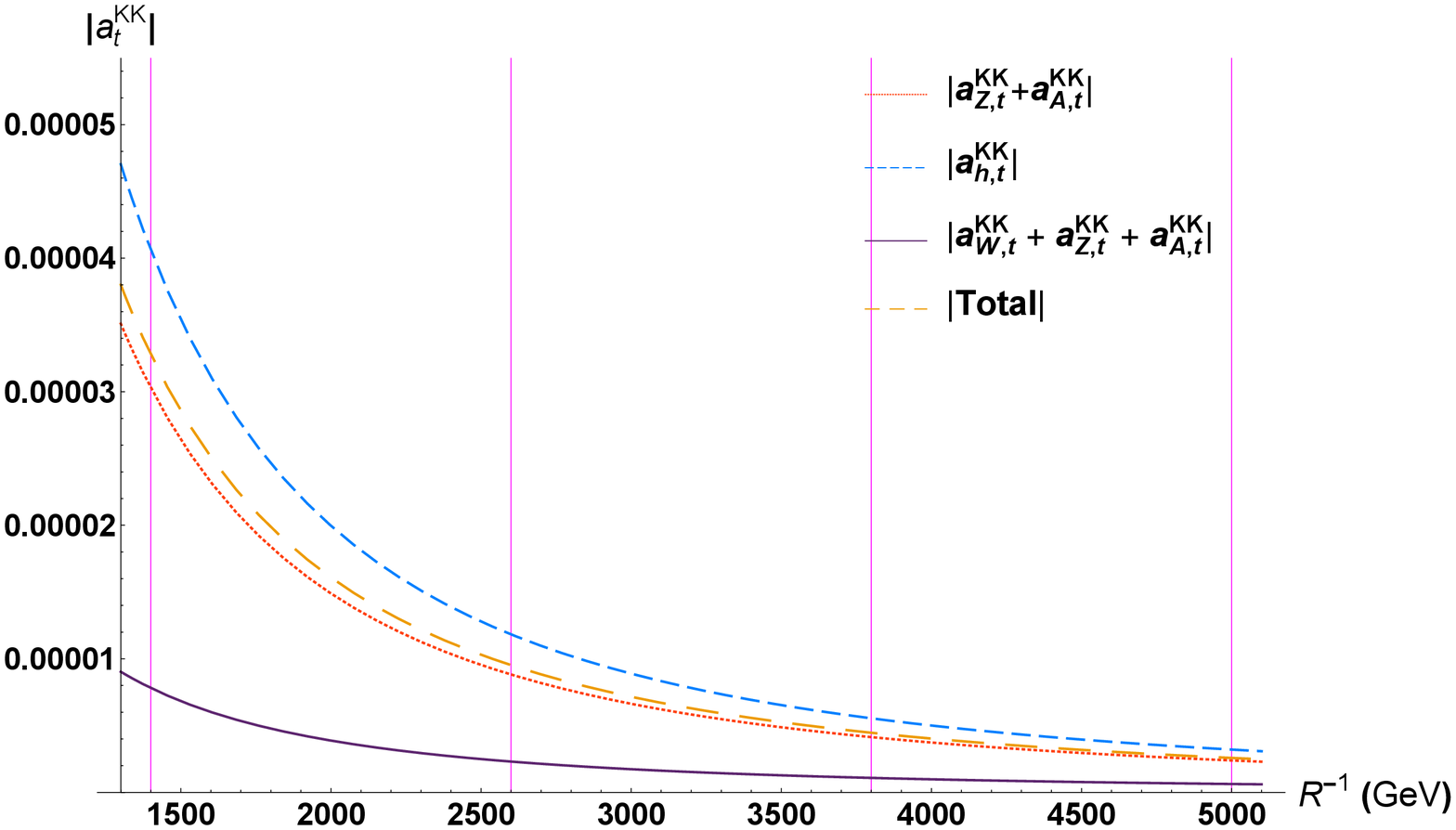}
\caption{\label{KKcontributiontotopamm} \footnotesize{Contributions from the KK theory to the top-quark AMM, as functions of $R^{-1}$. {\it Upper graph:}} Individual contributions $a^{\rm KK}_{W,t}$, $a^{\rm KK}_{Z,t}$, $a^{\rm KK}_{A,t}$, $a^{\rm KK}_{h,t}$, and total contribution. {\it Lower graph:} Absolute values of $a^{\rm KK}_{Z,t}+a^{\rm KK}_{A,t}$, $a^{\rm KK}_{h,t}$, $a^{\rm KK}_{W,t}+a^{\rm KK}_{Z,t}+a^{\rm KK}_{A,t}$ and $a^{\rm KK}_t=a^{\rm KK}_{W,t}+a^{\rm KK}_{Z,t}+a^{\rm KK}_{A,t}+a^{\rm KK}_{h,t}$. The two graphs have been plotted in GeV units.}
\end{figure}

The variety of different contributions from the KK excited modes to the AMM of the top quark are displayed, as functions of the compactification scale $R^{-1}$, in Fig.~\ref{KKcontributiontotopamm}, within the energy range $1.4\,{\rm TeV}<R^{-1}<5\,{\rm TeV}$.
The upper graph of this figure shows curves for the AMM contributions $a^{\rm KK}_{W,t}$, $a^{\rm KK}_{Z,t}$, $a^{\rm KK}_{A,t}$, and $a^{\rm KK}_{h,t}$. A curve for the total contribution $a^{\rm KK}_t=a^{\rm KK}_{W,t}+a^{\rm KK}_{Z,t}+a^{\rm KK}_{A,t}+a^{\rm KK}_{h,t}$ has been included as well. The sum of positive contributions $a^{\rm KK}_{Z,t}+a^{\rm KK}_{A,t}$, produced by KK excited modes of the $Z$ boson and the photon, is attenuated by the negative contribution $a^{\rm KK}_{W,t}$, associated to $W$-boson KK-excited-mode fields. Therefore, $a^{\rm KK}_{h,t}$, which is the largest individual contribution, determines $a^{\rm KK}_t$ to be negative. This is better illustrated by the lower graph of Fig.~\ref{KKcontributiontotopamm}, where, for comparison purposes, the absolute values $|a^{\rm KK}_{Z,t}+a^{\rm KK}_{A,t}|$, $|a^{\rm KK}_{h,t}|$, $|a^{\rm KK}_{W,t}+a^{\rm KK}_{Z,t}+a^{\rm KK}_{A,t}|$, and the absolute value of the total contribution, $|a^{\rm KK}_t|$, have been plotted. Without taking absolute values, Tab.~\ref{AMMtabcontcomp} provides quantitative instances of such contributions, each established by a specific choice of the compactification scale $R^{-1}$.
\begin{table}[!ht]
\center
\begin{tabular}{|c|c|c|c|}
\hline
$R^{-1}$ & $a^{\rm KK}_{W,t}+a^{\rm KK}_{Z,t}+a^{\rm KK}_{A,t}$ & $a^{\rm KK}_{h,t}$ & $a^{\rm KK}_t$
\\ \hline
1.4 TeV & $+7.8\times10^{-6}$ & $-4.1\times10^{-5}$ & $-3.3\times10^{-5}$
\\ \hline
2.6 TeV & $+2.3\times10^{-6}$ & $-1.2\times10^{-5}$ & $-9.5\times10^{-6}$
\\ \hline
3.8 TeV & $+1.1\times10^{-6}$ & $-5.5\times10^{-6}$ & $-4.5\times10^{-6}$
\\ \hline
5.0 TeV & $+6.3\times10^{-7}$ & $-3.2\times10^{-6}$ & $-2.6\times10^{-6}$
\\ \hline
\end{tabular}
\caption{\label{AMMtabcontcomp} \footnotesize{Values of the contributions $a^{\rm KK}_{W,t}+a^{\rm KK}_{Z,t}+a^{\rm KK}_{A,t}$ (KK excited modes of the $W$ boson, the $Z$ boson and the photon), $a^{\rm KK}_h$ (KK excited modes of the Higgs boson), and $a^{\rm KK}_t$ (sum of the two previous contributions) for various choices of the compactification scale $R^{-1}$.} }
\end{table}
The vertical solid lines in the lower graph of Fig.~\ref{KKcontributiontotopamm} correspond to the values of $R^{-1}$ that have been placed in the entries of the first column of Tab.~\ref{AMMtabcontcomp}, which means that points where the curves and these vertical lines cross each other correspond to values shown in other columns of this table.\\

\begin{figure}[!ht]
\center
\includegraphics[width=8.5cm]{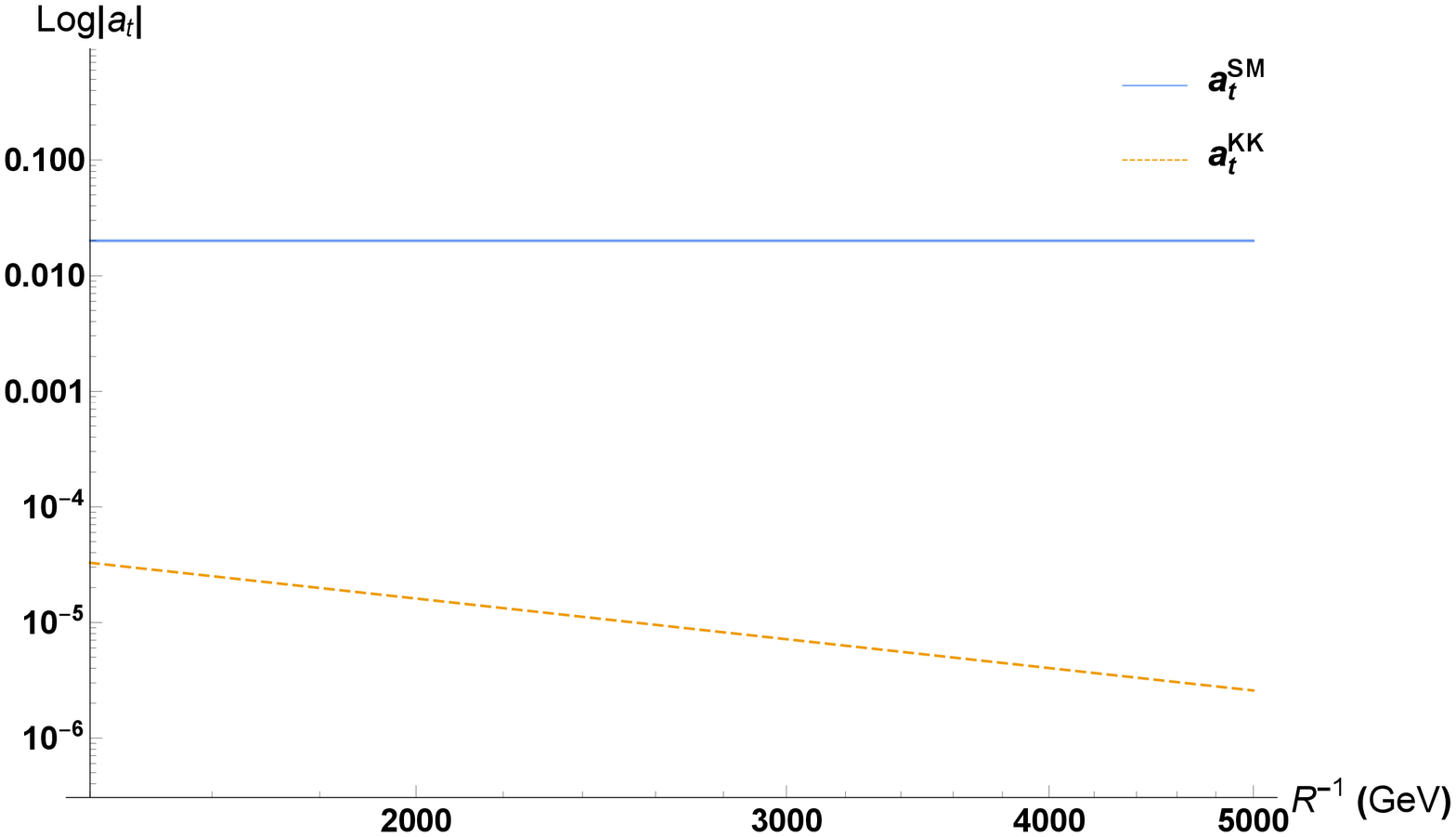}
\\ \vspace{0.7cm}
\includegraphics[width=8.5cm]{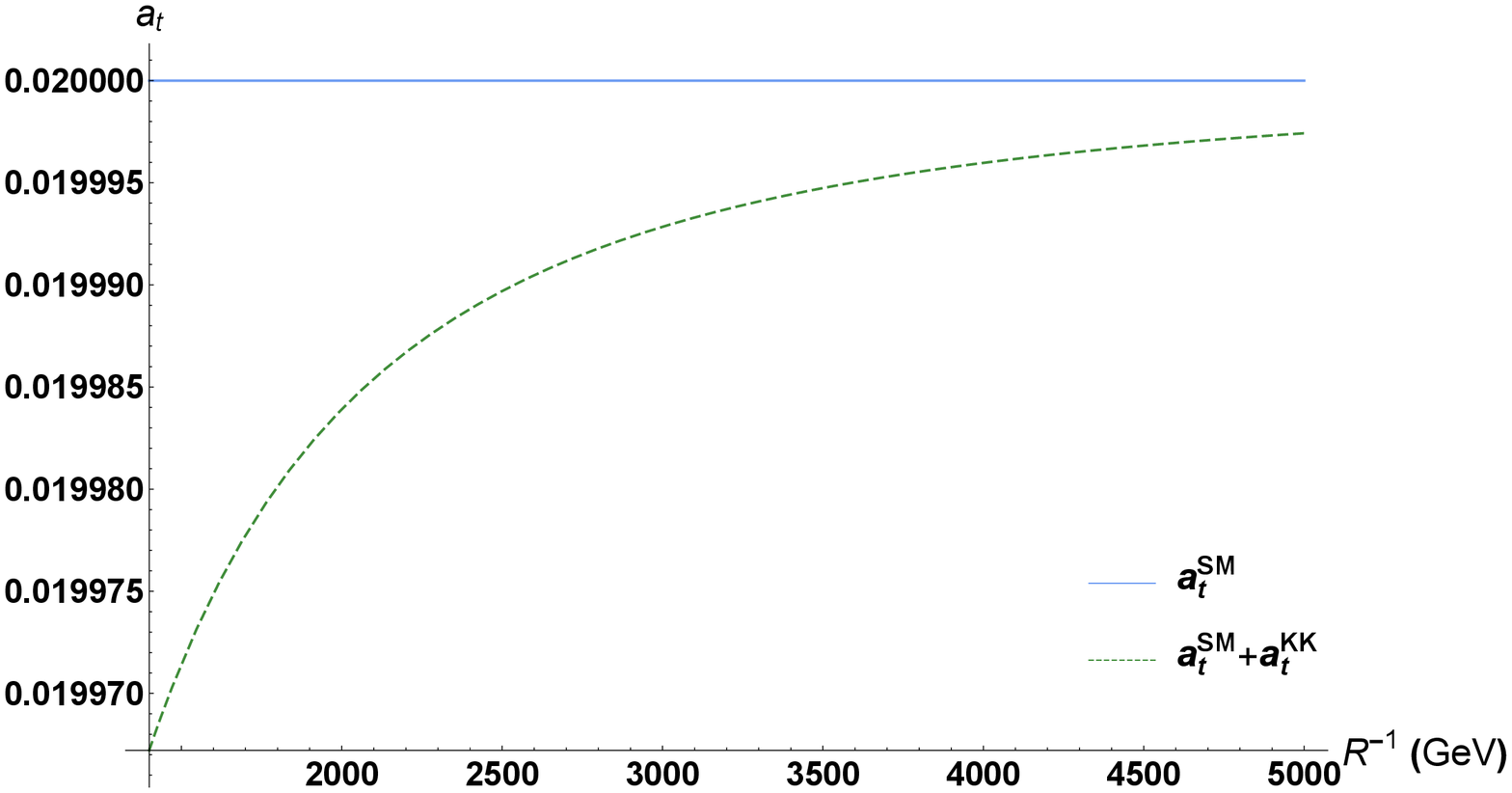}
\caption{\label{ammSMvsKK} \footnotesize{Prediction for the top-quark AMM from the whole set of KK fields, both zero and excited modes. {\it Upper graph:} contribution $a^{\rm SM}_t$ (horizontal line) versus $a^{\rm KK}_t$ (dashed curve). {\it Lower graph:} Contribution $a^{\rm SM}_t$ (horizontal line) versus the total contribution $a^{\rm SM}_t+a^{\rm KK}_t$ (dashed curve). In both graphs, GeVs have been taken as mass units.}}
\end{figure}

With the inclusion of the 4DSM prediction for the AMM of the top quark~\cite{BBGHLMR}, the graphs of Fig.~\ref{ammSMvsKK} are meant to provide a quantitative idea of how the contributions from KK zero and excited modes compare to each other. 
From here on, we denote the contribution from the 4DSM as $a^{\rm SM}_t$. The upper graph of Fig.~\ref{ammSMvsKK} displays plots for the absolute values of the contributions $a^{\rm SM}_t$ and $a^{\rm KK}_t$, which have been carried out in logarithmic scale. According to such a graph, the impact of this extra-dimensional physics on the AMM of the top quark is 3 to 4 orders of magnitude smaller than the numbers produced by the 4DSM, for a compactification scale within the range $1.4\,{\rm TeV}<R^{-1}<5\,{\rm TeV}$. The lower graph of Fig.~\ref{ammSMvsKK} exhibits the low-energy AMM contribution $a^{\rm SM}_t$, which corresponds to the horizontal line, together with the total contribution $a^{\rm SM}_t+a^{\rm KK}_t$, from both KK zero and excited modes, which has been represented by the dashed curve.

\subsection{Flavor-changing decays}
In this subsection, we implement results from the previous section to estimate the branching ratios for the decay processes $t^{(0)}\to A_\mu^{(0)}c^{(0)}$, $t^{(0)}\to A_\mu^{(0)}u^{(0)}$, and $c^{(0)}\to A_\mu^{(0)}u^{(0)}$, within the framework of the 5-dimensional Standard Model and its KK effective Lagrangian. 
The one-loop contribution from the 4DSM to the decay $t^{(0)}\to A_\mu^{(0)}c^{(0)}$ was calculated in Ref.~\cite{EHS}, with the branching ratio ${\rm Br}(t^{(0)}\to A_\mu^{(0)}c^{(0)})_{\rm SM}\sim10^{-12}-10^{-11}$ reported for a variety of values for the top-quark mass, not yet measured at the time. By using the {\it looptools} package~\cite{HaPe,OldVer}, we reproduced the results of this reference, which, according to up-to-date data reported by the {\it Particle Data Group} (2018)~\cite{PDG}, is given by ${\rm Br}(t^{(0)}\to A_\mu^{(0)}c^{(0)})_{\rm SM}\approx2.31\times10^{-13}$. Moreover, we have estimated the branching ratios ${\rm Br}(t^{(0)}\to A_\mu^{(0)}u^{(0)})_{\rm SM}\approx1.73\times10^{-15}$ and ${\rm Br}(c^{(0)}\to\ A_\mu^{(0)}u^{(0)})_{\rm SM}\approx9.59\times10^{-15}$ as well.\\

\begin{figure}[!ht]
\center
\includegraphics[width=8.5cm]{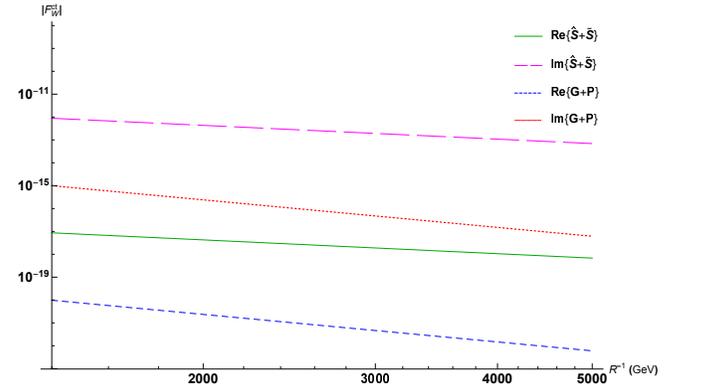}
\caption{\label{ttocampsM} \footnotesize{Real and imaginary parts of absolute values of contributions to form factors $M^{ct}_W$ and $E^{ct}_W$. Plots have been realized in logarithmic scale and GeVs have been taken as mass units. These contributions correspond to Eqs.~(\ref{GPMR4})-(\ref{S12ER2}). Compactification-scale values within $1.4\,{\rm TeV}<R^{-1}<5\,{\rm TeV}$ have been considered.}}
\end{figure}

For the moment, let us focus on the decay process $t^{(0)}\to A_\mu^{(0)}c^{(0)}$. A comparison among the contributions from Eqs.~(\ref{GPMR4})-(\ref{S12ER2}) to the magnetic and electric form factors $M^{ct}_W$ and $E^{ct}_W$ is presented in Fig.~\ref{ttocampsM}.
All the curves in this figure have been plotted in logarithmic scale.
The vertical axis corresponds to absolute values of either the real or the imaginary parts of vector and pseudo-Goldstone-boson contributions, Eqs.~(\ref{GPMR4}) and (\ref{GPER4}), or contributions from KK physical scalars, Eqs.~(\ref{S12MR2}) and (\ref{S12ER2}), with all these quantities plotted as functions of the compactification scale $R^{-1}$, in GeV units. Keeping in mind the case under consideration, $\alpha=t$ and $\beta=c$, we point out the presence of the factors $26m_{t^{(0)}}\pm7m_{c^{(0)}}$ and $5m_{t^{(0)}}\pm3m_{c^{(0)}}$ in the leading contributions given in Eqs.~(\ref{GPMR4})-(\ref{S12ER2}), from which we emphasize two aspects: (a) the only difference among magnetic and electric moments is the sign of charm-quark-mass terms in such factors; (b) these factors imply that terms proportional to the top-quark mass practically determine the contributions. Thus the magnetic- and electric-moment contributions are very similar to each other, and their plots look practically the same. In this practical sense we have represented both quantities by the very same curves in Fig.~\ref{ttocampsM}. These plots show that the imaginary part of the total scalar contribution, corresponding to Eqs.~(\ref{S12MR2}) and (\ref{S12ER2}) and represented by the long-dashed curve, introduces the dominant effects, which are larger than the leading contributions from KK vectors and pseudo-Goldstone bosons (dotted curve), produced by Eqs.~(\ref{GPMR4}) and (\ref{GPER4}), by about 3 orders of magnitude.  \\

\begin{figure}[!ht]
\center
\includegraphics[width=8.5cm]{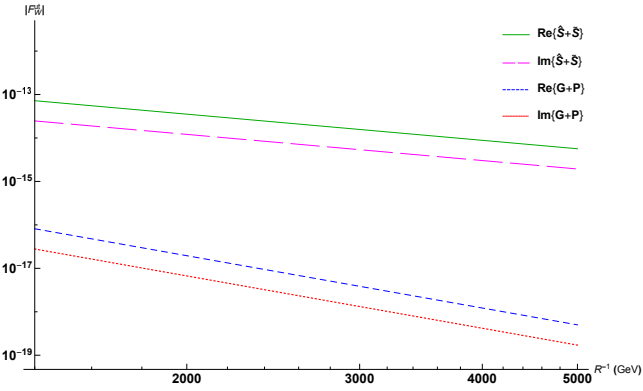}
\caption{\label{ttouampsM} \footnotesize{Real and imaginary parts of absolute values of contributions to form factors $M^{ut}_W$ and $E^{ut}_W$. Plots have been realized in logarithmic scale and GeVs have been taken as mass units. These contributions correspond to Eqs.~(\ref{GPMR4})-(\ref{S12ER2}). Compactification-scale values within $1.4\,{\rm TeV}<R^{-1}<5\,{\rm TeV}$ have been considered.}}
\end{figure}

Regarding the decay $t^{(0)}\to A_\mu^{(0)}u^{(0)}$, consider the contributions to $M^{ut}_W$ and $E^{ut}_W$, plotted in Fig.~\ref{ttouampsM}.
In this case, the two lower curves, which are the short-dashed and the dotted plots, represent the real and imaginary parts of Eqs.~(\ref{GPMR4}) and (\ref{GPER4}), corresponding to contributions from vector fields and pseudo-Goldstone bosons. On the other hand, the solid and the long-dashed plots, which are the upper curves in this figure, respectively represent the real and imaginary parts of Eqs.~(\ref{S12MR2}) and (\ref{S12ER2}), that is, the contributions from physical-scalar diagrams. Fig.~\ref{ttocampsM} then shows that the contributions to electric and magnetic form factors from physical scalars are larger that those from the vector fields and pseudo-Goldstone bosons by around 3 orders of magnitude. 
\\

\begin{figure}[!ht]
\center
\includegraphics[width=8.5cm]{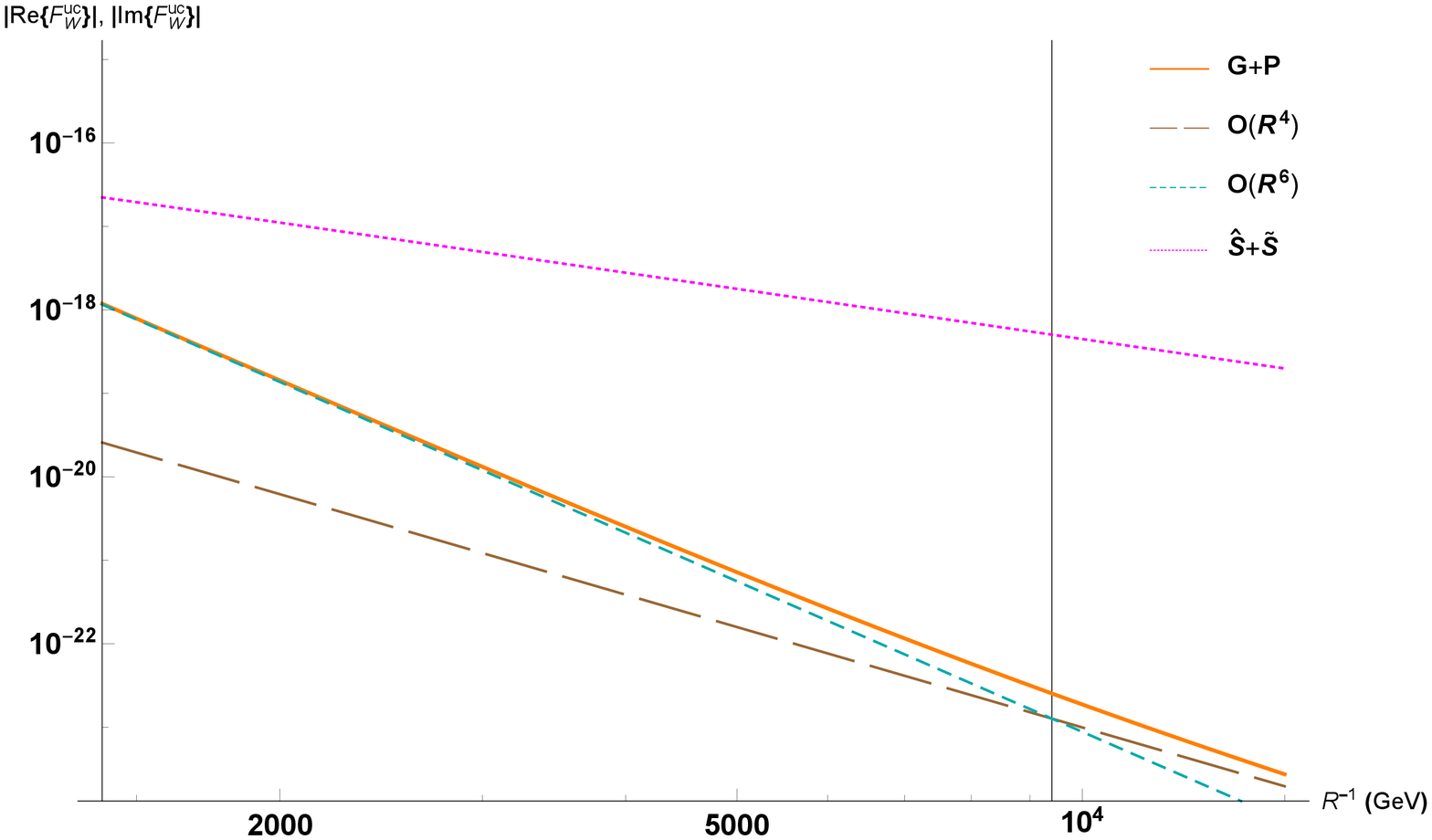}
\caption{\label{ctouamps} Real or imaginary parts of absolute values of contributions to form factors $M^{uc}_W$ and $E^{uc}_W$. Plots have been realized in logarithmic scale and GeVs have been taken as mass units. Compactification-scale values within $1.4\,{\rm TeV}<R^{-1}<15\,{\rm TeV}$ have been considered. The vertical solid line represents the value $R^{-1}\sim9.4\,{\rm TeV}$.}
\end{figure}

About the charm-quark decay $c^{(0)}\to A_\mu^{(0)}u^{(0)}$, it is worth commenting that for $R^{-1}\lesssim9.4\,{\rm TeV}$ the leading vector-field and pseudo-Goldstone-boson contributions do not come from $R^4$-order terms, shown explicitly in Eqs.~(\ref{GPMR4}) and (\ref{GPER4}). Terms of order $R^6$ lead this sort of contributions instead for such a range of compactification-scale values, which is displayed in Fig.~\ref{ctouamps}.
In the case of this decay process, real and imaginary parts of any contribution are very similar to each other, so either real or imaginary parts of contributions are represented by the same curves in Fig.~\ref{ctouamps}, with both types of quantities corresponding to the vertical axis. Moreover, magnetic and electric form factors are quite alike as well, so this figure also represents any of such cases. The dotted curve, which represents the dominant contributions, is produced by Feynman diagrams with virtual physical scalars, and is produced by the $R^2$-order terms given in Eqs.~(\ref{S12MR2}) and (\ref{S12ER2}). The solid plot depicts either real or imaginary parts of contributions from vector fields and pseudo-Goldstone bosons, taking terms of orders $R^4$ and $R^6$ at once. The long-dashed and the short-dashed curves correspond, respectively, to $R^4$- order and $R^6$-order contributions. Then notice that within $R^{-1}\lesssim9.4\,{\rm TeV}$ terms of order $R^6$ dominate, but at $R^{-1}\sim9.4\,{\rm TeV}$, indicated in the figure by the vertical solid line, such contributions are reached by $R^4$-order contributions, which then become dominant.\\

\begin{figure}[!ht]
\center
\includegraphics[width=8.5cm]{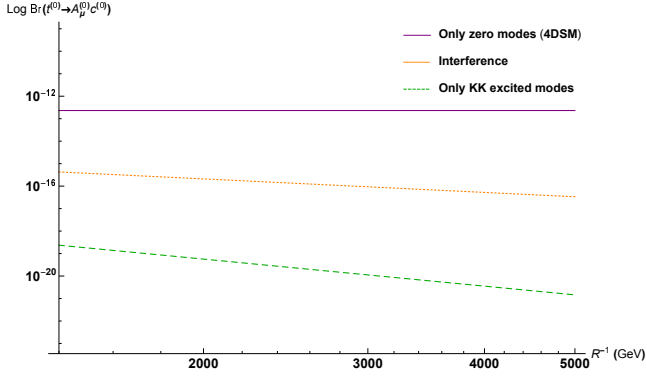}
\\ \vspace{0.7cm}
\includegraphics[width=8.3cm]{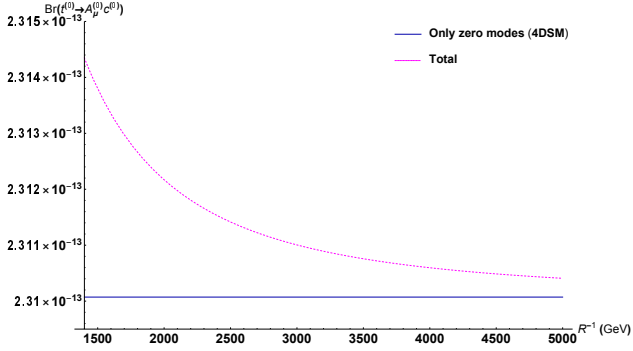}
\caption{\label{BRttoc} The branching ratio ${\rm BR}(t^{(0)}\to A_\mu^{(0)}c^{(0)})$. {\it Upper graph:} Comparison among contributions from the KK theory, in logarithmic scale. {\it Lower graph:} Total contribution VS the contribution from the 4DSM.}
\end{figure}

\begin{figure}[!ht]
\center
\includegraphics[width=8.5cm]{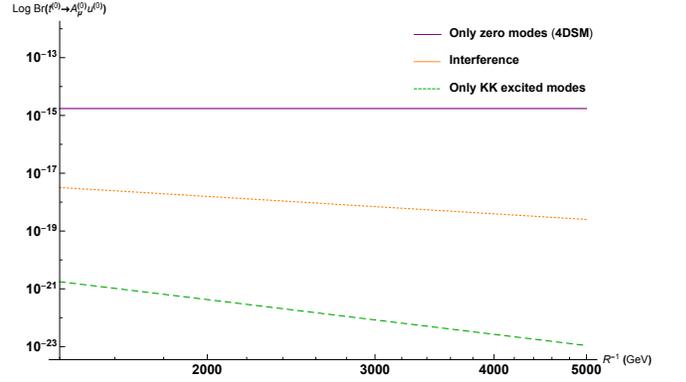}
\\ \vspace{0.7cm}
\includegraphics[width=8.3cm]{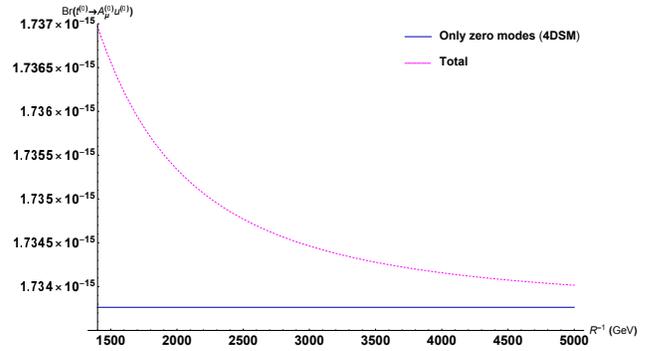}
\caption{\label{BRttou} The branching ratio ${\rm BR}(t^{(0)}\to A_\mu^{(0)}u^{(0)})$. {\it Upper graph:} Comparison among contributions from the KK theory, in logarithmic scale. {\it Lower graph:} Total contribution VS the contribution from the 4DSM.}
\end{figure}

\begin{figure}[!ht]
\center
\vspace{0.4cm}
\includegraphics[width=8.5cm]{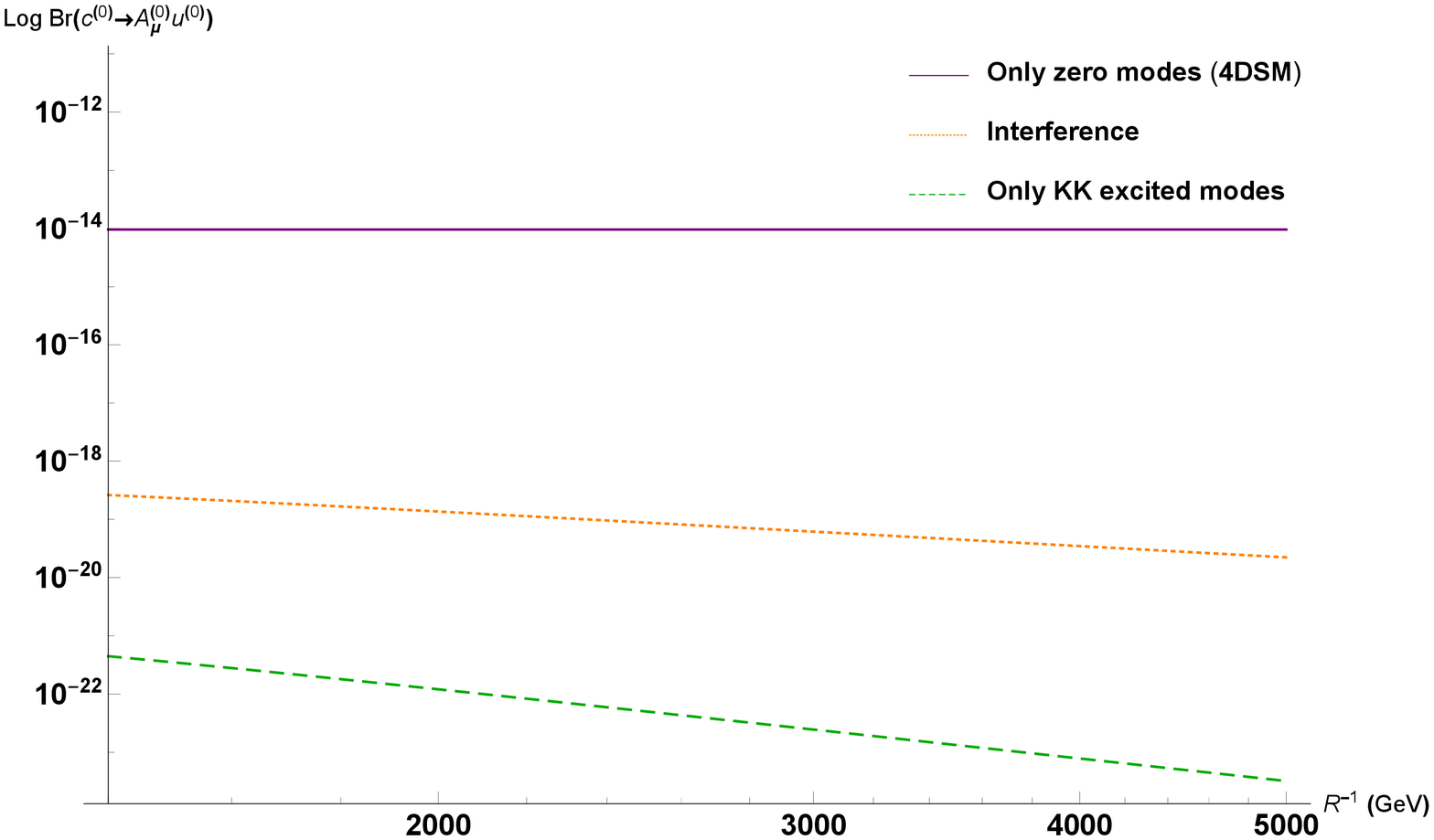}
\\ \vspace{0.7cm}
\includegraphics[width=8.3cm]{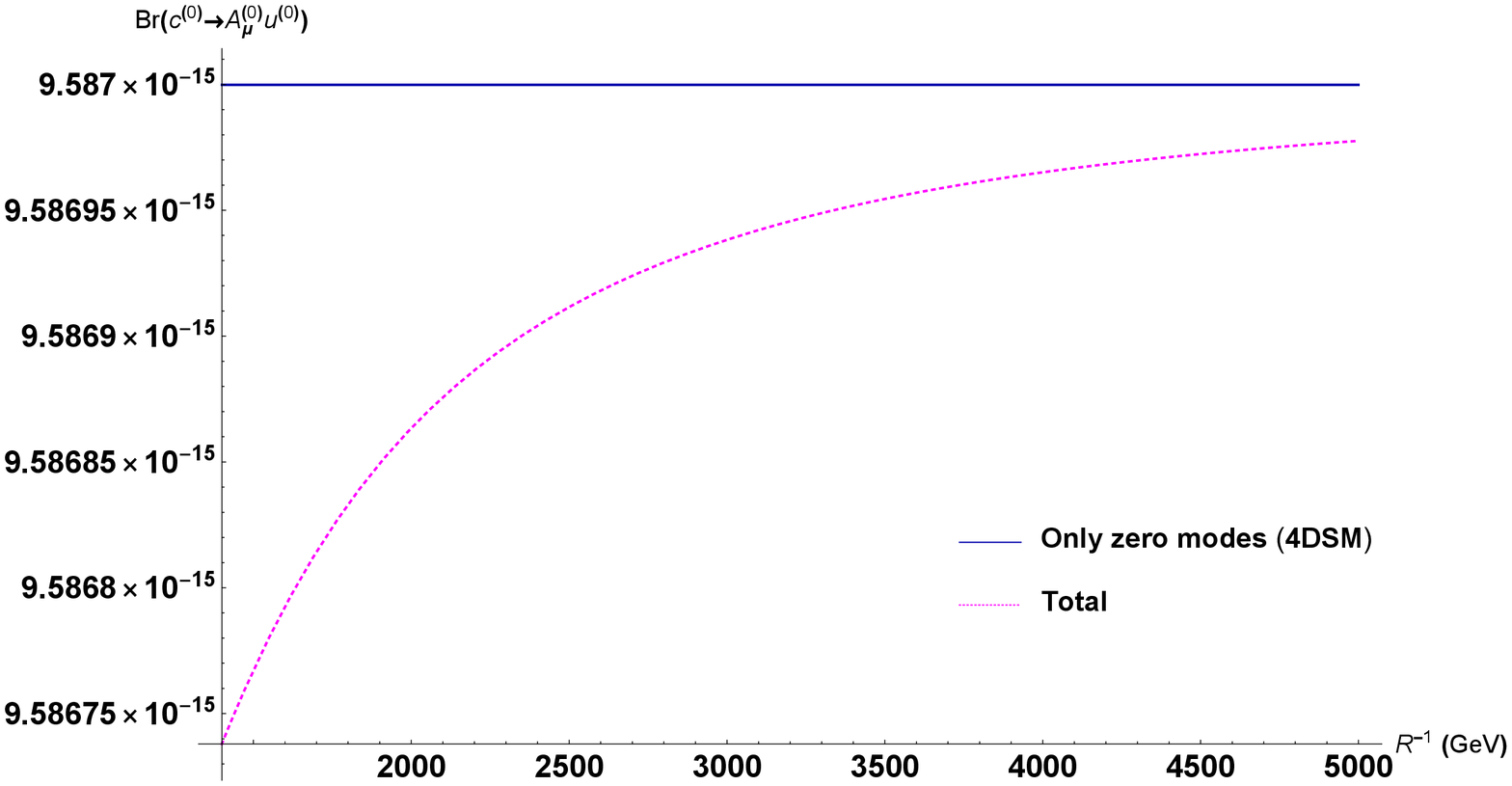}
\caption{\label{BRctou} The branching ratio ${\rm BR}(c^{(0)}\to A_\mu^{(0)}u^{(0)})$. {\it Upper graph:} Comparison among contributions from the KK theory, in logarithmic scale. {\it Lower graph:} Total contribution VS the contribution from the 4DSM.}
\end{figure}

In the next step, the branching ratios for the decays $u^{(0)}_\alpha\to A_\mu^{(0)}u^{(0)}_\beta$ are estimated, for which we use the decay-rate expression given in Eq.~(\ref{DR1to2}). Fig.~\ref{BRttoc} displays all the KK contributions, from both zero- and excited-mode fields, to ${\rm Br}(t^{(0)}\to A_\mu^{(0)}c^{(0)})$.
The upper graph, plotted in logarithmic scale, provides, within the compactification-scale range $1.4\,{\rm TeV}<R^{-1}<5\,{\rm TeV}$, our estimations of contributions from the sole 4DSM (solid plot), from the interference of zero and excited modes (dotted plot), and from KK excited-mode fields only (dashed plot). The quantitative difference between contributions from the 4DSM and from extra-dimensional physics amounts to about 3 to 4 orders of magnitude. Interference effects, which produce a positive branching-ratio contribution, dominate over contributions exclusively generated by KK excited modes. This was expected, since the interference term, corresponding to the second line of Eq.~(\ref{DR1to2}), involves the lowest power of the compactification radius $R$ in the whole contribution, in this case $R^2$, which translates into less-suppressed contributions than those given only by the set of KK excited modes, where the lowest compactification-radius suppression is $R^4$. The lower graph of Fig.~\ref{BRttoc} displays the 4DSM contribution, there represented by the solid horizontal line, and the contribution generated by the complete KK theory (dotted curve), which includes such low-energy-physics effects and which shows decoupling of new physics as $R^{-1}\to\infty$. In relation with the lower graph of Fig.~\ref{BRttoc}, let us mention Ref.~\cite{KDandJ}, where a calculation of ${\rm Br}(t^{(0)}\to A_\mu^{(0)}c^{(0)})$ has been reported. We find agreement with the results from this reference.
Regarding the two other decay processes, the corresponding estimations are given in Figs.~\ref{BRttou}, for ${\rm Br}(t^{(0)}\to A_\mu^{(0)}u^{(0)})$,
and \ref{BRctou}, for ${\rm Br}(c^{(0)}\to A_\mu^{(0)}u^{(0)})$.
The descriptions corresponding to the graphs of such figures are similar to that of Fig~\ref{BRttoc}.


\section{Conclusions}
\label{concs}
The Standard Model defined on 5 spacetime dimensions has been addressed in the present paper. 
The discussion performed in this work has included diverse aspects of the model, among which we emphasize its definition, its connection to the 4-dimensional Kaluza-Klein description, the determination of its complete set of physical Kaluza-Klein mass eigenfields, and gauge fixing. 
The main ingredients constituting the model are the same dynamic variables and symmetries as the ordinary 4-dimensional Standard Model, but with both elements defined on the extra-dimensional spacetime. Under the assumption that the extra dimension is orbifold compactified, a change of perspective, from 5 to 4 spacetime dimensions, was implemented on the model  through canonical maps that allow for the integration of the extra-dimensional coordinates in the action and thus yield the emergence of the 4-dimensional Kaluza-Klein effective theory, whose dynamic variables are the Kaluza-Klein modes, and which is manifestly governed by 4-dimensional Poincar\'e symmetry and by the gauge group of the 4-dimensional Standard Model. The gauge, scalar and fermion sectors of the Kaluza-Klein theory were revised in some detail. In particular, the generation of mass-term contributions by means of the Kaluza-Klein and Englert-Higgs mechanisms were discussed. Moreover, every transformation aimed at the definition of mass eigenfields was established and implemented. The emergence, in the Kaluza-Klein description, of two independent sets of gauge transformations, namely the standard and the nonstandard gauge transformations, made it possible to fix the gauge for Kaluza-Klein zero and excited modes by means of procedures that are independent of each other. A set of non-linear gauge-fixing functions were given in order to remove invariance with respect to the nonstandard gauge transformations, whereas the unitary gauge was utilized to fix gauge symmetry of the 4-dimensional Standard Model. \\

Phenomenological applications of the model have been investigated, analyzed and discussed in the present paper as well. Some extra attention was devoted to the quark sector of the Kaluza-Klein theory, from which Lagrangian terms, ready for the calculation of Feynman rules, were determined and explicitly shown, with the objective of calculating the whole set of Feynman diagrams that produce one-loop contributions to the $u$-quark electromagnetic vertex $A^{(0)}u^{(0)}_\alpha u^{(0)}_\alpha$ and to the quark-flavor-changing decay process $u^{(0)}_\alpha\to A^{(0)}_\mu u^{(0)}_\beta$. From the general parametrization of the electromagnetic vertex, new-physics contributions from Kaluza-Klein modes to the anomalous magnetic moments of $u$-type quarks and to the decay rate for $u^{(0)}_\alpha\to A^{(0)}_\mu u^{(0)}_\beta$ were determined, first by an exact calculation and then with the derivation of leading contributions from approximate expressions, valid in a scenario of large compactification scale, which is supported by current lower bounds. The results so obtained were implemented to particular cases of physical interest. Concretely, the Kaluza-Klein contributions to the anomalous magnetic moment of the top quark were estimated and compared with the prediction given by the 4-dimensional Standard Model, which turned out to be larger by 3 to 4 orders of magnitude for compactification scales between $1.4\,{\rm TeV}$ and $5\,{\rm TeV}$. The branching ratios ${\rm Br}\big(u^{(0)}_\alpha\to A^{(0)}_\mu u^{(0)}_\beta\big)$, suppressed by the Glashow-Iliopoulos-Maiani mechanism, were estimated. We determined that the extra-dimensional new-physics contribution to the branching ratio of $t^{(0)}\to A^{(0)}_\mu c^{(0)}$ is smaller than its 4-dimensional-Standard-Model counterpart by about 3 to 4 orders of magnitude for $1.4\,{\rm TeV}<R^{-1}<5\,{\rm TeV}$.


\begin{acknowledgments}
The authors acknowledge financial support from CONACYT and SNI (M\'exico). J. M. thanks C\'atedras CONACYT project 1753.
\end{acknowledgments}


\appendix

\section{Explicit expressions of Kaluza-Klein contributions to anomalous magnetic moments}
\label{AppffAMM}
Following the parametrization of the electromagnetic vertex, displayed in Eq.~(\ref{genAffpar}), the total contribution from the full set of KK excited modes to the AMM of $u$-type zero-mode quarks, which we denote by $a^{\rm KK}_\alpha$, is determined. In turn, $a^{\rm KK}_\alpha$ can be expressed as a sum of individual contributions, each corresponding to a kind of Feynman diagrams among those shown in Figs.~\ref{NeutContMMAs}-\ref{sdiags}: $a^{\rm KK}_\alpha=a^{\rm KK}_{Z,\alpha}+a^{\rm KK}_{A,\alpha}+a^{\rm KK}_{h,\alpha}+a^{\rm KK}_{W,\alpha}$, with $a^{\rm KK}_{Z,\alpha}$, $a^{\rm KK}_{A,\alpha}$, and $a^{\rm KK}_{h,\alpha}$ respectively generated by diagrams with virtual KK excited-modes associated to the $Z$ boson, the photon and the Higgs boson, all of them shown in Fig.~\ref{NeutContMMAs}, whereas $a^{\rm KK}_W$ represents the total contribution emerged from diagrams with loop KK excited modes that are related to the $W$ boson, provided in Figs.~\ref{vdiags}-\ref{sdiags}. Taking a further step, we write such contributing terms as the sum of all the individual contributions produced by the KK excited modes, namely $a^{\rm KK}_{Z,\alpha}=\sum_{k=1}^\infty a^{(k)}_{Z,\alpha}$, $a^{\rm KK}_{A,\alpha}=\sum_{k=1}^\infty a^{(k)}_{A,\alpha}$, $a^{\rm KK}_{h,\alpha}=\sum_{k=1}^\infty a^{(k)}_{h,\alpha}$, and $a^{\rm KK}_{W,\alpha}=\sum_{k=1}^\infty\sum_{\gamma=d,s,b} a^{(k)}_{W,\alpha,\gamma}$. KK-mode neutral-field contributions $a^{(k)}_{Z,\alpha}$, $a^{(k)}_{A,\alpha}$, and $a^{(k)}_{h,\alpha}$ come from diagrams with KK index $(k)$ fixed. Individual contributions $a^{\rm KK}_{W,\alpha,\gamma}$, from charged-boson KK excitations, are produced by Feynman diagrams characterized by a specific $(k)$ and fixed $d$-type-quark flavor, with entries of the Cabibbo-Kobayashi-Maskawa matrix $\kappa$ participating in the sum.\\

Now we present the exact expressions for the individual AMM contributions from the KK excited modes, written in terms of Passarino-Veltman scalar functions. To this aim, the set of scalar functions involved in the corresponding equations are shown next.
\\ \\
{\it Z-boson-related contributions:}
\begin{eqnarray}
B_0^1&=&B_0(0,m^2_{u^{(k)}_\alpha},m^2_{Z^{(k)}}),
\\
B_0^2&=&B_0(m^2_{u^{(0)}_\alpha},m^2_{u^{(k)}_\alpha},m^2_{Z^{(k)}}),
\\
B_0^3&=&B_0(0,m^2_{u^{(k)}_\alpha},m^2_{u^{(k)}_\alpha}),
\\
C_0^1&=&C_0(m^2_{u^{(0)}_\alpha},m^2_{u^{(0)}_\alpha},0,m^2_{u^{(k)}_\alpha},m^2_{Z^{(k)}},m^2_{u^{(k)}_\alpha}).
\end{eqnarray}
\\
{\it Photon-related contributions:}
\begin{eqnarray}
B_0^4&=&B_0(0,m^2_{A^{(k)}},m^2_{u^{(k)}_\alpha}),
\\
B_0^5&=&B_0(m^2_{u^{(0)}_\alpha},m^2_{A^{(k)}},m^2_{u^{(k)}_\alpha}),
\\
B_0^6&=&B_0(0,m^2_{u^{(k)}_\alpha},m^2_{u^{(k)}_\alpha}),
\\
C_0^2&=&C_0(m^2_{u^{(0)}_\alpha},m^2_{u^{(0)}_\alpha},0,m^2_{u^{(k)}_\alpha},m^2_{A^{(k)}},m^2_{u^{(k)}_\alpha}).
\end{eqnarray}
\\
{\it Higgs-boson-related contributions:}
\begin{eqnarray}
B_0^7&=&B_0(0,m^2_{h^{(k)}},m^2_{u^{(k)}_\alpha}),
\\
B_0^8&=&B_0(m^2_{u^{(0)}_\alpha},m^2_{h^{(k)}},m^2_{u^{(k)}_\alpha}),
\\
B_0^9&=&B_0(0,m^2_{u^{(k)}_\alpha},m^2_{u^{(k)}_\alpha}),
\\
C_0^3&=&C_0(m^2_{u^{(0)}_\alpha},m^2_{u^{(0)}_\alpha},0,m^2_{u^{(k)}_\alpha},m^2_{h^{(k)}},m^2_{u^{(k)}_\alpha}).
\end{eqnarray}
\\
{\it W-boson-related contributions:}
\begin{eqnarray}
B_0^{10}&=&B_0(0,m^2_{d^{(k)}_\gamma},m^2_{d^{(k)}_\gamma}),
\\
B_0^{11}&=&B_0(0,m^2_{d^{(k)}_\gamma},m^2_{W^{(k)}}),
\label{PaVeB11}
\\
B_0^{12}&=&B_0(0,m^2_{W^{(k)}},m^2_{W^{(k)}}),
\\
B_0^{13}&=&B_0(m^2_{u^{(0)}_\alpha},m^2_{d^{(k)}_\gamma},m^2_{W^{(k)}}),
\label{PaVeB13}
\\
C_0^4&=&C_0(m^2_{u^{(0)}_\alpha},m^2_{u^{(0)}_\alpha},0,m^2_{d^{(k)}_\gamma},m^2_{W^{(k)}},m^2_{d^{(k)}_\gamma}),
\\
C_0^5&=&C_0(m^2_{u^{(0)}_\alpha},m^2_{u^{(0)}_\alpha},0,m^2_{W^{(k)}},m^2_{d^{(k)}_\gamma},m^2_{W^{(k)}}).
\nonumber \\
\end{eqnarray}
Then the contributions from the KK excited modes read
\begin{widetext}
\begin{eqnarray}
&&
a^{(k)}_{Z,\alpha}=\frac{\alpha }{1728\,\pi s_W^2 m_{u^{(0)}_\alpha}^2 m_{W^{(0)}}^2m_{Z^{(k)}}^2}
\Big\{
\Big[
-2 \big(m_{u^{(k)}_\alpha}-m_{Z^{(k)}}\big)\big(m_{u^{(k)}_\alpha}+m_{Z^{(k)}}\big)
\nonumber \\ &&
\times\big(9\big(-\big(m_{(k)}^2-2 m_{u^{(k)}_\alpha}^2\big)\big(m_{(k)}^2+m_{Z^{(0)}}^2\big)-2m_{Z^{(k)}}^2 (m^2_{(k)}-m^2_{Z^{(0)}})+m_{Z^{(k)}}^4\big)
\nonumber \\ &&
+32m_{Z^{(0)}}^2 s_W^4\left(m_{(k)}^2+m_{Z^{(0)}}^2+2m_{Z^{(k)}}^2\right)-72 m_{Z^{(0)}}^2m_{Z^{(k)}}^2 s_W^2\big)
\Big]
(B_0^1-B_0^2)
\nonumber \\ &&
+
\Big[
9 m_{(k)}^4 \big(m_{u^{(0)}_\alpha}^2-3m_{u^{(k)}_\alpha}^2+3m_{Z^{(k)}}^2\big)+m_{(k)}^2\big(m_{u^{(0)}_\alpha}^2 \big(-90m_{u^{(k)}_\alpha}^2+m_{Z^{(0)}}^2 (9-160s_W^4)
\nonumber \\ &&
+90 m_{Z^{(k)}}^2\big)+3(m_{u^{(k)}_\alpha}-m_{Z^{(k)}})(m_{u^{(k)}_\alpha}+m_{Z^{(k)}}) \big(18m_{u^{(k)}_\alpha}^2+m_{Z^{(0)}}^2 (32s_W^4-9)-18m_{Z^{(k)}}^2\big)\big)
\nonumber \\ &&
-m_{u^{(0)}_\alpha}^2 \big(90 m_{u^{(k)}_\alpha}^2m_{Z^{(0)}}^2+160 m_{Z^{(0)}}^4 s_W^4+2m_{Z^{(0)}}^2 m_{Z^{(k)}}^2 (4 s_W^2 (8s_W^2-21)-63)+9 m_{Z^{(k)}}^4\big)
\nonumber \\ &&
+3(m_{u^{(k)}_\alpha}-m_{Z^{(k)}})(m_{u^{(k)}_\alpha}+m_{Z^{(k)}}) \big(18m_{u^{(k)}_\alpha}^2 m_{Z^{(0)}}^2+32m_{Z^{(0)}}^4 s_W^4
\nonumber \\ &&
+2 m_{Z^{(0)}}^2m_{Z^{(k)}}^2 (4 s_W^2-3) (8s_W^2-3)+9 m_{Z^{(k)}}^4\big)
\Big]
(B_0^2-B_0^3)
+
\Big[
9 m_{(k)}^4 \big(m_{u^{(0)}_\alpha}^4
\nonumber \\ &&
+2m_{u^{(0)}_\alpha}^2 \big(m_{u^{(k)}_\alpha}^2+m_{Z^{(k)}}^2\big)-3\big(m_{u^{(k)}_\alpha}^2-m_{Z^{(k)}}^2\big)^2\big)+m_{(k)}^2\big(m_{u^{(0)}_\alpha}^4 \big(54m_{u^{(k)}_\alpha}^2+m_{Z^{(0)}}^2(96s_W^4+9)
\nonumber \\ &&
-54 m_{Z^{(k)}}^2\big)+2m_{u^{(0)}_\alpha}^2 \big(32 m_{Z^{(0)}}^2s_W^4 \big(m_{Z^{(k)}}^2-3m_{u^{(k)}_\alpha}^2\big)+9 \big(-6m_{u^{(k)}_\alpha}^4+m_{u^{(k)}_\alpha}^2\big(m_{Z^{(0)}}^2
\nonumber \\ &&
+8m_{Z^{(k)}}^2\big)+m_{Z^{(k)}}^2\big(m_{Z^{(0)}}^2-2m_{Z^{(k)}}^2\big)\big)\big)+3(m_{u^{(k)}_\alpha}-m_{Z^{(k)}})^2(m_{u^{(k)}_\alpha}+m_{Z^{(k)}})^2 \big(18m_{u^{(k)}_\alpha}^2
\nonumber \\ &&
+m_{Z^{(0)}}^2(32s_W^4-9)-18m_{Z^{(k)}}^2\big)\big)+\text{mu0$\alpha$}^4 \big(54 m_{u^{(k)}_\alpha}^2m_{Z^{(0)}}^2+96 m_{Z^{(0)}}^4 s_W4-2m_{Z^{(0)}}^2 m_{Z^{(k)}}^2 
\nonumber \\ &&
\times(4 s_W^2 (8s_W^2+3)-63)-9 m_{Z^{(k)}}^4\big)-2m_{u^{(0)}_\alpha}^2 \big(54m_{u^{(k)}_\alpha}^4m_{Z^{(0)}}^2
+m_{u^{(k)}_\alpha}^2 \big(96m_{Z^{(0)}}^4 s_W^4
\nonumber \\ &&
+8 m_{Z^{(0)}}^2m_{Z^{(k)}}^2 (s_W^2 (8s_W^2-15)-9)+9 m_{Z^{(k)}}^4\big)-32m_{Z^{(0)}}^4 m_{Z^{(k)}}^2 s_W4+2m_{Z^{(0)}}^2 m_{Z^{(k)}}^4 
\nonumber \\ &&
\times(4 s_W^2 (8s_W^2-3)+45)+9 m_{Z^{(k)}}^6\big)+3(m_{u^{(k)}_\alpha}-m_{Z^{(k)}})^2(m_{u^{(k)}_\alpha}+m_{Z^{(k)}})^2 \big(18m_{u^{(k)}_\alpha}^2 m_{Z^{(0)}}^2
\nonumber \\ &&
+32m_{Z^{(0)}}^4 s_W4+2 m_{Z^{(0)}}^2m_{Z^{(k)}}^2 (4 s_W^2-3) (8s_W^2-3)+9 m_{Z^{(k)}}^4\big)
\Big]
C_0^1
+2 m_{u^{(0)}_\alpha}^2 \big(9\big(m_{(k)}^2
\nonumber \\ &&
-2 m_{u^{(k)}_\alpha}^2\big)
\big(m_{(k)}^2+m_{Z^{(0)}}^2\big)-32m_{Z^{(0)}}^2 s_W4\big(m_{(k)}^2+m_{Z^{(0)}}^2+2m_{Z^{(k)}}^2\big)+18 m_{Z^{(k)}}^2(m_{(k)}
\nonumber \\ &&
-m_{Z^{(0)}})(m_{(k)}+m_{Z^{(0)}})+72 m_{Z^{(0)}}^2m_{Z^{(k)}}^2 s_W^2-9 m_{Z^{(k)}}^4\big)
\Big\},
\end{eqnarray}
\begin{eqnarray}
&&
a^{(k)}_{A,\alpha}=\frac{\alpha }{54 \pi\,m_{u^{(0)}_\alpha}^2}
\Big\{
-6 \big(m_{A^{(k)}}-m_{u^{(k)}_\alpha}\big) \big(m_{A^{(k)}}+m_{u^{(k)}_\alpha}\big)
(B_0^4-B_0^5)
\nonumber \\ &&
+\big(
9 m_{A^{(k)}}^2+7 m_{u^{(0)}_\alpha}^2-9 m_{u^{(k)}_\alpha}^2
\big)
(B_0^5-B_0^6)
+\Big[
-9 m_{A^{(k)}}^4+2 m_{A^{(k)}}^2 \big(m_{u^{(0)}_\alpha}^2+9 m_{u^{(k)}_\alpha}^2\big)
\nonumber \\ &&
-m_{u^{(0)}_\alpha}^4+10 m_{u^{(0)}_\alpha}^2 m_{u^{(k)}_\alpha}^2-9m_{u^{(k)}_\alpha}^4
\Big]
C_0^2
+6 m_{u^{(0)}_\alpha}^2
\Big\},
\end{eqnarray}
\begin{eqnarray}
&&a^{(k)}_{h,\alpha}=\frac{-\alpha}{96 \pi  m_{W^{(0)}}^2 s_W^2}
\Big\{
2 \big(m^2_{h^{(k)}}-m^2_{u^{(k)}_\alpha}\big)
(B_0^7-B_0^8)
+
3 \big(-m_{h^{(k)}}^2+m_{u^{(0)}_\alpha}^2
\nonumber \\ &&
+m_{u^{(k)}_\alpha}^2\big)
(B_0^8-B_0^9)
+\Big[
3 m_{h^{(k)}}^4-6 m_{h^{(k)}}^2 \big(m_{u^{(0)}_\alpha}^2+m_{u^{(k)}_\alpha}^2\big)-5 m_{u^{(0)}_\alpha}^4
\nonumber \\ &&
+2 m_{u^{(0)}_\alpha}^2 m_{u^{(k)}_\alpha}^2+3m_{u^{(k)}_\alpha}^4
\Big]
C_0^3
-2 m_{u^{(0)}_\alpha}^2
\Big\},
\end{eqnarray}
\begin{eqnarray}
&&
a^{(k)}_{W,\alpha,\gamma}=
\frac{\alpha }{192 \pi  m_{u^{(0)}_\alpha}^2 m_{W^{(0)}}^2
   m_{W^{(k)}}^2 s_W^2}
\Big\{
\Big[
m_{W^{(k)}}^4 \big(-3 m_{d^{(k)}_\gamma}^2-6m_{(k)}^2+m_{u^{(0)}_\alpha}^2+6m_{W^{(0)}}^2\big)
\nonumber \\ &&
+m_{W^{(k)}}^2\big(m_{d^{(k)}_\gamma}^2 \big(9m_{(k)}^2-3m_{W^{(0)}}^2\big)-m_{u^{(0)}_\alpha}^2\big(7 m_{(k)}^2+11m_{W^{(0)}}^2\big)\big)+\big(-3m_{d^{(k)}_\gamma}^4
\nonumber \\ &&
+6 m_{d^{(k)}_\gamma}^2m_{u^{(0)}_\alpha}^2+m_{u^{(0)}_\alpha}^4\big)\big(m_{(k)}^2+m_{W^{(0)}}^2\big)+3m_{W^{(k)}}^6
\Big]
(B_0^{10}-B_0^{11})
+
\Big[
m_{W^{(k)}}^4 \big(m_{d^{(k)}_\gamma}^2+2m_{(k)}^2
\nonumber \\ &&
+m_{u^{(0)}_\alpha}^2-2m_{W^{(0)}}^2\big)+m_{W^{(k)}}^2\big(m_{d^{(k)}_\gamma}^2\big(m_{W^{(0)}}^2-3m_{(k)}^2\big)-m_{u^{(0)}_\alpha}^2\big(11 m_{(k)}^2+15m_{W^{(0)}}^2\big)\big)
\nonumber \\ &&
+\big(m_{d^{(k)}_\gamma}^4+10 m_{d^{(k)}_\gamma}^2m_{u^{(0)}_\alpha}^2+m_{u^{(0)}_\alpha}^4\big)\big(m_{(k)}^2+m_{W^{(0)}}^2\big)-m_{W^{(k)}}^6
\Big]
(B_0^{11}-B_0^{12})
\nonumber \\&&
+
2\Big[
 m_{W^{(k)}}^4 \big(5 m_{d^{(k)}_\gamma}^2+2 \big(5 m_{(k)}^2+m_{u^{(0)}_\alpha}^2-5 m_{W^{(0)}}^2\big)\big)+m_{W^{(k)}}^2\big(5 m_{W^{(0)}}^2 \big(m_{d^{(k)}_\gamma}^2+3 m_{u^{(0)}_\alpha}^2\big)
\nonumber \\ &&
-m_{(k)}^2 \big(15m_{d^{(k)}_\gamma}^2+m_{u^{(0)}_\alpha}^2\big)\big)+\big(5 m_{d^{(k)}_\gamma}^4-m_{d^{(k)}_\gamma}^2 m_{u^{(0)}_\alpha}^2+2 m_{u^{(0)}_\alpha}^4\big)\big(m_{(k)}^2+m_{W^{(0)}}^2\big)
\nonumber \\ &&
-5m_{W^{(k)}}^6
\Big]
(B_0^{12}-B_0^{13})
+
\Big[
-2 m_{W^{(k)}}^6 \big(3 m_{d^{(k)}_\gamma}^2+3m_{(k)}^2+m_{u^{(0)}_\alpha}^2-3m_{W^{(0)}}^2\big)
\nonumber \\ &&
+\big(m_{d^{(k)}_\gamma}^2-m_{u^{(0)}_\alpha}^2\big)^2 \big(3m_{d^{(k)}_\gamma}^2-m_{u^{(0)}_\alpha}^2\big)\big(m_{(k)}^2+m_{W^{(0)}}^2\big)+m_{W^{(k)}}^4 \big(3 m_{d^{(k)}_\gamma}^4+m_{d^{(k)}_\gamma}^2 \big(15m_{(k)}^2
\nonumber \\ &&
-2 m_{u^{(0)}_\alpha}^2-9m_{W^{(0)}}^2\big)-m_{u^{(0)}_\alpha}^2\big(m_{(k)}^2+m_{u^{(0)}_\alpha}^2+17m_{W^{(0)}}^2\big)\big)+4 m_{W^{(k)}}^2\big(-3 m_{d^{(k)}_\gamma}^4 m_{(k)}^2
\nonumber \\ &&
+3m_{d^{(k)}_\gamma}^2 m_{u^{(0)}_\alpha}^2\big(m_{(k)}^2+m_{W^{(0)}}^2\big)+m_{u^{(0)}_\alpha}^4 \big(3 m_{W^{(0)}}^2-2m_{(k)}^2\big)\big)+3 m_{W^{(k)}}^8
\Big]
C_0^4
\nonumber \\&&
+
3\Big[
-2 m_{W^{(k)}}^6 \big(3 m_{d^{(k)}_\gamma}^2+3 m_{(k)}^2+m_{u^{(0)}_\alpha}^2-3m_{W^{(0)}}^2\big)+\big(m_{d^{(k)}_\gamma}^2-m_{u^{(0)}_\alpha}^2\big)^2 \big(3m_{d^{(k)}_\gamma}^2
\nonumber \\ &&
-m_{u^{(0)}_\alpha}^2\big)\big(m_{(k)}^2+m_{W^{(0)}}^2\big)+m_{W^{(k)}}^4 \big(3 m_{d^{(k)}_\gamma}^4+m_{d^{(k)}_\gamma}^2 \big(15m_{(k)}^2-2 m_{u^{(0)}_\alpha}^2-9m_{W^{(0)}}^2\big)
\nonumber \\ &&
-m_{u^{(0)}_\alpha}^2\big(m_{(k)}^2+m_{u^{(0)}_\alpha}^2+17m_{W^{(0)}}^2\big)\big)
+4 m_{W^{(k)}}^2\big(-3 m_{d^{(k)}_\gamma}^4 m_{(k)}^2+3m_{d^{(k)}_\gamma}^2 m_{u^{(0)}_\alpha}^2\big(m_{(k)}^2
\nonumber \\ &&
+m_{W^{(0)}}^2\big)+m_{u^{(0)}_\alpha}^4 \big(3 m_{W^{(0)}}^2-2m_{(k)}^2\big)\big)+3m_{W^{(k)}}^8
\Big]
C_0^5
-8 m_{u^{(0)}_\alpha}^2\big(\big(m_{d^{(k)}_\gamma}^2+m_{u^{(0)}_\alpha}^2\big)
\nonumber \\ &&
\times\big(m_{(k)}^2+m_{W^{(0)}}^2\big)+2m_{W^{(k)}}^2\big(m_{W^{(0)}}^2-m_{(k)}^2\big)+m_{W^{(k)}}^4\big)
\Big\}.
\end{eqnarray}
\end{widetext}


\section{Explicit expressions of Kaluza-Klein contributions to $u^{(0)}_\alpha\to u^{(0)}_\beta A^{(0)}_\mu$}
\label{AppffFCD}
In the present appendix, we provide explicit expressions for the KK form-factor contributions $G^{(k)\beta\alpha}_{M,\gamma}$, $P^{(k)\beta\alpha}_{M,\gamma}$, $\hat{S}^{(k)\beta\alpha}_{M,\gamma}$, $\tilde{S}^{(k)\beta\alpha}_{M,\gamma}$, $G^{(k)\beta\alpha}_{E,\gamma}$, $P^{(k)\beta\alpha}_{E,\gamma}$, $\hat{S}^{(k)\beta\alpha}_{E,\gamma}$, and $\tilde{S}^{(k)\beta\alpha}_{E,\gamma}$, which determine the form factors $M^{(k)\beta\alpha}_{W,\gamma}$ and $E^{(k)\beta\alpha}_{W,\gamma}$, as stated by Eqs.~(\ref{contsfsM}) and (\ref{contsfsE}). The form-factor contributions $M^{(k)\beta\alpha}_{W,\gamma}$ and $E^{(k)\beta\alpha}_{W,\gamma}$ are then KK summed to define the magnetic and electric form factors $M^{\beta\alpha}_W$ and $E^{\beta\alpha}_W$, in accordance with Eqs.~(\ref{fMff}) and (\ref{fEff}). We define
\begin{equation}
\zeta_{\alpha\beta}=\frac{i\alpha^{3/2}}{12\sqrt{\pi}m_{u^{(0)}_\alpha}m_{u^{(0)}_\beta} \big(m_{u^{(0)}_\alpha}^2-m_{u^{(0)}_\beta}^2\big)^2\sin^2\theta_W}.
\end{equation}

The set of Passarino-Veltman scalar functions involved in this calculation are $B_0^{11}$ and $B_0^{13}$, defined in Eqs.~(\ref{PaVeB11}) and (\ref{PaVeB13}), and also
\begin{eqnarray}
B_0^{14}&=&B_0(m^2_{u^{(0)}_\beta},m^2_{d^{(k)}_\gamma},m^2_{W^{(k)}}),
\\
C_0^6&=&C_0\big(m^2_{u^{(0)}_\alpha},m^2_{u^{(0)}_\beta},0,m_{d^{(k)}_\gamma}^2,m_{W^{(k)}}^2,m_{d^{(k)}_\gamma}^2\big),
\\
C_0^7&=&C_0\big(m^2_{u^{(0)}_\alpha},m^2_{u^{(0)}_\beta},0,m_{W^{(k)}}^2,m_{d^{(k)}_\gamma}^2,m_{W^{(k)}}^2\big),
\label{C2}
\end{eqnarray}
with the case $(k)=(0)$ included. Then the Kaluza-Klein excited-mode contributions are expressed explicitly as
\begin{widetext}
\begin{eqnarray}
&&
G^{(k)\beta\alpha}_{{\rm M},\gamma}=\zeta_{\alpha\beta}
\Big[
\big(m_{d^{(k)}_\gamma}^2-m_{W^{(k)}}^2\big)\big(m_{u^{(0)}_\alpha}-m_{u^{(0)}_\beta}\big)^2 \big(m_{u^{(0)}_\alpha}+m_{u^{(0)}_\beta}\big)
(B_0^{11}-B_0^{13})
\nonumber \\&&
+m_{u^{(0)}_\alpha}\big(m_{u^{(0)}_\alpha}-m_{u^{(0)}_\beta}\big) \Big(m_{d^{(k)}_\gamma}^2\big(m_{u^{(0)}_\alpha}+2m_{u^{(0)}_\beta}\big)+2m_{u^{(0)}_\beta}\big(m_{u^{(0)}_\alpha}^2+3m_{u^{(0)}_\alpha}m_{u^{(0)}_\beta}+m_{u^{(0)}_\beta}^2\big)
\nonumber \\ &&
-m_{W^{(k)}}^2\big(m_{u^{(0)}_\alpha}+2m_{u^{(0)}_\beta}\big)\Big)
(B_0^{13}-B_0^{14})
-m_{d^{(k)}_\gamma}^2m_{u^{(0)}_\alpha}m_{u^{(0)}_\beta}\big(m_{u^{(0)}_\alpha}-m_{u^{(0)}_\beta}\big)^2\big(m_{u^{(0)}_\alpha}+m_{u^{(0)}_\beta}\big)
C_0^6
\nonumber \\ &&
+3 m_{u^{(0)}_\alpha}m_{u^{(0)}_\beta}\big(m_{u^{(0)}_\alpha}-m_{u^{(0)}_\beta}\big)^2\big(m_{u^{(0)}_\alpha}+m_{u^{(0)}_\beta}\big)\Big(\big(m_{u^{(0)}_\alpha}+m_{u^{(0)}_\beta}\big)^2-m^2_{W^{(k)}}\Big)
C_0^7
\nonumber \\ &&
-2m_{u^{(0)}_\alpha}m_{u^{(0)}_\beta}\big(m_{u^{(0)}_\alpha}-m_{u^{(0)}_\beta}\big)^2\big(m_{u^{(0)}_\alpha}+m_{u^{(0)}_\beta}\big)
\Big],
\end{eqnarray}
\begin{eqnarray}
&&
P^{(k)\beta\alpha}_{{\rm M},\gamma}
=
\frac{\zeta_{\alpha\beta}}{2\,m^2_{W^{(k)}}}
\Big[
\big(m^2_{d^{(k)}_\gamma}-m^2_{W^{(k)}}\big)\big(m_{u^{(0)}_\alpha}-m_{u^{(0)}_\beta}\big)^2 \big(m_{u^{(0)}_\alpha}+m_{u^{(0)}_\beta}\big)\big(m_{d^{(k)}_\gamma}^2+m_{u^{(0)}_\alpha}m_{u^{(0)}_\beta}\big)
(B_0^{11}-B_0^{13})
\nonumber \\&&
+
m_{u^{(0)}_\alpha}\big(m_{u^{(0)}_\alpha}-m_{u^{(0)}_\beta}\big)\Big(m_{d^{(k)}_\gamma}^4m_{u^{(0)}_\alpha}+2m_{d^{(k)}_\gamma}^4m_{u^{(0)}_\beta}-m_{d^{(k)}_\gamma}^2m_{u^{(0)}_\alpha}^2m_{u^{(0)}_\beta}
\nonumber \\ &&
-m_{W^{(k)}}^2 \big(m_{u^{(0)}_\alpha}+2m_{u^{(0)}_\beta}\big) \big(m_{d^{(k)}_\gamma}^2+m_{u^{(0)}_\alpha}m_{u^{(0)}_\beta}\big)-2m_{d^{(k)}_\gamma}^2m_{u^{(0)}_\beta}^3+2m_{u^{(0)}_\alpha}^2m_{u^{(0)}_\beta}^3\Big)
(B_0^{13}-B_0^{14})
\nonumber \\ &&
+
m_{d^{(k)}_\gamma}^2m_{u^{(0)}_\alpha}m_{u^{(0)}_\beta}\big(m_{u^{(0)}_\alpha}-m_{u^{(0)}_\beta}\big)^2(m_{u^{(0)}_\alpha}+m_{u^{(0)}_\beta})\big(-m_{d^{(k)}_\gamma}^2+m_{u^{(0)}_\alpha}^2+m_{u^{(0)}_\alpha}m_{u^{(0)}_\beta}+m_{u^{(0)}_\beta}^2\big)
C_0^6
\nonumber \\ &&
-3\,m_{u^{(0)}_\alpha}m_{u^{(0)}_\beta}m_{W^{(k)}}^2\big(m_{u^{(0)}_\alpha}-m_{u^{(0)}_\beta}\big)^2 \big(m_{u^{(0)}_\alpha}+m_{u^{(0)}_\beta}\big)\big(m_{d^{(k)}_\gamma}^2+m_{u^{(0)}_\alpha}m_{u^{(0)}_\beta}\big)
C_0^7
\nonumber \\ &&
-2\,m_{u^{(0)}_\alpha}m_{u^{(0)}_\beta}\big(m_{u^{(0)}_\alpha}-m_{u^{(0)}_\beta}\big)^2\big(m_{u^{(0)}_\alpha}+m_{u^{(0)}_\beta}\big)\big(m_{d^{(k)}_\gamma}^2+m_{u^{(0)}_\alpha}m_{u^{(0)}_\beta}\big)
\Big],
\end{eqnarray}
\begin{eqnarray}
&&
\hat{S}^{(k)\beta\alpha}_{{\rm M},\gamma}
=
\frac{\zeta_{\alpha\beta}\sin^2\theta_{d^{(k)}_\gamma}}{2\,m^2_{W^{(0)}}m^2_{W^{(k)}}}
\Big[
\big(m^2_{d^{(k)}_\gamma}-m^2_{W^{(k)}}\big)\big(m_{u^{(0)}_\alpha}-m_{u^{(0)}_\beta}\big)^2\big(m_{u^{(0)}_\alpha}+m_{u^{(0)}_\beta}\big)
\nonumber \\ && \times
\Big(m^2_{(k)}\big(m_{d^{(k)}_\gamma}^2+m_{u^{(0)}_\alpha}m_{u^{(0)}_\beta}\big)-2m_{d^{(k)}_\gamma}m_{(k)}m_{W^{(k)}}^2+m_{W^{(k)}}^4\Big)
(B_0^{11}-B_0^{13})
\nonumber \\ &&
+
m_{u^{(0)}_\alpha}\big(m_{u^{(0)}_\alpha}-m_{u^{(0)}_\beta}\big)\bigg(m_{d^{(k)}_\gamma}^4m_{(k)}^2\big(m_{u^{(0)}_\alpha}+2m_{u^{(0)}_\beta}\big)-2m_{d^{(k)}_\gamma}^3m_{(k)}m_{W^{(k)}}^2 \big(m_{u^{(0)}_\alpha}+2m_{u^{(0)}_\beta}\big)
\nonumber \\ &&
+m_{d^{(k)}_\gamma}^2\Big(m_{W^{(k)}}^4\big(m_{u^{(0)}_\alpha}+2m_{u^{(0)}_\beta}\big)-m_{(k)}^2\Big(m_{u^{(0)}_\beta}\big(m_{u^{(0)}_\alpha}^2+2m_{u^{(0)}_\beta}^2\big)+m_{W^{(k)}}^2\big(m_{u^{(0)}_\alpha}+2m_{u^{(0)}_\beta}\big)\Big)\Big)
\nonumber \\ &&
+2m_{d^{(k)}_\gamma}m_{(k)}m_{W^{(k)}}^2\Big(m_{u^{(0)}_\beta}\big(m_{u^{(0)}_\alpha}^2+m_{u^{(0)}_\beta}^2\big)+m_{W^{(k)}}^2\big(m_{u^{(0)}_\alpha}+2m_{u^{(0)}_\beta}\big)\Big)
\nonumber \\ &&
+\Big(2m_{u^{(0)}_\alpha}m_{u^{(0)}_\beta}^2-m_{W^{(k)}}^2 \big(m_{u^{(0)}_\alpha}+2m_{u^{(0)}_\beta}\big)\Big)\big(m_{(k)}^2m_{u^{(0)}_\alpha}m_{u^{(0)}_\beta}+m_{W^{(k)}}^4\big)\bigg)
(B_0^{13}-B_0^{14})
\nonumber \\ &&
-m_{d^{(k)}_\gamma}m_{u^{(0)}_\alpha}m_{u^{(0)}_\beta}\big(m_{u^{(0)}_\alpha}-m_{u^{(0)}_\beta}\big)^2\big(m_{u^{(0)}_\alpha}+m_{u^{(0)}_\beta}\big)
\Big(m_{d^{(k)}_\gamma}m_{(k)}^2\big(m_{d^{(k)}_\gamma}^2-m_{u^{(0)}_\alpha}^2-m_{u^{(0)}_\alpha}m_{u^{(0)}_\beta}-m_{u^{(0)}_\beta}^2\big)
\nonumber \\ &&
+m_{(k)}m_{W^{(k)}}^2\Big(\big(m_{u^{(0)}_\alpha}+m_{u^{(0)}_\beta}\big)^2-2m_{d^{(k)}_\gamma}^2\Big)+m_{d^{(k)}_\gamma}m_{W^{(k)}}^4\Big)
C_0^6
\nonumber \\ &&
-3\,m_{u^{(0)}_\alpha}m_{u^{(0)}_\beta}m_{W^{(k)}}^2\big(m_{u^{(0)}_\alpha}-m_{u^{(0)}_\beta}\big)^2\big(m_{u^{(0)}_\alpha}+m_{u^{(0)}_\beta}\big)\Big(m_{(k)}^2\big(m_{d^{(k)}_\gamma}^2+m_{u^{(0)}_\alpha}m_{u^{(0)}_\beta}\big)
\nonumber \\ &&
-2m_{d^{(k)}_\gamma}m_{(k)}m_{W^{(k)}}^2+m_{W^{(k)}}^4\Big)
C_0^7
-2\,m_{u^{(0)}_\alpha}m_{u^{(0)}_\beta}\big(m_{u^{(0)}_\alpha}-m_{u^{(0)}_\beta}\big)^2\big(m_{u^{(0)}_\alpha}+m_{u^{(0)}_\beta}\big)
\nonumber \\ && \times
\Big(m_{(k)}^2\big(m_{d^{(k)}_\gamma}^2+m_{u^{(0)}_\alpha}m_{u^{(0)}_\beta}\big)-2m_{d^{(k)}_\gamma}m_{(k)}m_{W^{(k)}}^2+m_{W^{(k)}}^4\Big)
\Big],
\end{eqnarray}
\begin{eqnarray}
&&
\tilde{S}^{(k)\beta\alpha}_{{\rm M},\gamma}
=
\frac{\zeta_{\alpha\beta}\cos^2\theta_{d^{(k)}_\gamma}}{2\,m^2_{W^{(0)}}m^2_{W^{(k)}}}
\Big[
\big(m^2_{d^{(k)}_\gamma}-m^2_{W^{(k)}}\big)\big(m_{u^{(0)}_\alpha}-m_{u^{(0)}_\beta}\big)^2\big(m_{u^{(0)}_\alpha}+m_{u^{(0)}_\beta}\big)
\nonumber \\ &&\times
\Big(m_{(k)}^2\big(m_{d^{(k)}_\gamma}^2+m_{u^{(0)}_\alpha}m_{u^{(0)}_\beta}\big)+2m_{d^{(k)}_\gamma}m_{(k)}m_{W^{(k)}}^2+m_{W^{(k)}}^4\Big)
(B_0^{11}-B_0^{13})
\nonumber \\ &&
+
m_{u^{(0)}_\alpha}\big(m_{u^{(0)}_\alpha}-m_{u^{(0)}_\beta}\big)\bigg(m_{d^{(k)}_\gamma}^4m_{(k)}^2\big(m_{u^{(0)}_\alpha}+2m_{u^{(0)}_\beta}\big)+2m_{d^{(k)}_\gamma}^3m_{(k)}m_{W^{(k)}}^2 \big(m_{u^{(0)}_\alpha}+2m_{u^{(0)}_\beta}\big)
\nonumber \\ &&
+m_{d^{(k)}_\gamma}^2\Big(m_{W^{(k)}}^4\big(m_{u^{(0)}_\alpha}+2m_{u^{(0)}_\beta}\big)-m_{(k)}^2\Big(m_{u^{(0)}_\beta}\big(m_{u^{(0)}_\alpha}^2+2m_{u^{(0)}_\beta}^2\big)+m_{W^{(k)}}^2\big(m_{u^{(0)}_\alpha}+2m_{u^{(0)}_\beta}\big)\Big)\Big)
\nonumber \\ &&
-2m_{d^{(k)}_\gamma}m_{(k)}m_{W^{(k)}}^2 \Big(m_{u^{(0)}_\beta}\big(m_{u^{(0)}_\alpha}^2+m_{u^{(0)}_\beta}^2\big)+m_{W^{(k)}}^2\big(m_{u^{(0)}_\alpha}+2m_{u^{(0)}_\beta}\big)\Big)
\nonumber \\ &&
+\Big(2m_{u^{(0)}_\alpha}m_{u^{(0)}_\beta}^2-m_{W^{(k)}}^2\big(m_{u^{(0)}_\alpha}+2m_{u^{(0)}_\beta}\big)\Big)\big(m_{(k)}^2m_{u^{(0)}_\alpha}m_{u^{(0)}_\beta}+m_{W^{(k)}}^4\big)\bigg)
(B_0^{13}-B_0^{14})
\nonumber \\ &&
+
m_{d^{(k)}_\gamma}m_{u^{(0)}_\alpha}m_{u^{(0)}_\beta}\big(m_{u^{(0)}_\alpha}-m_{u^{(0)}_\beta}\big)^2\big(m_{u^{(0)}_\alpha}+m_{u^{(0)}_\beta}\big)\Big(m_{d^{(k)}_\gamma}m_{(k)}^2 \big(-m_{d^{(k)}_\gamma}^2+m_{u^{(0)}_\alpha}^2
\nonumber \\ &&
+m_{u^{(0)}_\alpha}m_{u^{(0)}_\beta}+m_{u^{(0)}_\beta}^2\big)+m_{(k)}m_{W^{(k)}}^2\Big(\big(m_{u^{(0)}_\alpha}+m_{u^{(0)}_\beta}\big)^2-2m_{d^{(k)}_\gamma}^2\Big)-m_{d^{(k)}_\gamma}m_{W^{(k)}}^4\Big)
C_0^6
\nonumber \\ &&
-3\,m_{u^{(0)}_\alpha}m_{u^{(0)}_\beta}m_{W^{(k)}}^2\big(m_{u^{(0)}_\alpha}-m_{u^{(0)}_\beta}\big)^2 \big(m_{u^{(0)}_\alpha}+m_{u^{(0)}_\beta}\big) \Big(m_{(k)}^2\big(m_{d^{(k)}_\gamma}^2+m_{u^{(0)}_\alpha}m_{u^{(0)}_\beta}\big)
\nonumber \\ &&
+2m_{d^{(k)}_\gamma}m_{(k)}m_{W^{(k)}}^2+m_{W^{(k)}}^4\Big)
C_0^7
-2\,m_{u^{(0)}_\alpha}m_{u^{(0)}_\beta}\big(m_{u^{(0)}_\alpha}-m_{u^{(0)}_\beta}\big)^2\big(m_{u^{(0)}_\alpha}+m_{u^{(0)}_\beta}\big)
\nonumber \\ && \times
\Big(m_{(k)}^2\big(m_{d^{(k)}_\gamma}^2+m_{u^{(0)}_\alpha}m_{u^{(0)}_\beta}\big)+2m_{d^{(k)}_\gamma}m_{(k)}m_{W^{(k)}}^2+m_{W^{(k)}}^4\Big)
\Big],
\end{eqnarray}
\begin{eqnarray}
&&
G^{(k)\beta\alpha}_{{\rm E},\gamma}
=
\zeta_{\alpha\beta}\Big[
-\big(m^2_{d^{(k)}_\gamma}-m^2_{W^{(k)}}\big)\big(m_{u^{(0)}_\alpha}-m_{u^{(0)}_\beta}\big)\big(m_{u^{(0)}_\alpha}+m_{u^{(0)}_\beta}\big)^2(B_0^{11}-B_0^{13})
\nonumber \\ &&
-
m_{u^{(0)}_\alpha}\big(m_{u^{(0)}_\alpha}+m_{u^{(0)}_\beta}\big)
\Big(
m_{d^{(k)}_\gamma}^2\big(m_{u^{(0)}_\alpha}-2m_{u^{(0)}_\beta}\big)-2m_{u^{(0)}_\beta}\big(m_{u^{(0)}_\alpha}^2-3m_{u^{(0)}_\alpha}m_{u^{(0)}_\beta}+m_{u^{(0)}_\beta}^2\big)
\nonumber \\ &&
-m_{W^{(k)}}^2\big(m_{u^{(0)}_\alpha}-2m_{u^{(0)}_\beta}\big)
\Big)
(B_0^{13}-B_0^{14})
-m^2_{d^{(k)}_\gamma}m_{u^{(0)}_\alpha}m_{u^{(0)}_\beta}\big(m_{u^{(0)}_\alpha}-m_{u^{(0)}_\beta}\big)\big(m_{u^{(0)}_\alpha}+m_{u^{(0)}_\beta}\big)^2
C_0^6
\nonumber \\ &&
+
3m_{u^{(0)}_\alpha}m_{u^{(0)}_\beta}\big(m_{u^{(0)}_\alpha}-m_{u^{(0)}_\beta}\big)\big(m_{u^{(0)}_\alpha}+m_{u^{(0)}_\beta}\big)^2 \Big(\big(m_{u^{(0)}_\alpha}-m_{u^{(0)}_\beta}\big)^2-m_{W^{(k)}}^2\Big)
C_0^7
\nonumber \\ &&
-2m_{u^{(0)}_\alpha}m_{u^{(0)}_\beta}\big(m_{u^{(0)}_\alpha}-m_{u^{(0)}_\beta}\big)\big(m_{u^{(0)}_\alpha}+m_{u^{(0)}_\beta}\big)^2
\Big],
\end{eqnarray}
\begin{eqnarray}
&&
P^{(k)\beta\alpha}_{{\rm E},\gamma}
=
\frac{\zeta_{\alpha\beta}}{2\,m_{W^{(k)}}^2}
\Big[
\big(m^2_{d^{(k)}_\gamma}-m^2_{W^{(k)}}\big)\big(m_{u^{(0)}_\alpha}-m_{u^{(0)}_\beta}\big)\big(m_{u^{(0)}_\alpha}+m_{u^{(0)}_\beta}\big)^2\big(m_{u^{(0)}_\alpha}m_{u^{(0)}_\beta}-m_{d^{(k)}_\gamma}^2\big)
(B_0^{11}-B_0^{13})
\nonumber \\ &&
-m_{u^{(0)}_\alpha}\big(m_{u^{(0)}_\alpha}+m_{u^{(0)}_\beta}\big)\Big(m_{d^{(k)}_\gamma}^4m_{u^{(0)}_\alpha}-2m_{d^{(k)}_\gamma}^4m_{u^{(0)}_\beta}+m_{d^{(k)}_\gamma}^2m_{u^{(0)}_\alpha}^2m_{u^{(0)}_\beta}
\nonumber \\ &&
+m_{W^{(k)}}^2\big(m_{u^{(0)}_\alpha}-2m_{u^{(0)}_\beta}\big)\big(m_{u^{(0)}_\alpha}m_{u^{(0)}_\beta}-m_{d^{(k)}_\gamma}^2\big)+2m_{d^{(k)}_\gamma}^2m_{u^{(0)}_\beta}^3-2m_{u^{(0)}_\alpha}^2m_{u^{(0)}_\beta}^3\Big)
(B_0^{13}-B_0^{14})
\nonumber \\ &&
+
m_{d^{(k)}_\gamma}^2m_{u^{(0)}_\alpha}m_{u^{(0)}_\beta}\big(m_{u^{(0)}_\alpha}-m_{u^{(0)}_\beta}\big)\big(m_{u^{(0)}_\alpha}+m_{u^{(0)}_\beta}\big)^2\big(-m_{d^{(k)}_\gamma}^2+m_{u^{(0)}_\alpha}^2-m_{u^{(0)}_\alpha}m_{u^{(0)}_\beta}+m_{u^{(0)}_\beta}^2\big)
C_0^6
\nonumber \\ &&
+
3\,m_{u^{(0)}_\alpha}m_{u^{(0)}_\beta}m_{W^{(k)}}^2\big(m_{u^{(0)}_\alpha}-m_{u^{(0)}_\beta}\big)\big(m_{u^{(0)}_\alpha}+m_{u^{(0)}_\beta}\big)^2\big(m_{u^{(0)}_\alpha}m_{u^{(0)}_\beta}-m_{d^{(k)}_\gamma}^2\big)
C_0^7
\nonumber \\ &&
+
2\,m_{u^{(0)}_\alpha}m_{u^{(0)}_\beta}\big(m_{u^{(0)}_\alpha}-m_{u^{(0)}_\beta}\big)\big(m_{u^{(0)}_\alpha}+m_{u^{(0)}_\beta}\big)^2 \big(m_{u^{(0)}_\alpha}m_{u^{(0)}_\beta}-m_{d^{(k)}_\gamma}^2\big)
\Big],
\end{eqnarray}
\begin{eqnarray}
&&
\hat{S}^{(k)\beta\alpha}_{{\rm E},\gamma}
=\frac{\zeta_{\alpha\beta}\sin^2\theta_{d^{(k)}_\gamma}}{2\,m_{W^{(0)}}^2m_{W^{(k)}}^2}
\Big[
\big(m^2_{d^{(k)}_\gamma}-m^2_{W^{(k)}}\big)\big(m_{u^{(0)}_\alpha}-m_{u^{(0)}_\beta}\big)\big(m_{u^{(0)}_\alpha}+m_{u^{(0)}_\beta}\big)^2
\nonumber \\ && \times
\Big(-m_{d^{(k)}_\gamma}^2m_{(k)}^2+2m_{d^{(k)}_\gamma}m_{(k)}m_{W^{(k)}}^2+m_{(k)}^2m_{u^{(0)}_\alpha}m_{u^{(0)}_\beta}-m_{W^{(k)}}^4\Big)
(B_0^{11}-B_0^{13})
\nonumber \\ &&
+
m_{u^{(0)}_\alpha}\big(m_{u^{(0)}_\alpha}+m_{u^{(0)}_\beta}\big)\bigg(-m_{(k)}^2\Big(m_{d^{(k)}_\gamma}^4\big(m_{u^{(0)}_\alpha}-2m_{u^{(0)}_\beta}\big)
+m_{d^{(k)}_\gamma}^2m_{u^{(0)}_\beta}\big(m_{u^{(0)}_\alpha}^2+2m_{u^{(0)}_\beta}^2\big)
\nonumber \\ &&
-2m_{u^{(0)}_\alpha}^2m_{u^{(0)}_\beta}^3\Big)+m_{(k)}m_{W^{(k)}}^2\Big(2m_{d^{(k)}_\gamma}^3 \big(m_{u^{(0)}_\alpha}-2m_{u^{(0)}_\beta}\big)+m_{d^{(k)}_\gamma}^2m_{(k)}\big(m_{u^{(0)}_\alpha}-2m_{u^{(0)}_\beta}\big)
\nonumber \\ &&
+2m_{d^{(k)}_\gamma}m_{u^{(0)}_\beta}\big(m_{u^{(0)}_\alpha}^2+m_{u^{(0)}_\beta}^2\big)-m_{(k)}m_{u^{(0)}_\alpha}m_{u^{(0)}_\beta}\big(m_{u^{(0)}_\alpha}-2m_{u^{(0)}_\beta}\big)\Big)
\nonumber \\ &&
-m_{W^{(k)}}^4\Big(m_{d^{(k)}_\gamma}m_{u^{(0)}_\alpha}\big(m_{d^{(k)}_\gamma}+2m_{(k)}\big)-2m_{d^{(k)}_\gamma}m_{u^{(0)}_\beta}\big(m_{d^{(k)}_\gamma}+2m_{(k)}\big)+2m_{u^{(0)}_\alpha}m_{u^{(0)}_\beta}^2\Big)
\nonumber \\ &&
+m_{W^{(k)}}^6\big(m_{u^{(0)}_\alpha}-2m_{u^{(0)}_\beta}\big)\bigg)
(B_0^{13}-B_0^{14})
-m_{d^{(k)}_\gamma}m_{u^{(0)}_\alpha}m_{u^{(0)}_\beta}\big(m_{u^{(0)}_\alpha}-m_{u^{(0)}_\beta}\big)\big(m_{u^{(0)}_\alpha}+m_{u^{(0)}_\beta}\big)^2
\nonumber \\ && \times
\Big(m_{d^{(k)}_\gamma}m_{(k)}^2\big(m_{d^{(k)}_\gamma}^2-m_{u^{(0)}_\alpha}^2+m_{u^{(0)}_\alpha}m_{u^{(0)}_\beta}-m_{u^{(0)}_\beta}^2\big)+m_{(k)}m_{W^{(k)}}^2\Big(\big(m_{u^{(0)}_\alpha}-m_{u^{(0)}_\beta}\big)^2-2m_{d^{(k)}_\gamma}^2\Big)
\nonumber \\ &&
+m_{d^{(k)}_\gamma}m_{W^{(k)}}^4\Big)
C_0^6
-3\,m_{u^{(0)}_\alpha}m_{u^{(0)}_\beta}m_{W^{(k)}}^2\big(m_{u^{(0)}_\alpha}-m_{u^{(0)}_\beta}\big)\big(m_{u^{(0)}_\alpha}+m_{u^{(0)}_\beta}\big)^2
\nonumber \\ && \times
\Big(m_{(k)}^2\big(m_{d^{(k)}_\gamma}^2-m_{u^{(0)}_\alpha}m_{u^{(0)}_\beta}\big)-2m_{d^{(k)}_\gamma}m_{(k)}m_{W^{(k)}}^2+m_{W^{(k)}}^4\Big)
C_0^7
\nonumber \\ &&
+
2\,m_{u^{(0)}_\alpha}m_{u^{(0)}_\beta}\big(m_{u^{(0)}_\alpha}-m_{u^{(0)}_\beta}\big)\big(m_{u^{(0)}_\alpha}+m_{u^{(0)}_\beta}\big)^2\big(-m_{d^{(k)}_\gamma}^2m_{(k)}^2
\nonumber \\ &&
+2m_{d^{(k)}_\gamma}m_{(k)}m_{W^{(k)}}^2+m_{(k)}^2m_{u^{(0)}_\alpha}m_{u^{(0)}_\beta}-m_{W^{(k)}}^4\big)
\Big],
\end{eqnarray}
\begin{eqnarray}
&&
\tilde{S}^{(k)\beta\alpha}_{{\rm E},\gamma}
=
\frac{\zeta_{\alpha\beta}\cos^2\theta_{d^{(k)}_\gamma}}{2\,m^2_{W^{(0)}}m^2_{W^{(k)}}}
\Big[
-\big(m^2_{d^{(k)}_\gamma}-m^2_{W^{(k)}}\big)\big(m_{u^{(0)}_\alpha}-m_{u^{(0)}_\beta}\big)(m_{u^{(0)}_\alpha}+m_{u^{(0)}_\beta})^2
\nonumber \\ && \times
\Big(m_{(k)}^2\big(m_{d^{(k)}_\gamma}^2-m_{u^{(0)}_\alpha}m_{u^{(0)}_\beta}\big)+2m_{d^{(k)}_\gamma}m_{(k)}m_{W^{(k)}}^2+m_{W^{(k)}}^4\Big)
(B_0^{11}-B_0^{13})
\nonumber \\ &&
-m_{u^{(0)}_\alpha} \big(m_{u^{(0)}_\alpha}+m_{u^{(0)}_\beta}\big) \bigg(m_{(k)}^2\Big(m_{d^{(k)}_\gamma}^4\big(m_{u^{(0)}_\alpha}-2m_{u^{(0)}_\beta}\big)+m_{d^{(k)}_\gamma}^2m_{u^{(0)}_\beta}\big(m_{u^{(0)}_\alpha}^2+2m_{u^{(0)}_\beta}^2\big)
\nonumber \\ &&
-2m_{u^{(0)}_\alpha}^2m_{u^{(0)}_\beta}^3\Big)+m_{(k)}m_{W^{(k)}}^2\Big(2m_{d^{(k)}_\gamma}^3\big(m_{u^{(0)}_\alpha}-2m_{u^{(0)}_\beta}\big)-m_{d^{(k)}_\gamma}^2m_{(k)}\big(m_{u^{(0)}_\alpha}-2m_{u^{(0)}_\beta}\big)
\nonumber \\ &&
+2m_{d^{(k)}_\gamma}m_{u^{(0)}_\beta}\big(m_{u^{(0)}_\alpha}^2+m_{u^{(0)}_\beta}^2\big)+m_{(k)}m_{u^{(0)}_\alpha}m_{u^{(0)}_\beta} \big(m_{u^{(0)}_\alpha}-2m_{u^{(0)}_\beta}\big)\Big)
\nonumber \\ &&
+m_{W^{(k)}}^4\Big(m_{d^{(k)}_\gamma}m_{u^{(0)}_\alpha}\big(m_{d^{(k)}_\gamma}-2m_{(k)}\big)-2m_{d^{(k)}_\gamma}m_{u^{(0)}_\beta}\big(m_{d^{(k)}_\gamma}-2m_{(k)}\big)+2m_{u^{(0)}_\alpha}m_{u^{(0)}_\beta}^2\Big)
\nonumber \\ &&
-m_{(k)}^6\big(m_{u^{(0)}_\alpha}-2m_{u^{(0)}_\beta}\big)\bigg)
(B_0^{13}-B_0^{14})
+
m_{d^{(k)}_\gamma}m_{u^{(0)}_\alpha}m_{u^{(0)}_\beta} \big(m_{u^{(0)}_\alpha}-m_{u^{(0)}_\beta}\big)\big(m_{u^{(0)}_\alpha}+m_{u^{(0)}_\beta}\big)^2
\nonumber \\ &&\times
\Big(m_{d^{(k)}_\gamma}m_{(k)}^2 \big(-m_{d^{(k)}_\gamma}^2+m_{u^{(0)}_\alpha}^2-m_{u^{(0)}_\alpha}m_{u^{(0)}_\beta}+m_{u^{(0)}_\beta}^2\big)
+m_{(k)}m_{W^{(k)}}^2\Big(\big(m_{u^{(0)}_\alpha}-m_{u^{(0)}_\beta}\big)^2
\nonumber \\ &&
-2m_{d^{(k)}_\gamma}^2\Big)-m_{d^{(k)}_\gamma}m_{(k)}^4\Big)
C_0^6
-3\,m_{u^{(0)}_\alpha}m_{u^{(0)}_\beta}m_{W^{(k)}}^2 \big(m_{u^{(0)}_\alpha}-m_{u^{(0)}_\beta}\big)\big(m_{u^{(0)}_\alpha}+m_{u^{(0)}_\beta}\big)^2
\nonumber \\ && \times
\Big(m_{(k)}^2\big(m_{d^{(k)}_\gamma}^2-m_{u^{(0)}_\alpha}m_{u^{(0)}_\beta}\big)+2m_{d^{(k)}_\gamma}m_{(k)}m_{W^{(k)}}^2+m_{W^{(k)}}^4\Big)
C_0^7
\nonumber \\ &&
-2\,m_{u^{(0)}_\alpha}m_{u^{(0)}_\beta}\big(m_{u^{(0)}_\alpha}-m_{u^{(0)}_\beta}\big)\big(m_{u^{(0)}_\alpha}+m_{u^{(0)}_\beta}\big)^2 
\nonumber \\ && \times
\Big(m_{(k)}^2\big(m_{d^{(k)}_\gamma}^2-m_{u^{(0)}_\alpha}m_{u^{(0)}_\beta}\big)+2m_{d^{(k)}_\gamma}m_{(k)}m_{W^{(k)}}^2+m_{W^{(k)}}^4\Big)
\Big].
\end{eqnarray}

\end{widetext}

\section*{References}


\end{document}